\documentclass{emulateapj}
\usepackage{subfigure}
\usepackage{natbib}
\usepackage{url}
\begin{document}

\shorttitle{Properties and Evolution of IRDC Clumps}
\shortauthors{Battersby et al.}

\def\Msun{\hbox{M$_{\odot}$}}
\def\Lsun{\hbox{L$_{\odot}$}}
\def\kms{km~s$^{\rm -1}$}
\def\hcop{HCO$^{+}$}
\def\n2hp{N$_{2}$H$^{+}$}
\def\micron{$\mu$m}
\def\13CO{$^{13}$CO}
\def\etamb{$\eta_{\rm mb}$}
\def\Inu{I$_{\nu}$}
\def\kapnu{$\kappa _{\nu}$}
\def\ffore{f$_{\rm{fore}}$}
\def\tastar{T$_{A}^{*}$}

\input epsf

\title{An Infrared through Radio Study of the Properties and Evolution of
  IRDC Clumps}

\author{Cara Battersby\altaffilmark{1,2}, John Bally\altaffilmark{1},
  James M. Jackson\altaffilmark{2}, Adam Ginsburg\altaffilmark{1}, 
  Yancy L. Shirley\altaffilmark{3},
  Wayne Schlingman\altaffilmark{3}, Jason Glenn\altaffilmark{1}}

\altaffiltext{1}{Center for Astrophysics and Space Astronomy --
University of Colorado Boulder}
\altaffiltext{2}{Institute for Astrophysical Research --
Boston University}
\altaffiltext{3}{Steward Observatory --
University of Arizona}

\begin{abstract}
We examine the physical properties and evolutionary stages of 
a sample of 17 clumps within 8 Infrared Dark Clouds (IRDCs) by combining
existing infrared, millimeter, and radio data with 
new Bolocam Galactic Plane Survey (BGPS) 1.1 mm data, VLA radio continuum
data, and HHT dense gas (\hcop~and \n2hp) spectroscopic data.  
We combine literature studies of star formation tracers and dust
temperatures within IRDCs with our search for 
ultra-compact (UC) H~II regions to discuss a possible evolutionary sequence for
IRDC clumps.  In addition, we perform an analysis of
mass tracers in IRDCs and find that 8~\micron~extinction masses and 1.1 mm 
Bolocam Galactic Plane Survey (BGPS) masses are complementary
mass tracers in IRDCs except for the most active clumps
(notably those containing UCH~II regions), for which both mass tracers
suffer biases.  We find that the measured virial masses in IRDC clumps 
are uniformly higher than the measured dust continuum masses on the scale
of $\sim$ 1 pc.
We use \13CO, \hcop, and \n2hp to study the molecular gas
properties of IRDCs and do not see any evidence of chemical differentiation
between hot and cold clumps on the scale of $\sim$ 1 pc.  However, both
\hcop~and \n2hp~are brighter in active clumps, due to an increase in
temperature and/or density.  We report the identification of four UCH~II 
regions embedded
within IRDC clumps and find that UCH~II regions are associated with 
bright ($\gtrsim$ 1 Jy) 24~\micron~point sources, and that the brightest 
UCH~II regions are associated with ``diffuse red clumps'' (an extended
enhancement at 8~\micron). 
The broad stages of the discussed evolutionary sequence
(from a quiescent clump to an embedded H~II region) are supported by
literature dust temperature estimates; however, no sequential nature can be
inferred between the individual star formation tracers. 
\end{abstract}


\section{Introduction}

Despite their profound impact on the galactic environment and enrichment, 
the protostellar evolution (and birth) of massive stars (M $>$ 8~\Msun) remains 
elusive.  Direct observations of the earliest evolutionary stages of
massive stars is a challenging task.  Massive stars evolve more 
quickly, are more rare, and consequently farther away on average than 
their low-mass counterparts.
In addition, massive stars, once formed,
quickly heat and ionize their environment, disrupting their natal
molecular cloud.  
Much like their low-mass counterparts, massive stars are thought to form in
cold, dense molecular clouds.  Massive
stars seem to form predominantly in clusters \citep{lad03, dew05} and therefore, their
natal molecular cloud should be more massive and denser than the natal
clouds of isolated low-mass stars.  Infrared Dark Clouds (IRDCs) have been
proposed to be the progenitors of massive stars and their host clusters
\citep{ega98, car98, rat06}.

IRDCs were identified as dark extinction features
against the bright mid-infrared galactic plane with the MSX and ISO space 
satellites \citep{ega98, per96, omo03}.  IRDCs were
catalogued from the MSX 8~\micron~data by \citet{sim06} and from the GLIMPSE
8~\micron~data more reccently by \citet{per09}.  
The strong continuum extinction and lack of infrared emission 
by these clouds belie their cold temperatures and high column
densitites.  Previous studies have confirmed that these clouds are indeed 
cold (T $<$ 25 K) and dense (n$_{H} >~$10$^{5}$ cm$^{-3}$) \citep{ega98, car98, car00}.
More recently, studies have found kinematic distances to 
these clouds and determined that their masses lie in the range of 10$^{2}$ to 
10$^{4}$ \Msun~\citep{rat06}.  \citet{rag09} find a clump mass spectra
(over the range of clump masses from 30 to 3000 \Msun) in
IRDCs with a slope of about 1.76 $\pm$ 0.05.

There is compelling evidence to suggest that IRDCs are the cold, dense
precursors to stellar clusters, and thus massive stars.  Millimeter 
continuum maps show extended cold dust emission, matching the morphology
of the IRDCs and surrounding compact clumps\footnote[1]{In this paper, we use the
term ``core'' to describe a small, dense
object within an IRDC clump, with a size of order 0.05 pc which will likely
form a single star or stellar system,
``clump'' to describe
a dense object within an IRDC with a size of order $\sim$ 1 pc
and a mass 10$^{2}$ - 10$^{3}$ \Msun, and ``cloud'' for an IRDC
which has a size scale of order 10 pc, and a mass of order 10$^{3}$ -
10$^{4}$ \Msun.}.  These clumps fall 
into two categories: active and quiescent.  The active clumps are
characterized by some or all of the following: 
1) luminous, embedded 24 \micron~emission, 2) ``green fuzzies," \citep{cha09} 
an extended 4.5 \micron~enhancement believed to be caused by shocked H$_{2}$, thus 
indicative of outflows 
\citep[also known as Extended Green Objects, EGOs][]{cyg08},
3) maser emission, and 4) ultra-compact (UC) H~II
regions.  A quiescent clump is characterized as a cold, dense millimeter
peak with none of the aforementioned signs of active star formation.  
It is believed that the active 
clumps are currently forming stars  
while the quiescent clumps are at an earlier stage of star formation: 
the cold, dense precursor before the star has ``turned on."

IRDCs are the densest clumps and filaments embedded within Giant Molecular Clouds
\citep[GMCs:][]{sim06b} with the favorable viewing condition of being on
the near-side of a bright mid-IR background.  They are preferentially found 
toward the Galaxy's largest star-forming region, the 5 kpc
ring.  IRDCs have typical sizes D $\sim$ 1-10 pc, densities n$_{H}$
$\gtrsim$ 10$^{4}$ cm$^{-3}$, and masses M $\sim$ 10$^{2}$-10$^{4}$ \Msun 
\citep{rat06}.
IRDCs are dense condesations embedded within GMCs (D
$\sim$ 50 pc, n $\sim$ 10$^{2}$ cm$^{-3}$, and M $\sim$ 10$^{5}$-10$^{6}$
\Msun) and are typically larger and more massive than Bok Globules
\citep[D $\sim$
0.1-2 pc, n $\sim$ 10$^{3}$-10$^{4}$ cm$^{-3}$, and M $\sim$ 1-100
\Msun;][]{rat06}.
Many have argued
that stars can form
anywhere in a GMC, but that the confining pressure and column density of an IRDC are
required for the formation of clusters and massive stars 
\citep[e.g.][]{mck03}.  The formation
mechanism of an IRDC within a GMC is not well understood.  Perhaps it is
simply the gravitational collapse of a high-density perturbation in a GMC
left on its own for many free-fall times.  Alternatively, IRDCs may be
the result of triggered collapse in a high-pressure environment, such as a
cloud-cloud collision, or the dynamic force from a bubble, such as an H~II
region, wind, supernova, or superbubble.  Converging flows are another 
possible trigger for the collapse of an IRDC from part of a GMC.
An IRDC, once
formed, has the properties of cluster-forming clumps \citep{lad03}
with the exception that IRDCs are much colder (10 - 20 K).
Do these IRDC clumps fragment into cores which evolve in isolation
or do the resulting cores undergo competitive accretion?  This is a
central question in massive star formation \citep{zin07, mck07}.   

This paper explores a sample of 17 IRDC clumps embedded within 8
IRDCs.  We compare various mass tracers in IRDCs, distinguish active and
quiescent clumps, and discuss a possible
evolutionary sequence.
This paper is organized as follows.  In \S \ref{sec:obs}, we describe the 
source selection and observations.  In \S \ref{sec:mass}, we present mass
estimation techniques, their results, and an analyis of the systematic
errors involved.  These masses are compared in \S \ref{sec:masscomp}.
In \S \ref{sec:dust} we compare the 8~\micron~dust extinction with the BGPS 1.1
mm dust emission.  In \S \ref{sec:virial} we address whether 
or not the IRDC clumps are bound, and in \S
\ref{sec:mol} we compare the molecular line tracers.
In \S \ref{sec:hiiregions}, we discuss the radio continuum sources and 
stellar type limits for the UCH~II regions found and in \S
\ref{sec:activity} we discuss star formation activity in the clumps.  
In \S \ref{sec:evseq}, we discuss a possible evolutionary sequence for
IRDC clumps.
\S \ref{sec:conclusion} is a summary of our conclusions.
Appendix A includes a discussion of each IRDC as well as images of each
source with all the data used in the paper.

%%%%%%%%%%%%%%%%%%%%%%%%%%%%%%%%%%%%%%%%%%%%%%%%%%%%%%%%%%%%%%%%%%%%%%%%%%%%%%%%%%%%%%%%%%%%%%%
% IRDC LOCATIONS
%%%%%%%%%%%%%%%%%%%%%%%%%%%%%%%%%%%%%%%%%%%%%%%%%%%%%%%%%%%%%%%%%%%%%%%%%%%%%%%%%%%%%%%%%%%%%%%
%\newpage
%\clearpage
\begin{deluxetable*}{lccccccccccccccccc}
\tabletypesize{\scriptsize}
\tablecaption{IRDC Clump Positions \label{table-names}}
\tablewidth{0pt}
%\rotate
\tablehead{
\colhead{IRDC Name\tablenotemark{a}} & 
\colhead{Glimpse} &
\colhead{Bolocat\tablenotemark{b}} & 
\colhead{Closest MAMBO\tablenotemark{c}} &
\colhead{R.A.} & \colhead{DEC.}\\
\colhead{}  & \colhead{Clump Name} & \colhead{Clump Name} &
\colhead{Clump Name} &
\colhead{(J2000)} & \colhead{(J2000)}}
\startdata
MSXDC G022.35+00.41   & GLM1  & 2860 &  MM1  &  18:30:24.2  &  -09:10:38.9  \\
                      & GLM2  & 2858 &  MM3  &  18:30:37.6  &  -09:12:54.1  \\
MSXDC G023.60+00.00   & GLM1  & 3125 &  MM2  &  18:34:21.3  &  -08:18:07.5  \\
                      & GLM2  & 3132 &  MM7  &  18:34:20.6  &  -08:17:21.7  \\
MSXDC G024.33+00.11   & GLM1  & 3284 &  MM1  &  18:35:08.1  &  -07:35:01.9  \\
                      & GLM2  & 3282 &  MM4  &  18:35:19.1  &  -07:37:19.7  \\
MSXDC G024.60+00.08   & GLM1  & 3382 &  MM2  &  18:35:35.8  &  -07:18:22.2  \\
                      & GLM2  & 3383 &  MM1  &  18:35:39.9  &  -07:18:46.0  \\
MSXDC G028.23-00.19   & GLM1  & 3923 &  MM1  &  18:43:30.3  &  -04:13:03.7  \\
MSXDC G028.37+00.07   & GLM1  & 3946 &  MM14 &  18:42:53.4  &  -04:02:23.8  \\
                      & GLM2  & 3939 &  MM4  &  18:42:50.6  &  -04:03:16.9  \\
                      & GLM3  & 3942 &  MM6  &  18:42:48.9  &  -04:02:05.4  \\
                      & GLM4  & 3955 &  MM1  &  18:42:52.5  &  -03:59:48.1  \\
MSXDC G028.53-00.25   & GLM1  & 3998 &  MM2  &  18:44:16.1  &  -04:00:09.9  \\
MSXDC G034.43+00.24   & GLM1  & 5373 &  MM1  &  18:53:18.0  &  +01:25:24.9  \\
                      & GLM2  & 5373 &  MM2  &  18:53:18.4  &  +01:24:51.3  \\
                      & GLM3  & 5385 &  MM3  &  18:53:19.6  &  +01:28:24.6  \\
\enddata
\tablenotetext{a}{from the MSX dark cloud catalog by \citet{sim06}.  
  For the remainder of the paper we drop the MSXDC prefix.}
\tablenotetext{b}{BGPS 1.1 mm clump source catalog, Bolocat, number \citep{ros10}}
\tablenotetext{c}{from \citet{rat06}}
\end{deluxetable*}

\section{Source Selection and Data}
\label{sec:obs}

\subsection{Source Selection}
In this paper, we investigate a sample of 17 IRDC clumps embedded
within 8 IRDCs.  These were selected from the MSX dark cloud catalog 
\citep{sim06},
and from the previous work of \citet{rat06}, who selected the 38
highest contrast IRDCs from the dark cloud catalog with known
kinematic distances.  These 38 IRDCs, therefore, are among the darkest and
densest in the
catalog by \citet{sim06}.  From this subset, and based upon the work by
\citet{cha09}, we selected mostly active clumps, with a few
quiescent clumps for comparison.  Our sample consists of 9 active, 4
intermediate, and 4 quiescent clumps.
We mostly selected clumps with signs of active star formation, motivated by our UCH~II
region survey, as this would maximize our detection rate.
We included a few quiescent clumps in our sample for comparison, although 
no UCH~II regions were expected in the quiescent clumps.  
Our sample of IRDC clumps is a subset of the darkest, most active IRDC
clumps in the First Galactic Quadrant (0 $\leq$ l $\leq$ 90 $^{o}$).  
In a sample of 106 IRDC clumps, \citet{cha09} find that 
65\% are quiescent and 35\% are active clumps.
We quantify the star-forming activity
by four criteria: 1) an embedded 24 \micron~point source, 
2) ``green fuzzies," \citep{cha09} 
an extended 4.5 \micron~enhancement believed to be caused by shocked H$_{2}$, thus 
indicative of outflows 
\citep[also known as Extended Green Objects, EGOs][]{cyg08}, 
3) H$_{2}$O or CH$_{3}$OH maser emission, and 4)
Ultra-compact (UC) H~II regions.  We assign the designation ``active'' to
clumps which have three or four of these signs of star formation, which
means that each ``active'' clump has either an H~II region or at least two
outflow tracers and a 24 \micron~point source, and is
actively forming stars.  The designation ``intermediate'' is for clumps 
which exhibit one or two signs of active star formation.  These clumps,
therefore, may have only shock/outflow signatures or a 24 \micron~point
source.  We reserve the designation ``quiescent'' for clumps with no signs
of active star formation.  

Selection of the clump positions from the sample discussed above required
simply that there be an IR-dark feature, which can be near but not
coincident with an IR-bright region,
which has corresponding millimeter emission.  The exact
clump positions were selected by eye based on the above criteria and the
overlap with existing data.  While the exact clump positions were not
selected by stringent threshholds, visual inspection of the images verifies
that the sample accurately represents the IRDCs.
The clump positions and the closest millimeter
clump, as identified by \citet{rat06}, are given in Table \ref{table-names}.

%%%%%%%%%%%%%%%%%%%%%%%%%%%%%%%%%%%%%%%%%%%%%%%%%%%%%%%%%%%%%%%%%%%%%%
% Observing Summary Table
%%%%%%%%%%%%%%%%%%%%%%%%%%%%%%%%%%%%%%%%%%%%%%%%%%%%%%%%%%%%%%%%%%%%%%

%\clearpage
\begin{deluxetable*}{cccc}
\tabletypesize{\scriptsize}
\tablecaption{Summary of Observations \label{table-obs}}
\tablewidth{0pt}
\tablehead{
\colhead{Data} & \colhead{$\lambda$}      & \colhead{Beam FWHM}      &
\colhead{1$\sigma$ Sensitivity}}
\startdata
GLIMPSE & 3.6 \micron                 & 1.7"        & 0.3 MJy/Sr\tablenotemark{a}  \\
GLIMPSE & 4.5 \micron                 & 1.7"        & 0.3 MJy/Sr                   \\
GLIMPSE & 5.8 \micron                 & 1.9"        & 0.7 MJy/Sr                    \\
GLIMPSE & 8.0 \micron                 & 1.9"        & 0.6 MJy/Sr                    \\
MIPSGAL & 24 \micron                  & 6"          & 0.67 mJy                    \\
HHT     & 1.12 mm\tablenotemark{b}    & 28"         & 0.04 K                      \\ 
HHT     & 1.07 mm\tablenotemark{b}    & 27"         & 0.04 K                      \\
GRS     & 2.86 mm\tablenotemark{b}    & 46"         & 0.27 K                      \\
VLA     & 3.6 cm                      & 2.8 x 2.4"  & 0.05 mJy/beam               \\
IRAM    & 1.2 mm                      & 11"         & 10 mJy/beam                 \\
BPGS    & 1.1 mm                      & 33"         & 30 mJy/beam                 \\

\enddata

\tablenotetext{a}{GLIMPSE raw 1$\sigma$ surface brightness 
  sensitivities from \citet{rea06}}
\tablenotetext{b}{~\hcop~J = 3-2, ~\n2hp~J = 3-2, and ~\13CO J = 1-0
  respectively.  For \hcop~and \n2hp, 0.04 K is the 1$\sigma$ sensitivity per 1.1
  \kms~channel and for \13CO, 0.27 K is the 1$\sigma$ sensitivity per 0.21
  \kms~channel}

\end{deluxetable*}

\subsection{Observations \& Data Description}
For this study, we examine a sample of 17 IRDC clumps within 8
IRDCs.
We utilize existing and new data to determine the physical 
properties and evolutionary stages of these clumps.  
A summary of the data used is found in Table \ref{table-obs}.

\subsubsection{Mid-Infrared Data}
We utilize mid-IR data taken as part of the Galactic Legacy Infrared
Mid-Plane Survey Extraordinaire \citep[GLIMPSE;][]{ben03} and 24
\micron~data taken as part of the MIPSGAL survey \citep{car09}.
Extinction at  8~\micron~gives 
an independent mass estimate, the 4.5 \micron~band
determines the presence of ``green fuzzies,'' and embedded stellar sources
appear at 8 and 24 \micron.
We utilize the ``green
fuzzy'' (extended 4.5 \micron~emission, indicative of shocks) catalog
by \citet{cha09}, as well as their 24 \micron~point source 
identification and H$_{2}$O and CH$_{3}$OH maser survey.

\subsubsection{Millimeter Continuum Dust Emission}
The millimeter continuum data traces the cold dust within the IRDCs and gives
cloud and clump mass estimates, as well as the locations of compact 
clumps within the IRDCs.  We 
utilize 1.2 mm data obtained at the Institut de Radioastronomie Millimetrique
(IRAM) 30~m telescope with MAMBO II from \citet{rat06}.  The FWHM angular 
resolution of each element in the array is 11", with a separation between
each element of 20".  The final
(1 $\sigma$) r.m.s. noise level in these maps is $\sim$ 10 mJy/beam.

We present additional millimeter data taken with Bolocam \citep{gle03}
on the Caltech Submillimeter
Observatory (CSO) 10 m diameter telescope at 1.1 mm as part of the 
Bolocam Galactic Plane Survey, BGPS v1.0
\footnote[1]{http://irsa.ipac.caltech.edu/data/BOLOCAM\_GPS/}
\citep{agu10, ros10}.
Bolocam is a 144-element (of which 115 are working) 
bolometer array arranged on a uniform hexagonal grid.  Each bolometer
has an effective FWHM of 33" and the array field of view is 7'.5.  The
data were taken with a 45 GHz bandwidth filter centered at 268 GHz,
which excludes the bright 230 GHz J=2-1 CO line.  
The data were reduced using a custom pipeline created for the BGPS using the
methods described in 
\citet{agu10}.  The dominant source of noise at these wavelengths, 
the atmospheric component, is removed using 
Principle Component Analysis (PCA).  This cleaning process
limits the spatial scale to which we are sensitive to between
about 30" to 400". 
The final (1 $\sigma$) r.m.s. noise level in these maps
is $\sim$ 30 mJy/beam.

\subsubsection{\13CO J=1-0 Molecular Gas Tracer}
As a tracer of column density, radial velocities, and for spectral
line comparison, we include \13CO J=1-0 data taken as part of the
Boston University-Five College Radio Astronomy Obsevatory Galactic
Ring Survey \citep[BU-FCRAO GRS;][]{jac06}.  These data was obtained
from 1998-2005 using the single sideband focal plane array receiver, 
SEcond QUabbin Optical Imaging Array (SEQUOIA) 
on the FCRAO 14 meter.  At a central frequency of 110.2 GHz, the beam 
FWHM is 46" and the spectral resolution is 0.21 \kms.  We correct the \13CO
antenna temperatures for the main beam efficiency of 0.48 through the
expression T$_{mb}$ = $T_{A}^{*}/\eta_{mb}$.  The 1 $\sigma$ sensitivity in
the GRS data is about 0.27 K.

\subsubsection{VLA Radio Continuum Data}
In order to unambiguously determine which massive clumps have entered 
the main sequence, we used the Very Large Array (VLA) in New Mexico to 
look for thermal bremsstrahlung from newly formed H~II 
regions.  We observed a total of twelve positions over eight 
IRDCs in the X band in the C configuration in the spring 2008.
We observed single channel continuum in the X band, centered at 3.6 cm with
a bandwidth of 50 MHz.
We performed flux calibrations
on 3C286 and phase calibrations on 1832-105 and 1851+005.  The data 
were reduced using
standard continuum interferometric techniques in the AIPS
software package.  The 
primary beam is 5.4', and the highest resolution element is about 2.8 x 2.4".
This translates to 0.06 (0.10) pc resolution for the nearest (farthest)
clump.  We had a total of three six-hour sessions.  We observed each
source three times per session, changing the order each night to obtain
good coverage of the u-v plane.  We integrated for an hour total on each
source achieving a 1 $\sigma$ r.m.s. flux density of $\sim$ 0.05 mJy/beam.

\subsubsection{\hcop~and \n2hp~Dense Molecular Gas Tracers}

To determine the radial velocities of the clumps and their molecular
structure, we probe the dense 
molecular gas using the J=3-2 transitions of \hcop~and \n2hp (rest
frequencies of 267.5576190 and 279.5117010 GHz respectively).
Using the 1 mm dual polarization ALMA-prototype Sideband Separating Receiver on the 
Heinrich Hertz Telescope (HHT), we 
observed \hcop~and \n2hp~simultaneously with 1.1 \kms~spectral resolution
and 28'' and 27'' spatial resolution ($\Theta_{FWHM}$).  We
performed two minute position-switched observations on the seventeen IRDC
clumps, which gave us an r.m.s. noise level of about 0.04 K.  We On-the-Fly mapped
\citep[see][]{man07} the five most compelling clouds in 
both molecular lines simultaneously.  We
mapped the clouds, scanning twice in RA and twice in DEC 
with a row spacing of 10" at a scan speed of 5" per second.  The separate maps 
were offset by about 4".  We pointed,
focused, and calibrated on Jupiter every 2 or 3 hours when it was available,
otherwise, we pointed and focused on G34.3+0.15.  
We used S140 as a check of the line 
position and intensity about every two hours, and also to determine the 
Vertical / Horizontal Polarization ratio (Vpol/Hpol)
and the typical sideband rejection measurement.  
Typical sideband rejection between the USB and LSB
for each polarization was between 13 and 15 dB.  
We correct for the main beam efficiency of
0.75 and the ratio Vpol/Hpol = 1.1 in our data reduction.

The single spectra were reduced using standard procedures in the 
Gildas CLASS\footnote[1]{http://www.iram.fr/IRAMFR/GILDAS}
software package.  The baselines were removed, the horizontal
polarization was multiplied by the Vpol/Hpol scaling factor and then
co-added.  We then corrected for the main beam efficiency, giving us
T$_{\rm mb}$.  The On-the-Fly maps were reduced using some custom and 
shared software in the Gildas CLASS package.  The same procedure as above
was followed for each individual spectrum.  The spectra were then 
interpolated onto a regular grid using a Gaussian tapered Bessel
function \citep[Reiter, M. private communication;][]{man07}.

%%%%%%%%%%%%%%%%%%%%%%%%%%%%%%%%%%%%%%%%%%%%%%%%%%%%%%%%%%%%%%%%%%%%%%%%%%%%%%%%%%%%%%%%%%%%%%%
% GENERAL IRDC INFO
%%%%%%%%%%%%%%%%%%%%%%%%%%%%%%%%%%%%%%%%%%%%%%%%%%%%%%%%%%%%%%%%%%%%%%%%%%%%%%%%%%%%%%%%%%%%%%%

%\clearpage
\begin{deluxetable*}{ccccccccccccccc}
\tabletypesize{\scriptsize}
\tablecaption{Properties of the IRDC clumps \label{table:properties}}
\tablewidth{0pt}
%\rotate
\tablehead{
\colhead{} & \colhead{} & \colhead{} & \colhead{}   &
\colhead{} & \colhead{} & \colhead{}  & \colhead{\13CO} & \colhead{\hcop} &
\colhead{\n2hp} & 
\multicolumn{3}{c}{Integrated Intensity (K \kms)} \\
\colhead{} & \colhead{Clump}        &	
\colhead{V$_{LSR}$} & \colhead{$\Delta$ V} & \colhead{Distance} & 
\colhead{\ffore\tablenotemark{a}} & 
\colhead{$\tau$$_{8\mu m}$} &
\colhead{Peak T$_{\rm{MB}}$} & \colhead{Peak T$_{\rm{MB}}$} & \colhead{Peak T$_{\rm{MB}}$} &
\colhead{\13CO} & \colhead{\hcop} & \colhead{\n2hp} \\ 
\colhead{} &\colhead{} &
\colhead{(\kms)}  & \colhead{(\kms)} & \colhead{(kpc)} & \colhead{} & \colhead{} & 
\colhead{(K)} & \colhead{(K)} & \colhead{(K)} & \colhead{} &
\colhead{} & \colhead{}  }

\startdata

\multicolumn{4}{l}{G022.35+00.41} &           &       &      &        &         &        &        &         &         & \\
& GLM1    	& 52.4  &  3.4 & 3.6  &  0.14 & 1.3 &  5.4   &   1.9   &  1.2   &  8.6   &   8.0   &   5.4   & \\
& GLM2    	& 83.9  &  3.4 & 4.8  &  0.23 & 0.7 &  8.2   &   0.8   &  0.3   &  18.0   &   2.5   &   0.8   & \\
\multicolumn{4}{l}{G023.60+00.00}&           &       &      &        &         &        &        &         &         &\\
& GLM1          & 53.3  &  6.7 & 3.6  &  0.14 & 0.7 &  4.3   &   1.1   &  1.3   &  12.0   &   7.1   &   6.0   & \\
& GLM2          & 53.4  &  4.5 & 3.6  &  0.14 & 0.8 &  4.7   &   0.6   &  0.6   &  12.1   &   2.4   &   2.4   & \\
\multicolumn{4}{l}{G024.33+00.11}&           &       &      &        &         &        &        &         &         & \\
& GLM1          & 113.4 &  6.7 & 5.9  & 0.33  & 0.7 &  6.8   &   1.6   &  3.4   & 30.0   &  12.1   &  20.0   & \\
& GLM2   	& 114.1 &  5.6 & 5.9  & 0.33  & 0.7 &  5.2   &   0.5   &  0.3   & 26.1   &   2.8   &   1.8   & \\
\multicolumn{4}{l}{G024.60+00.08}&            &       &      &        &         &        &        &         &         & \\
& GLM1     	& 114.7 &  3.4 & 6.0  &  0.33 & 1.3 &  4.6   &   0.6   &  0.5   &  7.7   &   2.2   &   2.2   & \\
& GLM2    	& 53.0  &  5.6 & 3.5  &  0.14 & 1.2 &  5.6   &   1.5   &  1.0   &  14.1   &   8.7   &   5.5   & \\
\multicolumn{4}{l}{G028.23-00.19}&            &       &      &        &         &        &        &         &         & \\
& GLM1      	& 79.7  &  4.5 & 4.6  &  0.23 & 1.5 &  5.8   &   0.3   &  0.3   & 28.4   &   1.7   &   1.5   & \\
\multicolumn{4}{l}{G028.37+00.07}&            &       &      &        &         &        &        &         &         & \\
& GLM1     	& 78.8  &  5.6 & 4.5  &  0.23 & 2.6 &  4.7   &   1.1   &  1.2   & 21.0   &   5.9   &   4.1   & \\
& GLM2      	& 77.9  &  4.5 & 4.5  &  0.23 & 2.5 &  6.3   &   2.1   &  1.6   & 29.0   &   9.9   &   6.8   & \\
& GLM3      	& 78.9  &  2.2 & 4.5  &  0.23 & 1.8 &  4.9   &   1.2   &  1.3   & 24.7   &   4.9   &   5.2   & \\
& GLM4      	& 76.7  &  6.7 & 4.4  &  0.23 & 1.4 &  7.0   &   1.4   &  1.8   & 25.9   &   7.4   &  12.3   & \\
\multicolumn{4}{l}{G028.53-00.25}&           &       &      &        &         &        &        &         &         & \\
& GLM1     	& 86.2  &  5.6 & 4.9  &  0.26 & 0.4 &  6.6   &   0.4   &  0.2   & 22.3   &   1.7   &   1.1   & \\
\multicolumn{4}{l}{G034.43+00.24}&           &       &      &        &         &        &        &         &         & \\
& GLM1      	& 58.0  &  6.7 & 3.6  &  0.18 & 0.7 &  6.7   &   5.4   &  6.4   & 23.9   &  37.4   &  37.0   & \\
& GLM2      	& 57.5  &  6.7 & 3.6  &  0.18 & 0.4 &  8.5   &   6.4   &  6.2   & 30.0   &  46.0   &  34.6   & \\
& GLM3          & 59.5  &  5.6 & 3.7  &  0.18 & 0.9 &  7.2   &   2.9   &  2.5   & 18.4   &  18.0   &  11.7   & \\ 

\enddata
\tablenotetext{a}{The fraction of foreground emission along the line of
  sight, see \S \ref{sec:extmasses}}
\end{deluxetable*}

\section{Mass Estimates}
\label{sec:mass}
Mass is one of the most fundamental physical properties necessary 
to understand 
the nature of a region.  However, there is no perfect tracer 
of mass.  Molecular line mass tracers, such as \13CO may suffer
from depletion, optical depth effects or varying excitation
conditions.  Virial masses are only reliable in the idealized
case of a gravitationally bound object in virial equilibrium,
where the line is broadened 
simply by virialized motion.  An extinction mass is limited
by \textit{a priori} knowledge of the foreground and background
radiation field, and the absorption properties of the 
obscuring dust.  Dust continuum massses are thought to be
a promising tracer of cold, dense environments, but are 
also sensitive to varying dust temperature, emissivity,  
and grain properties.  The distance estimate plays
a significant role in the mass uncertainty
via the assumption of a Galactic rotation curve, and the 
ambiguity associated with the velocity scatter of objects at any given
distance.  Each of these tracers is 
also affected by the statistical uncertainties in each
measured quantity.  For this study, we find that the
systematic uncertainties far outweigh the statistical
uncertainties and include only estimates of the 
systematic uncertainty in \S \ref{sec:masserror}.

\begin{figure*}
\centering
\includegraphics[scale=0.4]{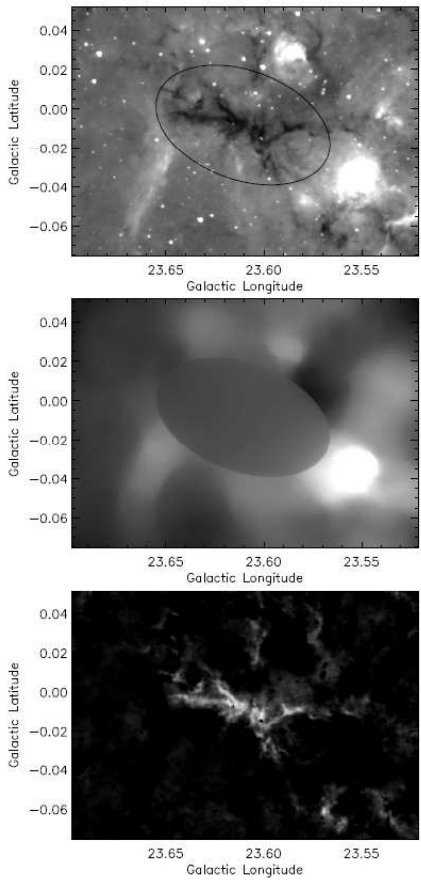}
\caption{G023.60+00.00 Top: 8 $\mu$m GLIMPSE image with ellipse approximating the
extent of the IRDC.  This image represents the values of \Inu$_{1,obs}$ in
Equation \ref{eq:inu1final}.  Middle: Circularly median filtered 8 $\mu$m image with
the points in the ellipse determined by interpolation from outside so they 
are not skewed by the IRDC.  This is a proxy for the diffuse Galactic
background, \Inu$_{0,obs}$ in
Equation \ref{eq:inu0final}  Bottom: Final surface density map in units of g cm$^{-2}$}
\label{fig:irdc33extmass}
\end{figure*}

A minimization and characterization of systematics
is vital for a fair comparison of IRDC mass tracers.
For this reason, we have 
convolved the extinction and MAMBO 1.2 mm maps to the BGPS
resolution using a Gaussian kernel with a width given by
$\sigma_{conv} = \sqrt{\sigma^{2}_{bgps} - \sigma^{2}_{org}}$
where $\sigma_{org}$ is the original width of the beam 
that is being convolved.  Each mass is then
retrieved within a circular aperture with the
equivalent area of the BGPS Gaussian beam,
  \begin{equation}
    r_{eff} = \frac{\Theta_{fwhm}}{2\sqrt{\rm ln2}} = 19.81''.
    \label{reff_eq}
  \end{equation} 
This allows us to compare these mass estimates independent of
beam filling factors and beam or source sizes.  Where possible,
we also calculate the total cloud mass, the mass within
the elliptical aperture used to define the extent of the
IRDC for extinction mapping (see Figure \ref{fig:irdc33extmass}).
The line widths
used to determine the virial masses were achieved from single
pointings of \hcop~on the HHT, which has an effective beam FWHM
of about 28'' at these frequencies, similar to the BGPS beam.  
In most cases the radius used to determine the virial mass (the 
``virial radius'') is very similar in size
to the circular aperture radius, so these are fair comparisons
in terms of source size.  The \13CO masses were retrieved 
from single GRS pointings, with an effective FWHM of about
46''.  The mass in this case is estimated over
a greater area than the other techniques, so it will include a wider region
and more emission, which we consider
when making mass comparisons.  We chose not to degrade the 
resolution of the other maps past the 33'' for the BGPS beam to match the
\13CO beam as the \13CO is considered a secondary mass estimator, and is
included only for comparison.  

We explain below the techniques for achieving the 
different mass estimates and the estimated systematic
uncertainty in Section \ref{sec:masserror}.  
We then compare these mass estimates and discuss their differences and
similarities.

\subsection{Distance Estimates}
\label{sec:dis}
CO is the second most abundant molecule in interstellar space.
It is found nearly everywhere, including IRDCs.  In these dense, cold 
environments, however, even the less abundant \13CO can become optically thick
and can deplete onto dust grains \citep[e.g.][]{taf02}.  \hcop~and \n2hp~are 
less abundant than \13CO and therefore become optically
thick at higher columns of H$_{2}$.  The critical density
is also much higher, 3.5 $\times$ 10$^{6}$ cm$^{-3}$ and 3.0
$\times$ 10$^{6}$ cm$^{-3}$ for the J=3-2 transitions of 
\hcop~and \n2hp~respectively, compared with  1.9 x 10$^{3}$ cm$^{-3}$ for \13CO
J=1-0.  Therefore, there is typically only one \hcop~or 
\n2hp~source along a given line-of-sight, which can then be
confidently associated with the dense IRDC, as opposed to CO which is
often seen at many velocities along a given line-of-sight 
through the Galactic plane.  For this reason,
we use the \hcop~velocity to determine the kinematic distance
to each IRDC.  A kinematic distance had already been determined
to each of these objects using \13CO morphology matching
by \citet{sim06b}, however, we find that in the confused
inner Galaxy, morphology matching can misassign distances.  Two
clouds of our sample of eight (G022.35+00.41 and G024.60+00.08) are 
spatially adjacent clumps along the line-of-sight at 
very different distances, as opposed to a contiguous object as
found by \citet{sim06b}. 

Distances were estimated kinematically using the \hcop~clump velocities
and the rotation curve of \citet{re09}, determined with
trigonometric parallaxes and proper
motions of masers using the VLBA and Japanese VERA project.  
In the First Galactic quadrant, a
radial velocity can correspond to two possible distances, a problem
known as the ``Kinematic Distance Ambiguity''.  Since IRDCs are
perceived as dark extinction features against the Galactic background,
it is assumed that all IRDCs are located at the near kinematic
distance.
If we assume an average velocity scatter at any given distance of about 
5\% (about 5-10 \kms), this translates to an approximate distance
uncertainty of 20\%.  The distances, line-widths, and observed properties
of the clumps are listed in Table \ref{table:properties}.

%\clearpage

%\clearpage
\subsection{Extinction Masses}
\label{sec:extmasses}
The mid-IR extinction by IRDCs provides 
a powerful way to obtain an independent measure of their column
density.  We restrict our analysis of extinction to the 
8~\micron~band of GLIMPSE, as this band is dominated
by emission from the diffuse ISM, rather than stellar sources.  
We apply the extinction mapping method put
forth by \citet{but09}.  In brief, this method 
estimates the Galactic background and foreground emission, 
finds the optical depth required to produce the 
observed extinction feature, assumes a dust opacity, 
and finally extracts the gas mass surface density.  
Our application of this method is described below.

To accurately map the
extinction, we need to estimate the intensity of radiation behind the
cloud of interest, \Inu$_{0}$, and the radiation in front of the
cloud, \Inu$_{1}$.  Assuming no emission from the cloud itself
at these wavelengths, we have 
\begin{equation}
\label{eq:inu1tau}
I_{\nu 1} =
I_{\nu 0} e^{-\tau_{\nu}}
\end{equation}
where $\tau_{\nu}$ = \kapnu $\Sigma$, \kapnu~is the dust 
opacity, and $\Sigma$ is the surface mass density.  We
adopt a value of $\kappa_{8 \mu \rm m}$ = 11.7 cm$^{2}$ g$^{-1}$, following the analysis
of \citet{but09} (they adopt a value of 7.5 cm$^{2}$ g$^{-1}$ as they use a
gas-to-dust ratio of 156, rather than the value of 100 used in this paper).
This value is closest to the model of \citet{oss94} of thin ice mantles
that have undergone coagulation for 10$^{5}$ years at a
density of n$_{\rm H_{2}}$ $\sim$ 10$^{6}$ cm$^{-3}$, which is a reasonable
model for the cold, dense environment of an IRDC.  This expression
assumes that the dust opacity remains constant along the line-of-sight,
which may not be true.  However, most of the extinction arises from the 
coldest, densest parts of the cloud.  In the less dense regions, the
opacity may vary by up to a factor of two, but we assume that the majority of
the extinction is caused by the coldest, densest regions of the cloud,
which are well-characterized by the adopted value of $\kappa_{8 \mu \rm m}$.

Before proceeding with the background determination, we must
first estimate the fraction of emission from
dust in the foreground.  We assume that the Galactic
distribution of hot dust (in the Galactic midplane) 
is given by the Galactic surface density of OB associations \citep{mck97}
\begin{equation}
\Sigma_{OB} \propto e^{-R/H_{R}}
\end{equation}
where R is the Galactocentric radius (in kpc) and H$_{R}$ = 3.5 kpc
is the Galactic radial scale length.  We then integrate the
column of dust from the sun \citep[][at R$_{0}$ = 8.4 kpc from the Galactic center]{re09}
to the cloud, and the column of dust along the same line
of sight, from the Sun out to a Galactocentric radius of 16 kpc (beyond
which there is a negligible contribution to the dust emission).
The ratio of the column to the cloud over the column
out to 16 kpc is called the `Foreground Intensity
Ratio', \ffore; the fraction of emission along the 
line-of-sight that is from the foreground.  Thus, the true radiation behind the cloud, 
\Inu$_{0}$, determined from that observed, \Inu$_{0,obs}$, is given by
\begin{equation}
\label{eq:inu0final}
I_{\nu 0} = (1-f_{fore}) I_{\nu 0, obs}
\end{equation}
and the true radiation in front of the cloud, \Inu$_{1}$, determined from that
observed, \Inu$_{1,obs}$ is given by
\begin{equation}
\label{eq:inu1final}
I_{\nu 1} = I_{\nu 1, obs}-f_{fore} I_{\nu 0, obs}.
\end{equation}
\citet{but09} note that \ffore~is uncertain because of
the inevitable small-scale variations in the Galactic 
background and the fact that the regions surrounding the 
IRDC are likely to be embedded in the same GMC that hosts the IRDC.
This leads to a higher 
extinction of the integrated Galactic background, creating a 
tendency to underestimate \ffore~and thus $\Sigma$.

An upper limit for \ffore~is provided by the minimum flux at 8~\micron~
divided by the average background.  This
value would be \ffore~if the cloud were totally opaque at 8~\micron.  This
upper limit is about 2 times higher in most cases, except in G028.37+00.07,
where it is about the same.  
The remainder of the IRDCs may or may not be optically thick at 8~\micron,
however, the good correlation (see Figure \ref{fig:masscomp})
with BGPS 1.1 mm masses (which is almost
certainly optically thin) is an indication that many of the IRDCs are
still optically thin at 8~\micron.

For the purpose of estimating the Galactic background, we 
employ the more accurate small-scale median filter (SMF) approach
from \citet{but09}.  This approach is designed to capture
the small-scale variations in the background, without altering the
estimate by the darkness of the IRDC itself.  Thus, we first exclude an ellipse
approximating the size and shape of the IRDC from the filtering 
process (Figure \ref{fig:irdc33extmass}, top).  To smooth the data, we first 
remove the high-end
tail of the distribution (all points greater than twice the mode) 
which is mostly due to stars.
We then perform a circular spatial median filter at each point, i.e.
we compute the median of all the data in a circle around a given
point, and that becomes the smoothed data value at that point.
A reasonable size for the filter was empirically determined to 
be a radius of one arcminute.  This smoothed map is our estimate of 
the diffuse Galactic background.  We then estimate the `background' 
inside the ellipse by taking the average of the smoothed
data values from outside the ellipse, weighted by the inverse 
square of their distance from the point.  This is depicted in 
the middle frame of Figure \ref{fig:irdc33extmass}.  This image is now our estimate
of \Inu$_{0,obs}$, and our original image is \Inu$_{1,obs}$.  Using
Equation \ref{eq:sd} we then create a surface density
map, shown in Figure \ref{fig:irdc33extmass} in the bottom panel.  
\begin{equation}
\label{eq:sd}
\Sigma = 
-\frac{1}{\kappa _{\nu}}
\rm{ln}\bigg[\frac{I_{\nu 1, obs}-f_{fore} I_{\nu 0, obs}}{(1-f_{fore}) I_{\nu 0, obs}}
\bigg]
\end{equation}

There is an additional correction that must be taken into account for
quantities derived from extended sources in the IRAC data.  Scattering from extended
sources within the IRAC array systemtically increases the measured surface
brightness.  For a uniformly distributed extended source, this scattering
increases the measured surface brightness by about $\sim$ 30\% at 8 \micron~
(S. Carey, private communication and
http://ssc.spitzer.caltech.edu/irac/iracinstrumenthandbook/33/).
This means that the \textit{measured} background (\Inu$_{0,obs}$) is
systematically brighter than the \textit{true} background, which we will
call T$_{0}$ for simplicity.  Where s = 0.3, the scattering correction, the
\textit{true} background is given by $\rm{T}_{0} = (1 + \rm{s})$\Inu$_{0,obs}$.
The \textit{measured} foreground (\Inu$_{1,obs}$) is also contaminated by
scattered light from the background, so the \textit{true} foreground
(T$_{1}$) is given by $\rm{T}_{1} =$ \Inu$_{1,obs} + \rm{s} $
\Inu$_{0,obs}$.  Plugging these values into Equation \ref{eq:sd}, simplifying, and
writing the surface density in terms of our measured quantities, our final
expression for the surface density is:
\begin{equation}
\label{eq:sd_final}
\Sigma = 
-\frac{1}{\kappa _{\nu}}
\rm{ln}\bigg[\frac{(\rm{s} + 1)I_{\nu 1, obs}-(\rm{s} + \rm{f})f_{fore} I_{\nu 0, obs}}{(1-f_{fore}) I_{\nu 0, obs}}
\bigg]
\end{equation}

Using distances
determined from the \hcop~velocities and the \citet{re09} rotation
curve, we are then able to sum the surface density over the BGPS beam-sized
apertures to obtain the mass (see Table \ref{table-masses} and Figure 
\ref{fig:masscomp}) in the clump.  
The extinction masses for the clumps range from 60 to 620 \Msun~and from 1200 to 
26,020 \Msun~for the whole IRDC.

%%%%%%%%%%%%%%%%%%%%%%%%%%%%%%%%%%%%%%%%%%%%%%%%%%%%%%%%%%%%%%%%%%%%%%%%%%
% NEW MASS COLUMN COMPARISON
%%%%%%%%%%%%%%%%%%%%%%%%%%%%%%%%%%%%%%%%%%%%%%%%%%%%%%%%%%%%%%%%%%%%%%%%%%

%\begin{deluxetable*}{lcccccccccccccc}
\begin{deluxetable*}{lcccccccccccc}
%\rotate
\tabletypesize{\scriptsize} \tablecaption{Mass Comparison \label{table-masses}}
\tablewidth{0pt}
\setlength{\tabcolsep}{0.07in}
\tablehead{
\colhead{} & \colhead{Aperture} &
\multicolumn{3}{c}{Column Density in 33'' BGPS Beam} &
\colhead{\13CO} &
\multicolumn{3}{c}{Mass inside 33'' BGPS Beam}  & \colhead{\13CO} &
\colhead{} \\%& \colhead{} & \colhead{} \\
\cline{3-5} & \cline{6-8}
\colhead{Clump\tablenotemark{a}}  & \colhead{Size\tablenotemark{b}}  
& \colhead{Extinction} & \colhead{BGPS} & 
\colhead{MAMBO} & \colhead{Column} 
& \colhead{Extinction}   & 
\colhead{BGPS}      & \colhead{MAMBO} & \colhead{Mass} &
\colhead{Virial Mass} & \colhead{R$_{vir}$} \\
%& \colhead{n$_{BGPS}$\tablenotemark{c}} 
%& \colhead{M$_{Jean's}$\tablenotemark{d}} \\
\colhead{} & \colhead{(pc)} & 
\colhead{} & \colhead{(10$^{22}$ cm$^{-2}$)} & \colhead{}
& \colhead{} &
\colhead{(\Msun)} 
& \colhead{(\Msun)} & \colhead{(\Msun)} & \colhead{(\Msun)} & \colhead{(\Msun)} 
& \colhead{(pc)} }%& \colhead{(10$^{3}$ cm$^{-3}$)} & \colhead{(\Msun)} }

\startdata

G022.35+00.41 &           &     &      &      &      &        &       &       &       &       &      \\%&       &      \\
 ~~~GLM1           &  0.7      & 2.3 & 1.1  & 2.9  & 0.4  &  190   &  95   &  250  &  70   &  340  &  0.2 \\%&  5.2  & 9   \\
 ~~~GLM2           &  0.9      & 1.3 & 1.2  & 1.4  & 0.9  &  200   &  180  &  210  & 240   &  2300 &  1.3 \\%&  4.3  & 11   \\   
G023.60+00.00 & 2.9 x 1.8 &     &      &      &      &  1200   & 1200  & 1800  &       &       &      \\%&       &      \\
 ~~~GLM1           &  0.7      & 1.2 & 1.7  & 3.0  & 0.6  &   100   &  140  &  250  & 100   &  1800 &  0.3 \\%&  7.8  & 8  \\
 ~~~GLM2           &  0.7      & 1.4 & 1.0  & 1.7  & 0.6  &   120   &  81   &  140  &  90   &  660  &  0.2 \\%&  4.4  & 10  \\
G024.33+00.11 & 6.0 x 8.0 &     &      &      &      &  12010  & 13000 & 14000 &       &       &      \\%&       &     \\
 ~~~GLM1           &  1.1      & 1.4 & 4.8  & 8.6  & 1.5  &  310   &  1200 & 1950  & 610   &  2500 &  0.4 \\%&  14.0 & 6  \\
 ~~~GLM2           &  1.1      & 1.3 & 1.4  & 1.9  & 1.3  &  300   &  310  &  440  & 530   &  3100 &  0.7 \\%&  4.0  & 11  \\
G024.60+00.08 &           &     &      &      &      &        &       &       &       &       &      \\%&       &     \\
 ~~~GLM1           &  1.2      & 2.4 & 1.3  & 2.3  & 0.4  &  560   &  310  &  550  & 160   &  850  &  0.5 \\%&  3.5  & 12  \\
 ~~~GLM2           &  0.7      & 2.1 & 1.6  & 3.2  & 0.7  &  170   &  130  &  260  & 100   &  1000 &  0.2 \\%&  7.5  & 8  \\
G028.23-00.19 & 5.2 x 3.3 &     &      &      &      &  5520  & 3700  & 1900  &       &       &      \\%&       &     \\
 ~~~GLM1           &  0.9      & 2.8 & 1.4  & 2.0  & 1.4  &  380   &  190  &  270  & 350   &  910  &  0.3 \\%&  4.9  & 10  \\
G028.37+00.07 & 5.8 x 4.5 &     &      &      &      &  26020 & 13000 & 18000 &       &       &      \\%&       &     \\
 ~~~GLM1           &  0.9      & 4.7 & 1.6  & 3.0  & 1.0  &  620   &  220  &  400  & 250   &  2100 &  0.5 \\%&  5.9  & 9  \\
 ~~~GLM2           &  0.9      & 4.6 & 2.2  & 3.9  & 1.4  &  610   &  290  &  520  & 350   &  330  &  0.1 \\%&  7.9  & 8  \\
 ~~~GLM3           &  0.9      & 3.3 & 1.4  & 2.6  & 1.2  &  430   &  190  &  340  & 290   &  370  &  0.5 \\%&  5.1  & 10  \\
 ~~~GLM4           &  0.9      & 2.6 & 5.3  &10.3  & 1.3  &  350   &  700  & 1400  & 290   &  4400 &  0.7 \\%&  19.1 & 5  \\
G028.53-00.25   & 7.3 x 4.6 &     &      &      &      &  4800  &  7300 & 11000 &       &       &      \\%&       &     \\
 ~~~GLM1           &  0.9      & 0.7 & 1.6  & 3.5  & 1.1  &  110    &  250  &  540  & 310   &  3500 &  0.8 \\%&  5.7  & 9  \\
G034.43+00.24 & 6.9 x 2.3 &     &      &      &      &  5550  & 7500  & 14000 &       &       &      \\%&       &     \\
 ~~~GLM1           &  0.7      & 1.3 &10.1  &17.8  & 1.2  &  110    &  850  &  1500 & 180   &  3000 &  0.4 \\%&  46.7 & 3  \\
 ~~~GLM2           &  0.7      & 0.7 & 9.3  &15.9  & 1.5  &  60    &  790  &  1300 & 230   &  2500 &  0.4 \\%&  43.1 & 3  \\
 ~~~GLM3           &  0.7      & 1.7 & 2.2  & 3.9  & 0.9  &  140    &  190  &  330  & 150   &  1700 &  0.4 \\%&  10.3 & 7  \\

\enddata
\tablenotetext{a}{The clump names are from the MSX dark cloud catalog
  \citep{sim06}, and have had the ``MSXDC'' prefix removed}
\tablenotetext{b}{For the clumps, the aperture size is the physical size of 
 the 40'' diameter circular BGPS aperture.  For the cloud, it is the
 physical major and minor axes of the ellipse approximating the extent of
 the IRDC.}
%\tablenotetext{c}{The number density from the BGPS column density, assuming
%  that the clump has the same line-of-sight thickness as its extent on
%  the plane of the sky (the aperture size).}
%\tablenotetext{d}{The Jeans mass using the given density and the assumed
%  temperature, 15 K.~~
%  $M_{Jeans} = [\frac{3}{4\pi\rho}]^{1/2} 
%  [\frac{5kT}{\mu m_{H}G}]^{3/2}$  }

\end{deluxetable*}
%\clearpage
%\end{landscape}

%\clearpage
\subsection{Dust Continuum Masses}
Obtaining mass estimates of IRDCs with molecular lines is often
problematic due to high column densities and molecular freezeout onto
dust grains.  The cold dust emission, however, as observed in the 
millimeter, provides more reliable mass estimates as it is
optically thin and does not deplete.  We use both the 1.2 mm and 1.1 mm 
dust continuum emission from MAMBO and the BGPS, to achieve isothermal
mass estimates using the expression
\begin{equation}
M = \frac{S_{\nu} D^{2}}{\kappa_{\nu}B_{\nu}(T)}
\end{equation}
where S$_{\nu}$ is the source flux density, D is the distance,
$\kappa$$_{\nu}$ is the dust opacity and B$_{\nu}$(T) is the
Planck function at dust temperature, T.  We adopt a value of
$\kappa_{1.1 mm}$ = 0.0114 cm$^{2}$g$^{-1}$ from \citet{eno06}, in which
we have assumed a gas-to-dust ratio of 100.  This opacity is consistent
with the \citet{oss94} model used for the 8~\micron~opacity in the
extinction masses.
The mass equation reduces to
\begin{equation}
\label{eq:bgpsmass}
M_{1.1 mm} = 14.32~(e^{13.0/T} - 1)~\Bigg(\frac{S_{\nu}}{1~\rm Jy}\Bigg)
~\Bigg(\frac{D}{1~\rm kpc}\Bigg)^{2}~\Msun
\end{equation}
\begin{equation}
\label{eq:mambomass}
M_{1.2 mm} = 20.82~(e^{12.0/T} - 1)~\Bigg(\frac{S_{\nu}}{1~\rm Jy}\Bigg)
~\Bigg(\frac{D}{1~\rm kpc}\Bigg)^{2}~\Msun.
\end{equation}
We assume a dust temperature T = 15 K.
We can also find the column density of H$_{2}$, as given by 
\begin{equation}
N(H_{2}) = \frac{S_{\nu} }
{\Omega_{B}\kappa_{\nu}B_{\nu}(T) \mu_{H_{2}}m_{H}}
\end{equation}
\begin{equation}
N(H_{2})_{1.1 mm}= 2.20\times10^{22}(e^{13.0/T}-1)S_{\nu}~\rm cm^{-2}
\end{equation}
\begin{equation}
N(H_{2})_{1.2 mm}= 3.20\times10^{22}(e^{12.0/T}-1)S_{\nu}~\rm cm^{-2}
\end{equation}
where $\Omega_{B}$ is the beam size, $\mu_{H_{2}}$ is the mean molecular 
weight for which we adopt a value of $\mu_{H_{2}}$ = 2.8 \citep{kau08}. 
The dust continuum masses and column densities are listed in Table 
\ref{table-masses}.  The total BGPS 1.1 mm cloud masses range from 1200 to 
13,000 \Msun~while the clump masses range from 80 to 1100 \Msun.

%\clearpage
\subsection{Virial Masses}
The molecular line width of these objects provides an independent
measure of mass.  The virial mass is based on the assumption
that the object is gravitationally bound and in virial
equilibrium.  Therefore, the virial mass estimate, in 
combination with other mass estimates, may allow discussion of
whether or not the clump is bound and virialized.  For a density profile of
$\rho(\rm r) = \rm r^{-\alpha}$, the virial mass is given by
\begin{equation}
M_{\rm vir} = 3 \Bigg[\frac{5-2\alpha}{3-\alpha}\Bigg]\frac{R\sigma^{2}}{G}
\end{equation}
where $\sigma$ is the line-of-sight velocity dispersion, R is the virial
radius, and G is the gravitational constant.  We adopt a spectral index of
$\alpha$ = 1.8, which was found to
be the mean in a sample of 31 massive star-forming regions by 
\citet{mue02} with a standard deviation of 0.4.  However, this spectral 
index was derived from a sample of ``active'' clumps, so this is an added
source of uncertainty for quiescent clumps.  We do not include a
correction for ellipticity, as these clumps are compact enough to be
well-approximated by spheres.
This expression reduces to
\begin{equation}
\label{eq:virmass}
M_{\rm vir} = 147~\Bigg(\frac{R}{1~\rm pc}\Bigg)
\Bigg(\frac{\Delta  v_{fwhm}}{1~\rm km~s^{-1}}\Bigg)^{2}~\Msun
\end{equation}
where R is the virial radius and $\Delta \rm v_{fwhm}$ is the line width
at half maximum.  In order to determine the best estimate for the virial
radius, a two-dimensional circular Gaussian was fit to each 
BGPS clump.  The Gaussian fit was then deconvolved from the BGPS beam.
The effective radius (see Eq. \ref{reff_eq})
of the deconvolved Gaussian fit is the virial radius, given in
Table \ref{table-masses}.  
In this study, we assign a virial `radius' to 
a Gaussian beam which is likely convolved with a power-law distribution 
of cores within
the beam.  The meaning of the virial `radius' in this case is tricky as the 
deconvolved source size can vary by up to a factor of two
depending on the slope of the power law distribution of sources within the beam
\citep[Figure 16]{shi03}.  However, this method is consistent between
the clumps in our sample, and is a reasonable proxy for the size scale of
the diffuse emission.  

We use the FWHM of the \hcop~line for the virial
masses presented here.  Using the \n2hp~line width instead does not change
the results qualitatively, as the two tracers provide consistent line
widths across different activity levels.  \n2hp~has been shown to be well 
correlated with submm dust column in starless cores \citep{taf02, wyr00},
meaning that these virial masses are most likely based on the same material
which is producing the dust continuum.  The virial masses of the clumps were 
found to range from 330 to 4400 \Msun.

\begin{figure*}
\centering
\includegraphics[scale=1]{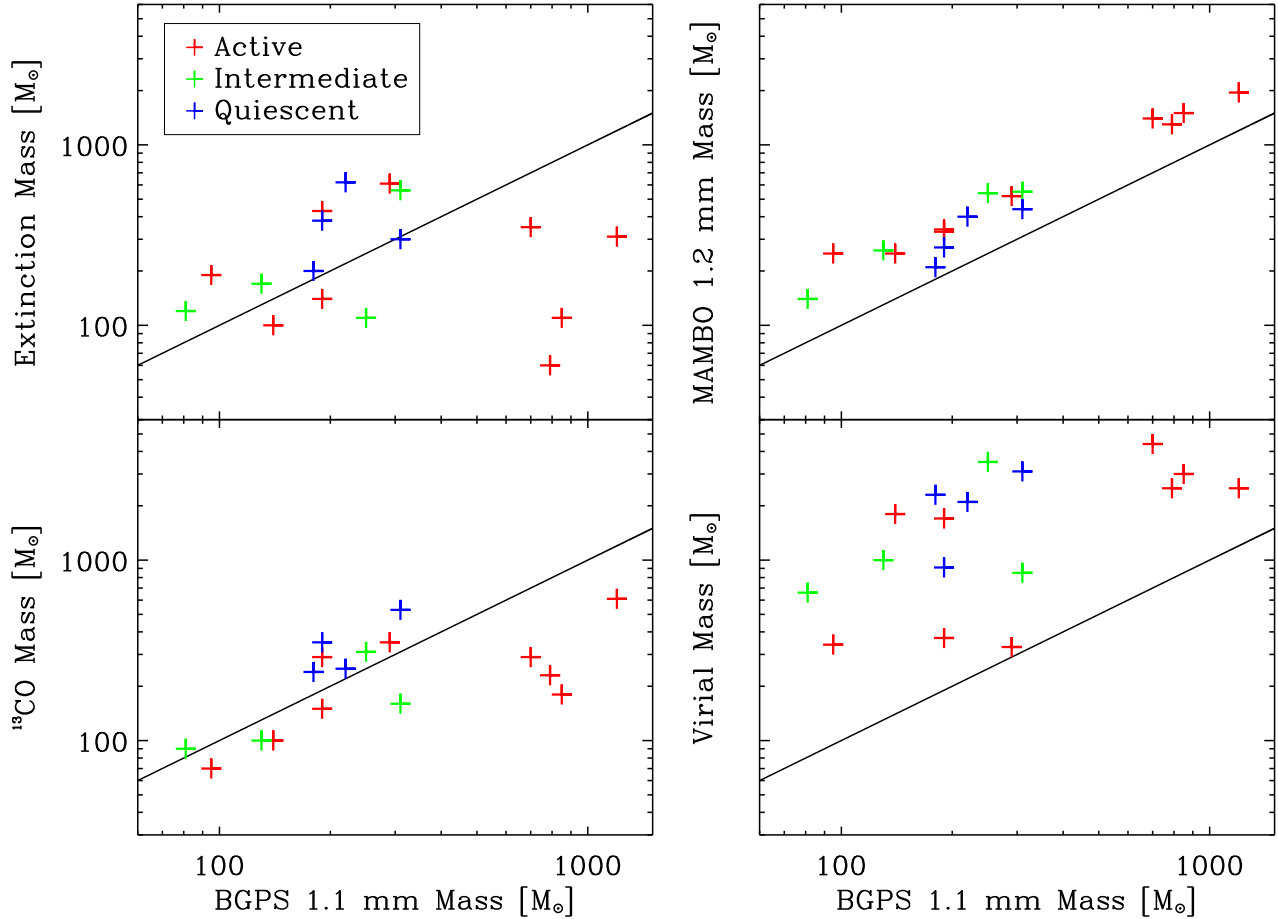}
\caption 
{A comparison of mass estimates for the clumps (see Table \ref{table-masses}).  The line drawn 
has a slope of 1, representing equality.  The red crosses are
``active'' clumps, the green are ``intermediate,'' and the blue are
``quiescent.''  Note that the four highest BGPS 1.1 mm mass objects (where there
is significant discrepancy) are all associated with UCH~II Regions.}
\label{fig:masscomp}
\end{figure*}

%\clearpage
\subsection{\13CO Masses}
\label{sec:13comass}
\13CO is commonly used to estimate the column density and mass
of molecular clouds in the Galaxy.  However, due to depletion onto dust 
grains \citep{taf02} and optical depth effects in cold, dense environments,
\13CO is not the most reliable tracer of column in IRDCs.  
While there is good morphological agreement between the
8~\micron~extinction and the \13CO toward the dense clumps in IRDCs,
\citet{duy08} show that \13CO is optically thick toward many IRDCs, so we
caution the reader that these tracers are probably not probing to the same
depth in dense IRDC clumps.  \citet{duy08} also demonstrate that typcial
excitation temperatures of \13CO in IRDCs are of the order of $\sim$ 10 K.
This indicates that the \13CO is at the very least tracing to some depth in
the cold, dense region of the clump, however the \13CO is also tracing the
surrounding GMC and IRDC envelope.
We include column density and mass estimates from \13CO
as a secondary mass tracer, and as a comparison
to other tracers.  In the optically thin, thermalized
limit, the \13CO column density is given by
\begin{equation}
N(H_{2}) = \frac{8\pi k\nu^{3}X_{^{13}CO}}{3c^{3}hB_{J}A_{10}}
(1-e^{-h\nu/kT_{ex}})^{-1}\int \! T_{mb} \, dv
\end{equation}
where $\nu$ is the frequency of the \13CO J=1-0 transition,
 A$_{10}$ is the Einstein A coefficient of \13CO from state
J=1 to J=0, T$_{ex}$ is the excitation temperature, B$_{J}$ is the 
rotation constant, and X$_{^{13}CO}$
is the fraction of \13CO to H$_{2}$.  We adopt a value of
$^{12}$CO / \13CO of 58 from \citet{luc98}, and a value of
$^{12}$CO / H$_{2}$ of 10$^{-4}$, a value of
55.101038 GHz for B$_{J}$, and standard NIST values for all
constants and spectral transition values.  This expression
then reduces to 
\begin{equation}
N(H_{2}) = 1.45 \times 10^{17} \frac{\int \! T_{mb} \, dv}{1 -
  e^{-5.29/T}}\rm cm^{-2}
\end{equation}
This column density can then be converted to a mass via
\begin{equation}
M = N(H_{2})\mu_{H_{2}}m_{H}A
\end{equation}
where $\mu_{H_{2}}$ is the mean molecular weight of H$_{2}$, m$_{H}$ 
is the mass of hydrogen, and A is the physical area over which the 
mass is summed.  We adopt a value of $\mu_{H_{2}}$ = 2.8 \citep{kau08}.
This expression reduces to
\begin{equation}
\label{eq:comass}
M = 5.00\times10^{-25}~N(H_{2})~\Bigg(\frac{A}{1~\rm arcsec^{2}}\Bigg)
~\Bigg(\frac{D}{1~\rm kpc}\Bigg)^{2} \Msun
\end{equation}
where A is the area in arcsec$^{2}$, over which the
mass is summed, and D is the distance in kpc.  The \13CO clump 
masses range from 70 to 610 \Msun.

%\newpage
%\clearpage
\begin{figure*}
  \centering
  \label{fig:extcolvsbgpscol_all}
  \includegraphics[scale=0.6]{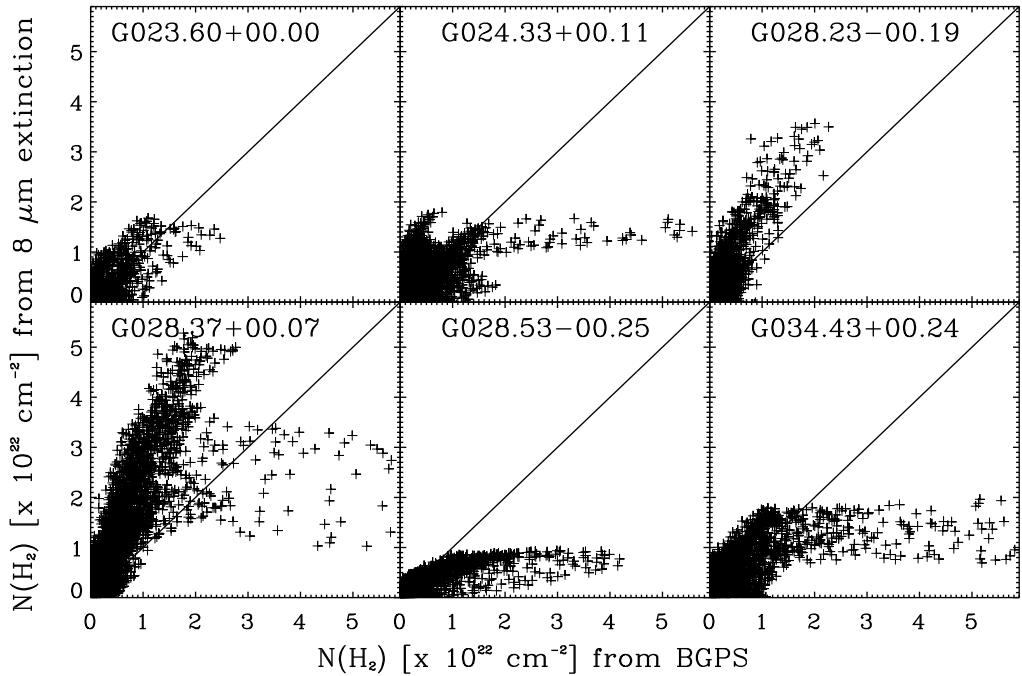}
 \caption{A pixel-by-pixel comparison of  8~\micron~extinction column density vs. BGPS 1.1 mm 
   column density for points within the IRDC ellipse.  The black line is
   the 1:1 line.}
%   Notice the continuum of sources, from a high extinction column
%   density relative to the BGPS 1.1 mm column density to a 
%   flattened extinction column density
%   relative to BGPS 1.1 mm column density.}
 \label{fig:extcolvsbgpscol}
 \end{figure*}

%\clearpage
\subsection{Mass Error Analysis}
\label{sec:masserror}
For this study, we find that the
systematic uncertainties far outweigh the statistical
uncertainties.  We present below our estimated systematic uncertainties and
the method used to derive them.

We assign each variable in the mass equation (Eqns. \ref{eq:sd},
\ref{eq:bgpsmass}, \ref{eq:mambomass}, \ref{eq:virmass}, and \ref{eq:comass})
a fractional uncertainty, perform a
Monte-Carlo simulation randomly drawing the variables from a Gaussian distribution, 
and report the 68\% confidence interval over which
the mass varies.  Since we have reason to believe that each of the
variables is centered at our adopted value, we assign each variable a
randomly populated Gaussian distribution, centered at the adopted value and
with a width such that 99.7\% of the points are within the lower and upper
bounds of the fractional uncertainty.  For example, we assign an
uncertainty to the distance of 20\%.  Therefore, for each run, we randomly
sample the distance value from a Gaussian which is centered at the adopted
value and has a width such that 99.7\% of the points lie
between 0.8 and 1.2 times the adopted distance value.  For each mass
estimate, we ran at least 10$^{7}$ points in the simulation.  We note that
several of the mass probability distributions below are asymmetric.
To be consistent with the literature, we present the mass estimates at 
the adopted variable values (which is equal to the mean of the mass
probability distribution), but note that this is not the same as the 
median or most probable value of the distribution. The 68\%
confidence intervals were drawn from the median of the distribution
outwards.  We divide the high and low mass ends of the confidence interval 
by the mass at the peak of the distribution to achieve the 
asymmetric 68\% confidence intervals. 

The dominant sources of uncertainty in the extinction mass estimate are the
background and foreground estimates, the scattering coefficient, 
as well as the distance, opacity, and \ffore.
We find that typical fluctuations in the 8~\micron~emission near
IRDCs, on the scale of the IRDCs, is around 40\%.  This background
fluctuation is the best measure we have for how much the background behind
the IRDC will vary from what we measure, so we assign a 40\% uncertainty to
the background estimate.  In calculating the foreground uncertainty, we
assign the same 40 \% uncertainty to the measured background and an
additional 50\% uncertainty to the value of \ffore.  We include a 15\%
uncertainty in the value of the scattering coefficient as the color
variation is about 5 \% and not being able to characterize the spatial
dependency of the scattering has about a 10-15\% effect (S. Carey, private
communication).   
We also include a 20\% uncertainty in the distance determination (as
discussed in Section \ref{sec:dis}) and factor
of two uncertainty in the value of the opacity as
recommended by \citet{oss94}.
The 68\% confidence intervals for the extinction mass are from 80\% to
2.2 times the quoted mass.

For the dust continuum (BGPS 1.1 mm and MAMBO 1.2 mm) mass estimates, the
dominant uncertainties are the temperature, the flux density, the opacity, and
the distance.  For both the 1.1 mm and 1.2 mm data, we assign 
an uncertainty of 50\% to the temperature \citep[for most clumps the temperature
is between 10-20 K;][]{dun10}, 10\% to the flux density \citep{agu10}, a factor of two
uncertainty in the opacity (as above), and 20\% to the distance.  The 68\%
confidence intervals for both dust continuum masses are from 80\% to 180\% of
the quoted mass. 

The systematic uncertainty associated with the virial mass is difficult to
characterize.  The basic assumption that the clump is a
gravitationally bound object in virial equilibrium introduces an inherent
uncertainty that cannot be quantified.  Another major uncertainty is the 
virial radius.  The true radius of these clumps is in most cases unknown, so 
assigning an uncertainty is somewhat questionable.  Our typical virial
radii are around 0.4 pc and some interferometric millimeter studies of 
IRDC clumps \citep[e.g][]{beu05} find sizes of about 0.1 pc.
Therefore, we assign an uncertainty of a factor of 4 to the virial 
radius.  The line-width is uncertain by about 20\% for these data.  These
uncertainties give us a 68\% confidence interval for the virial masses from
50\% to 170\% of the quoted value.  In cases where the adopted virial
radius is off by a wide margin, the virial mass will also be off by
that same wide margin (as the calculated mass scales linearly with the
adopted virial radius).

The quantifiable systematic uncertainties in the \13CO mass are 
the excitation temperature, integrated line intensity, and distance.  
However, there is a large uncertainty that we are unable to 
quantify associated with the optical depth, freezeout, excitation
conditions, and variable abundance of \13CO in cold, dense clumps
\citep[e.g.][]{taf02}.  These will have the effect of 
lowering the \13CO mass and column density estimates.  We assign
an uncertainty to the excitation temperature of 50\% (as above), an
uncertainty in the integrated intensity of 20\%, and an uncertainty
in the distance of 20\%.  We find that the 68\% confidence interval for the
\13CO masses is from 80\% to 120\% of the quoted mass, neglecting
additional uncertainties introduced by optical depth and freezeout effects.

\section{Analysis}
\label{sec:analysis}
\subsection{Mass Comparison}
\label{sec:masscomp}
Given the systematic uncertainties, the agreement between the various mass
tracers is reasonable, as shown in Table \ref{table-masses} and Figure
\ref{fig:masscomp}.  All of the mass tracers increase with increasing BGPS 1.1
mm dust continuum mass, except for extinction and \13CO at the highest BGPS
1.1 mm masses, as these are associated with UCHII Regions
(See Table \ref{table-activity}).  
The virial masses are uniformly higher than the BGPS 1.1 mm masses, the 
ratio of the two ranges from 1.1 to 14.4 with a median of 
6.4 $\pm$ 4.2.  A detailed comparison of the BGPS 1.1 mm mass
with the virial mass is given in \S \ref{sec:virial}. 
The MAMBO 1.2 mm
masses are also uniformly higher than the BGPS 1.1 mm masses, though 
not nearly as high as the virial masses.  The ratio of MAMBO 1.2 mm to BGPS 1.1 mm masses
ranges from 1.2 to 2.6 with a median of 1.8 $\pm$ 0.3.  
This offset is intrinsic to
data taken with MAMBO at 1.2 mm vs. Bolocam at 1.1 mm.  The reason
for this offset is being investigated by the BGPS team and will be
discussed in \citet{agu10}.
 The \13CO
masses agree surprisingly well with the BGPS 1.1 mm masses, the ratio between the 
two ranges from 0.2 to 1.9 with an median of 0.8 $\pm$ 0.5.  Recall, however, 
that the \13CO mass is taken over a 46'' beam rather than the 33'' effective beam
size of the rest, so we are including more of the GMC and the IRDC
envelope, 
and therefore more emission.
This means that the true comparable \13CO mass would be generally
\textit{lower} than the BGPS 1.1 mm mass.  The agreement between the two is
best at lower BGPS 1.1 mm masses.  At higher BGPS column there is likely a
higher than assumed BGPS temperature, as well as optical depth effects or
freezout of \13CO at high column densities \citep{taf02}.  
The extinction mass and BGPS 1.1 mm mass agree 
well, except at the highest BGPS 1.1 mm masses (M$_{BGPS}$ $>$ 600 \Msun), where
the sources are associated with UCH~II Regions.  The ratio of 
extinction to BGPS 1.1 mm mass ranges from 0.1 to 2.8 with an median of
1.1 $\pm$ 0.8.  A more detailed comparison of the extinction mass to
the BGPS 1.1 mm mass is given below.

%\clearpage
\begin{figure*}
\centering
\plotone{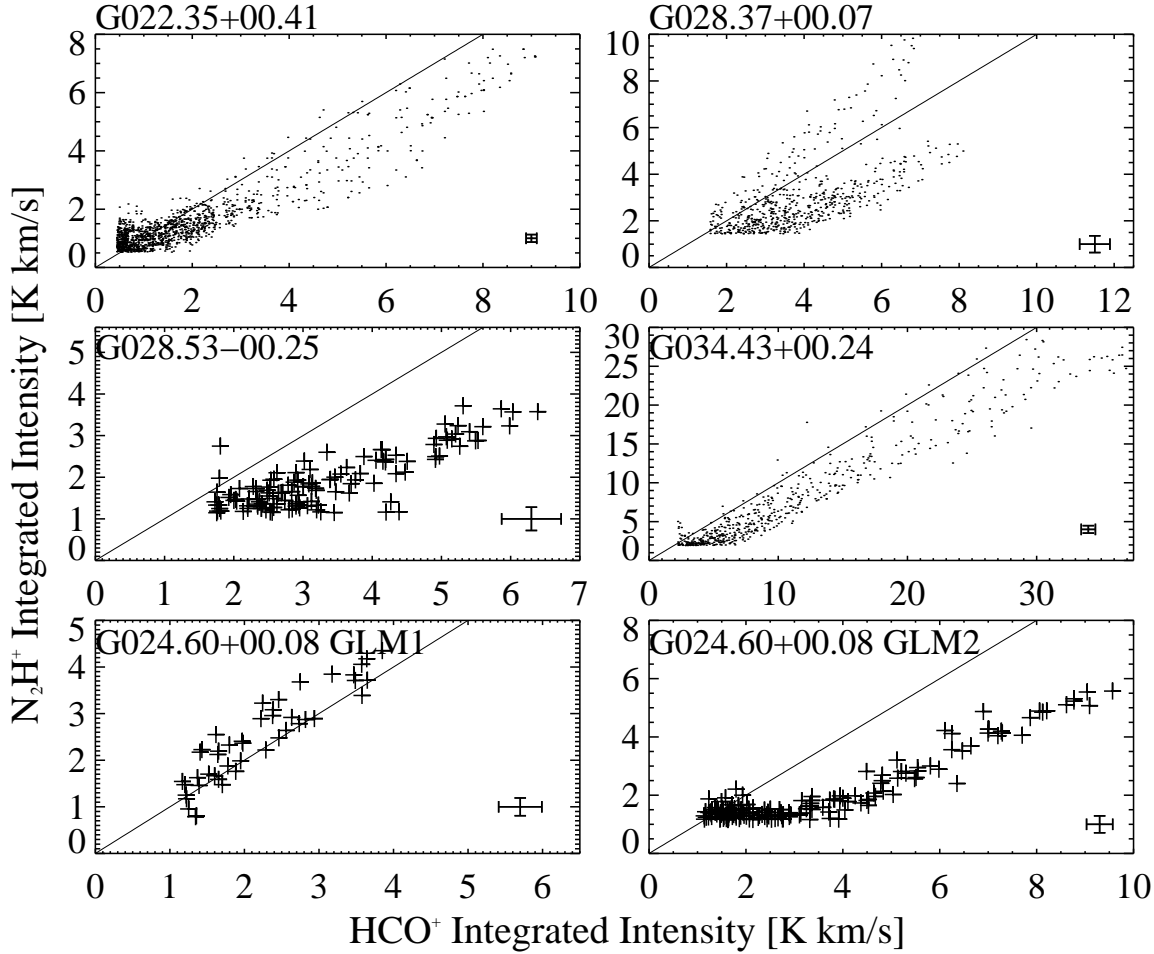}
\caption{Ratio of the integrated intensities of \n2hp~(J=3-2) to \hcop~(J=3-2).
Only pixels greater than 4$\sigma$ in \hcop~and \n2hp~are 
shown.  This is a pixel-by-pixel comparison of the \hcop~and \n2hp~points
within the IRDC ellipse.  The line drawn has a slope of one representing
equality.  The point in the lower right corner of each plot represents
the approximate 1 $\sigma$ error bars for the points in that plot.}

\label{fig:hcopvsn2hp_intint}
\end{figure*}

\subsection{Dust extinction and emission}
\label{sec:dust}
The relationship between 8~\micron~extinction and 1.1 mm dust emission
is reasonably good, and where it is not, there is good indication that
warm gas/dust due to a hot clump is skewing both extinction and
emission measures of mass.  Recent dust temperature estimates from
\citet{rat10} corroborate that active clumps are indeed warmer than
quiescent/intermediate clumps.  The agreement between 8~\micron~extinction
and 1.1 mm dust emission masses is poor in the four highest BGPS
1.1 mm mass sources, which are associated with UCH~II Regions.
In these regions, the temperature is almost certainly warmer than the
assumed 15 K, which contributes to the very large BGPS fluxes.  Additionally, the
UCH~II Region illuminates PAH features in the 8~\micron~band, boosting the
8~\micron~flux and reducing the measured extinction mass.

Figure \ref{fig:extcolvsbgpscol} shows the extinction column density
vs. BGPS 1.1 mm column density and we note that while the 
column density derived from extinction and the BGPS 1.1 mm
dust continuum generally agree there are some peculiarities.
Note that the clumps indicated in Table \ref{table-masses} are only a 
small portion of what is plotted in Figure \ref{fig:extcolvsbgpscol}.  
Figure \ref{fig:extcolvsbgpscol} plots the entire IRDC, while the
table only includes the subsections of the cloud chosen
for this study as clumps. In some sources the extinction flattens out as
the 1.1 mm column increases, while in others the extinction column is
higher than that 1.1 mm column.  These indicate a hot clump (with imperfect
temperature assumptions for the 1.1 mm dust emission and
8~\micron~contamination by PAH illumination) and either a cold clump or
imperfect foreground/background estimation, respectively.
For the individual clumps, the median of the ratio of extinction column
to BGPS column is 1.1 $\pm$ 0.8 overall, 0.7 $\pm$ 0.9 for
active clumps, 1.4 $\pm$ 0.6 for intermediate clumps and
2.0 $\pm$ 0.9 for quiescent clumps.  This is in agreement with 
our assessment that high extinction per 1.1 mm column is 
characteristic of quiescent clumps, low extinction per 1.1 mm column is
characteristic of active clumps, however, there is still a lot of scatter.
Though there are some exceptions, in general, dust extinction and emission 
are well correlated for
quiescent/intermediate clumps, but often diverge for active clumps,
particularly ones associated with UCHII Regions.

%\clearpage
\begin{figure*}
\centering
\plotone{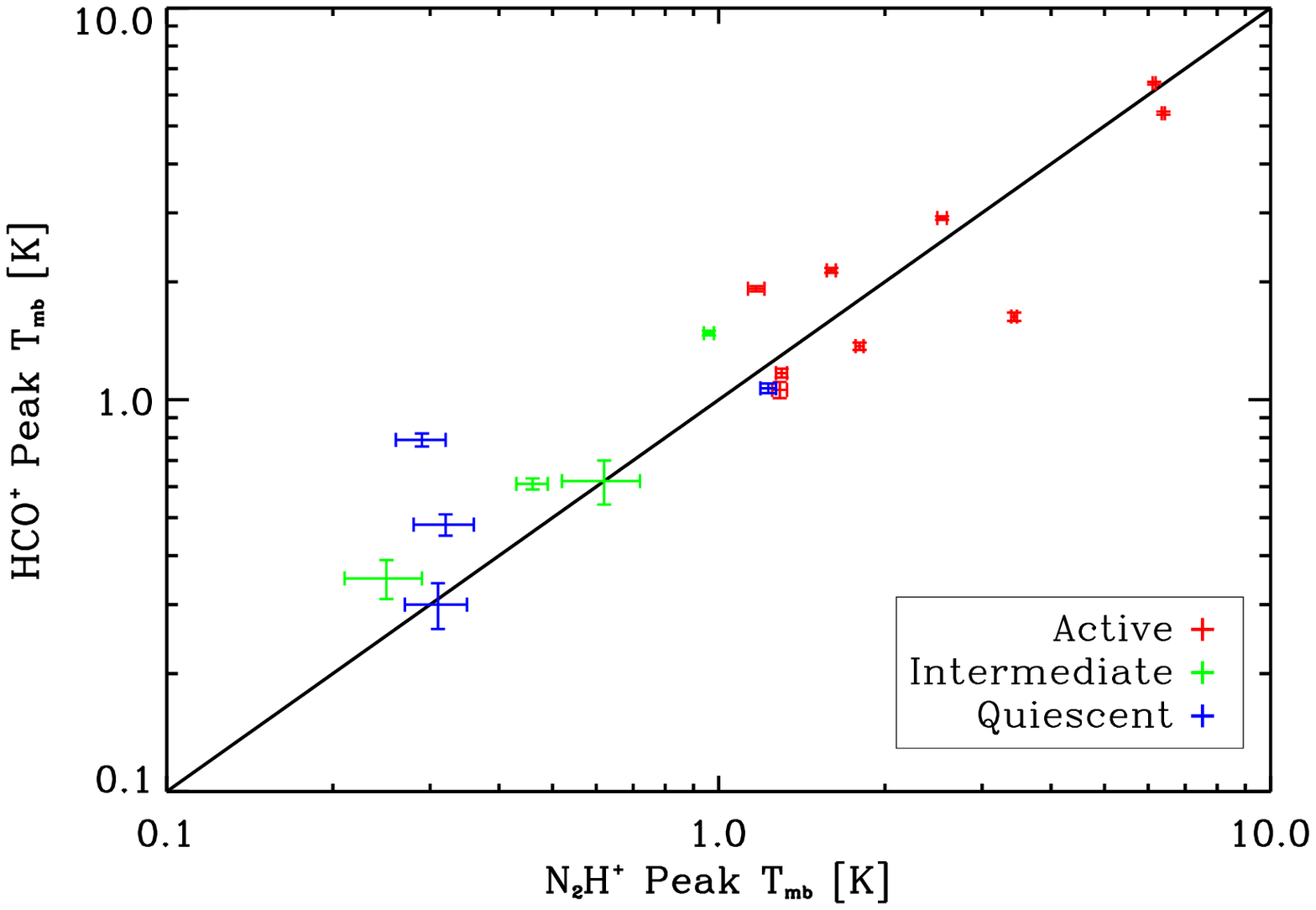}
\caption{\hcop~vs. \n2hp~Peak T$_{\rm mb}$ on a log-log plot.
  Red is active clumps,
green is intermediate and blue is quiescent (See Table 
\ref{table-activity}).  The line drawn has a slope of one, representing
  equality.  Note that there is no noticeable distinction in the 
  ratio of \hcop~to
  \n2hp~for active vs. quiescent clumps, but that active clumps are much
  brighter in \hcop~and \n2hp~than quiescent clumps.}
\label{fig:Peak_AIQ}
\end{figure*}

%\clearpage
\subsection{A Comparison of Virial and Dust Masses}
\label{sec:virial}
Simple calculations show that the measured virial masses are uniformly
higher than the masses measured through other techniques, such as the dust
continuum emission.  The median of the ratio
of the virial mass to the BGPS 1.1 mm mass is 6.4 $\pm$ 4.2, and 
ranges from 1.1 to 14.4.  
Interferometric observations \citep[for more details on specific sources,
please see the Appendix]{rat07, rat08, wan08, zha09} of several
of these sources find that most clumps fragment into smaller cores at
higher resolution, with sizes of about 0.03 pc or less.
\citet{zha09}
find that on the scale of $\sim$ 0.02 pc the virial and dust mass estimates
agree quite well.  On the scale of about 1 pc, we find that the virial
masses are much higher than the dust masses.  The virial radius required 
for our measured virial masses to match our measured BGPS
1.1 mm masses ranges from 0.02 to 0.25 pc with a mean of 0.09 $\pm$ 0.06
pc, similar to the interferometric size scales noted above.
The information presented is consistent with the picture of small bound
cores (about 0.01 pc in size), surrounded by a dense envelope (about 0.1 pc
in size) which is also bound, engulfed in a diffuse envelope (size 
scale of about 1 pc).  This
dense envelope is where the majority of dense gas is emitting and
therefore dominates the line width we measure, while the diffuse
envelope extends out to about 1 pc and comprises the extended emission
we see in the dust continuum.  In this picture, the line widths we measure
are primarily from a region that is smaller than the virial radius we
assign to it, thus explaining the very large virial to dust mass ratio.
Alternatively, if the material producing the large line widths and the 
dust continuum emission are the same, it is possible that
something more than virialized motion (e.g. magnetic fields, turbulence) 
are playing a role in the line widths or that the clumps are not
graviationally bound, and possibly ram-pressure confined by converging flows
\citep[e.g.][]{hei08}.
The question of whether these clouds, clumps, and cores are bound is of
great importance in understanding cluster formation.

\begin{figure*}
  \centering
  \includegraphics[width=1\textwidth]{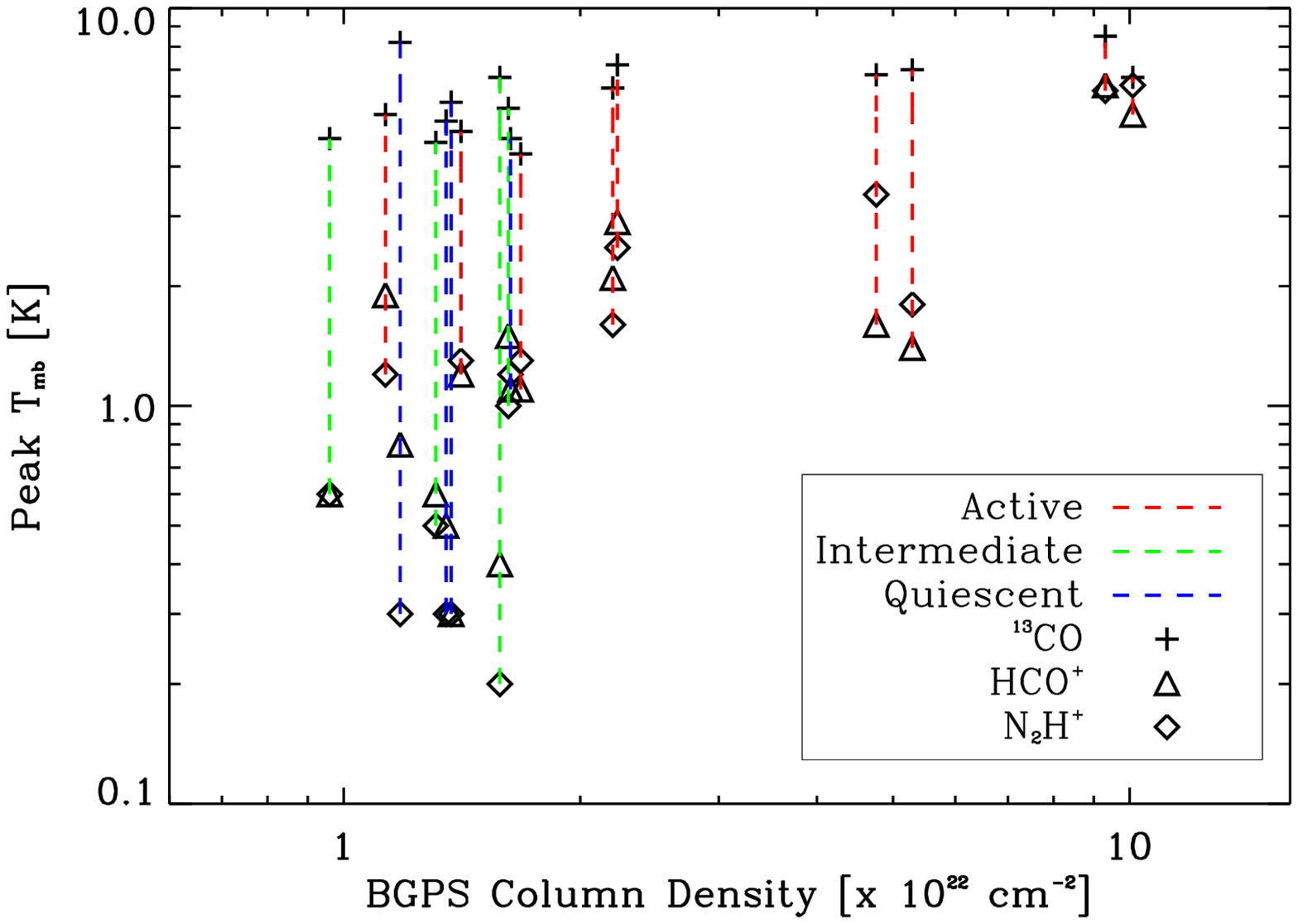}
  \caption{A comparison of the peak main beam temperatures for \13CO (cross),
    \hcop~(triangle), and \n2hp~(diamond) vs. BGPS column density.  
    Each dashed line represents
    a single clump, and the color of the line denotes whether it is active
    (red), intermediate (green), or quiescent (blue).  Note that \13CO is
    relatively flat with increasing column density, 
    corroborating that it is optically thick.  The
    variation in line strength between the various tracers 
    (the length of the dashed line drawn) is
    much smaller for active vs. quiescent clumps and is a potentially
    useful diagnostic for understanding star formation activity.}
  \label{fig:peaklineplot}
\end{figure*}

%\clearpage
\subsection{A Molecular Line Tracer Comparison}
\label{sec:mol}
The ratio of \hcop~to \n2hp~in clumps of varying activity levels is of special
interest, as it is thought that \hcop~will trace ``hot''
clumps while \n2hp~will trace ``cold'' clumps \citep{jor04}.  \hcop~is created by
CO in the gas phase, and will also deplete onto dust grains at cold
temperatures, so it should be more abundant in hotter environments.
\n2hp, on the other hand, is destroyed by CO in the gas phase, 
and should be more abundant in colder environments.  Figure
\ref{fig:hcopvsn2hp_intint} is a pixel-by-pixel comparison of \n2hp~to
\hcop~integrated intensities, and the line is of slope one, representing
equality.  Throughout the majority of the clouds, the ratio of \hcop~to
\n2hp~is greater than one.  In a few cases (most notably G028.37+00.07),
however, the \n2hp~integrated intensity is greater than the \hcop.  
This is due to a high column of gas and dust, in 
which \hcop~is optically thick, and self-absorbed which lessens the peak
(e.g. Figure \ref{fig:irdc51_allspectra}),
making \n2hp~brighter by comparison.  The general trend is that \hcop~and
\n2hp~are well correlated and that \hcop~is slightly brighter.
In Figure \ref{fig:Peak_AIQ} we see that our present sample shows
no trend between clump activity and the ratio of \hcop~
to \n2hp.  
It is important to note that we are
not resolving individual cores, so adjacent cores with differing 
chemistry will result in an average, rather than 
resembling distinct core chemistries.  Additionally, we are
comparing line strengths, not abundances.

While we see no evidence for a chemical differentiation between
\hcop~and \n2hp~on a $\sim$ 1 pc scale between ``active'' and ``quiescent''
clumps, we note that active clumps are significantly brighter in
\hcop~and \n2hp~than intermediate or quiescent clumps.
The increased brightness of \hcop~and \n2hp~in active clumps could be the
result of an increased column density, higher volume density, higher
temperature, and/or a larger beam filling factor.
We argue that an increase in column is
not the primary effect as not all active clumps have higher dust column, yet all
active clumps show an increased brightness in \hcop~and \n2hp.  The change
in filling factor is a possible explanation, however the \hcop~and \n2hp~maps
show extended emission around all the clumps.
Since at the typical density of an IRDC (n $\sim$ 10$^{5}$ cm$^{-3}$) we
are approaching the critical densities of \hcop~and \n2hp
\citep[see][]{dan07}, we suggest that
the active clumps may have begun to compress and become denser, 
increasing the intensity of \hcop~and \n2hp.  This increase in
density is also responsible for the increased optical depth of \hcop~in
active clumps, as evidenced by the self-absorption of \hcop~in some active clumps.
Recent dust temperature estimates from \citet{rat10} suggest that active
clumps are warmer, which would also increase the observed brightness
of \hcop~and \n2hp.  We observe that \hcop~and \n2hp~are brighter in active
than quiescent clumps and suggest that this is because the active clumps
are either warmer, have higher volume densities, or both.

We also compare the \hcop~and \n2hp~peak line intensities with those of
\13CO in Figure \ref{fig:peaklineplot}.  As discussed in Section
\ref{sec:13comass}, while there is good evidence that \13CO is optically
thick toward many of these clumps, we see that it at least traces to
some depth in the cold, dense portion of the clump.  However, the \13CO is
almost certainly tracing some part of the GMC and IRDC envelope as well.
Therefore, we include
it as an optically thick comparison point for the dense gas tracers.
The \13CO peak varies very little among the 
different clump types and shows no noticeable increase with column, again
corroborating that it is optically thick. The \hcop~and \n2hp
peak temperatures increase somewhat with column density, but most notably
with activity level.  This means that the active clumps all have much
smaller variation in line strength than the quiescent clumps (the line
connecting the peak T$_{mb}$'s is shorter).  The variation in line 
strength among the three tracers can be quantified by the difference 
between the brightest peak line strength of the three tracers and 
the weakest peak line strength of the three tracers.  The mean of the
variation in line strength is 4.5 $\pm$ 0.8 for active clumps, 4.8 $\pm$ 1.1
for intermediate clumps, and 5.5 $\pm$ 1.8 for quiescent clumps.  While
the numbers are uncertain, the trend is clear, and this is a potentially
useful tool for identifying active clumps.

%%%%%%%%%%%%%%%%%%%%%%%%%%%%%%%%%%%%%%%%%%%%%%%%%%%%%%%%%%%%%%%%%%%%%%%%%%
% Clump Activity
%%%%%%%%%%%%%%%%%%%%%%%%%%%%%%%%%%%%%%%%%%%%%%%%%%%%%%%%%%%%%%%%%%%%%%%%%%

%\clearpage
\begin{deluxetable*}{lcccccccc}
%\rotate
\tabletypesize{\scriptsize}
\setlength{\tabcolsep}{0.05in}
\tablecaption{Clump Activity \label{table-activity}}
\tablewidth{0pt}
\tablehead{
\colhead{Clump} & \colhead{3.6 cm Flux\tablenotemark{a}} & 
\colhead{H~II Region} & \colhead{``Green Fuzzy"?\tablenotemark{b}} &
\colhead{24 \micron flux\tablenotemark{b}} 
& \colhead{H$_{2}$O maser?\tablenotemark{b}} &
\colhead{CH$_{3}$OH maser?\tablenotemark{b}} &
\colhead{Clump\tablenotemark{c}} & \colhead{Evolutionary\tablenotemark{d}} \\
\colhead{} & \colhead{(mJy)} & \colhead{Type} &
\colhead{Y/N} & \colhead{(mJy)} & 
\colhead{(Y/N)} & \colhead{(Y/N)} & 
\colhead{Activity} & \colhead{Stage}}

\startdata

G022.35+00.41                    &       &        &     &      &     &     &    &    \\
  ~~ GLM1                  &  0.06 &  --    &  Y  &  13  &  Y  &  Y  & A  &  2\\
  ~~ GLM2                  &  0.36 &  --    &  N  &  --  &  N  &  N  & Q  &  1 \\
G023.60+00.00                   &       &        &     &      &     &     &    &    \\
  ~~ GLM1(VLA4)            &  0.30 &  UCH~II &  Y  & 1058 &  Y  &  N  & A  &  3 \\
  ~~ GLM2                  &  0.12 &  --    &  N  & 198  &  Y  &  N  & I  &  2\\
G024.33+00.11                   &       &        &     &      &     &     &    &    \\
  ~~ GLM1(VLA2)            &  0.29 &  UCH~II &  N  &  999 &  Y  &  Y  & A  &  3 \\
  ~~ GLM2                  &  0.33 &  --    &  N  &  --  &  N  &  N  & Q  &  1 \\
G024.60+00.08                    &       &        &     &      &     &     &    &    \\
  ~~ GLM1                  &  0.09 &  --    &  N  &  13  &  Y  &  N  & I  &  2\\
  ~~ GLM2                  &  0.09 &  --    &  Y  &  342 &  N  &  N  & I  &  2\\
G028.23-00.19                    &       &        &     &      &     &     &    &    \\
  ~~ GLM1                  &  0.03 &  --    &  N  &  --  &   N &  N  & Q  &  1 \\
G028.37+00.07                    &       &        &     &      &     &     &    &    \\
  ~~ GLM1                  &  0.18 &  --    &  N  &  --  &   N &  N  & Q  &  1 \\
  ~~ GLM2                  &  0.12 &  --    &  Y  &  45  &   Y &  N  & A  &  2\\
  ~~ GLM3                  &  0.15 &  --    &  Y  &  4   &   Y &  N  & A  &  2\\
  ~~ GLM4\tablenotemark{e} &  0.12 &  --    &  Y  &  22  &   Y &  Y  & A  &  2\\
G028.53-00.25                   &       &        &     &      &     &     &    &    \\
  ~~ GLM1                  &  0.06 &  --    &  Y  &  16  &   N &  N  & I  &  2\\
G034.43+00.24                   &       &        &     &      &     &     &    &    \\
  ~~ GLM1(VLA3)            &  0.31 &  UCH~II\tablenotemark{f} &  Y  & 1718 &   Y &  Y  & A &  3    \\
  ~~ GLM2(VLA1)            &  10.07&  UCH~II &  N  & 1401 &   Y &  Y  & A  &  4 \\
  ~~ GLM3                  &  0.09 &  --    &  Y  &  --  &   Y &  Y  & A  &  2\\

\enddata
\tablenotetext{a}{For point sources this is the peak flux, for
the extended source, G034.43+00.24:~GLM2, the flux is the integrated flux, and
for no detection\\
 (as marked by a -- in the 3rd column) it is the 3$\sigma$
upper limit.}
\tablenotetext{b}{\citet{cha09}}
\tablenotetext{c}{A=Active, I=Intermediate, Q=Quiescent}
\tablenotetext{d}{See \S \ref{sec:evseq} for details.}
\tablenotetext{e}{G028.37+00.07:~GLM4 slightly overlaps with the resolved H~II
  region, G028.37+00.07 VLA1, see Table \ref{table-vla}}
\tablenotetext{f}{G034.43+00.24:~GLM1 is a marginal detection.  It may or may not
  be an UCH~II region.  See \S \ref{sec:hiiregions} for details.}
\end{deluxetable*}

%%%%%%%%%%%%%%%%%%%%%%%%%%%%%%%%%%%%%%%%%%%%%%%%%%%%%%%%%%%%%%%%%%%%%%%%%%
% VLA Sources
%%%%%%%%%%%%%%%%%%%%%%%%%%%%%%%%%%%%%%%%%%%%%%%%%%%%%%%%%%%%%%%%%%%%%%%%%%

%\clearpage
\begin{deluxetable*}{lcccccc}
\tabletypesize{\scriptsize}
\tablecaption{Radio Continuum Sources \label{table-vla}}
\tablewidth{0pt}
\tablehead{
\colhead{VLA Source} & \colhead{R.A.} & \colhead{DECL.} &
\colhead{Peak 3.6 cm Flux} & 
\colhead{Total 3.6 cm Flux} & \colhead{$\sigma$} & %\tablenotemark{a}} &
\colhead{Point Source\tablenotemark{a}} \\
\colhead{} & \colhead{(J2000)} & \colhead{(J2000)} &
\colhead{(mJy/beam)} & \colhead{(mJy)} & \colhead{(mJy/beam)} &
\colhead{(Y/N)}}

\startdata

G022.35+00.41        &            &             &       &       &      &    \\
G023.60+00.00       &            &             &       &       &      &    \\
 ~~ VLA1       & 18:34:33.1 & -08:15:26.8 & 24.3  & 28.4  & 0.3  & Y  \\
 ~~ VLA2       & 18:34:09.6 & -08:17:49.4 & 1.5   & 68.9  & 0.1  & N  \\           
 ~~ VLA3       & 18:34:12.4 & -08:19.01.5 & 1.1   & 1.3   & 0.1  & Y  \\
 ~~ VLA4 (GLM1) & 18:34:21.1 & -08:18:12.4 & 0.3   & 0.3   & 0.1  & Y  \\
G024.33+00.11       &            &             &       &       &      &    \\
 ~~ VLA1       & 18:35:24.0 & -07:37:37.9 & 2.8   & 3.6   & 0.5  & Y  \\
 ~~ VLA2 (GLM1) & 18:35:08.1 & -07:35:04.0 & 0.29  & 0.4  & 0.02 & Y  \\
G024.60+00.08        &            &             &       &       &      &    \\
G028.23-00.19        &            &             &       &       &      &    \\
 ~~ VLA1       & 18:43:33.7 & -04:11:48.2 & 0.24  & 0.4  & 0.03 & Y  \\
G028.37+00.07        &            &             &       &       &      &    \\
 ~~ VLA1       & 18:42:52.9 & -04:00:09.2 & 3.0   & 49.9  & 0.5  & N  \\
 ~~ VLA2       & 18:42:41.4 & -04:02:16.9 & 1.7   & 9.6   & 0.4  & N  \\
 ~~ VLA3       & 18:43:03.5 & -04:00:13.2 & 1.98  & 2.2  & 0.08 & Y  \\
 ~~ VLA4       & 18:42:37.1 & -04:02:02.3 & 1.5   & 1.7   & 0.2  & Y  \\
 ~~ VLA5       & 18:43:04.5 & -04:00:15.0 & 1.16  & 1.5  & 0.08 & Y  \\
 ~~ VLA6       & 18:42:52.4 & -03:59:07.0 & 0.38  & 0.4  & 0.04 & Y  \\
G028.53-00.25       &            &             &       &       &      &    \\
G034.43+00.24       &            &             &       &       &      &    \\
 ~~ VLA1 (GLM2) & 18:53:18.7 & +01:24:47.2 & 8.79  & 10.1 & 0.07 & N  \\
 ~~ VLA2       & 18:53:08.4 & +01:29:33.9 & 1.00  & 1.1  & 0.08 & Y  \\
 ~~ VLA3 (GLM1) & 18:53:18.1 & +01:25:25.4 & 0.31  & 0.4  & 0.06 & Y  \\

\enddata
\tablenotetext{a}{Extended sources are plotted in Figures
  \ref{fig:irdc33_hcop_n2hp}-\ref{fig:irdc43_hcop_n2hp}.}
\end{deluxetable*}

%\clearpage
\subsection{H~II Regions and Stellar Type Limits}
\label{sec:hiiregions}
We detect radio continuum sources toward four of the clumps.  
Two of these are unresolved, optically thick UCH~II regions.
One is a slightly resolved, optically thin UCH~II region.
The final source is an unresolved, marginal detection which may 
be an UCH~II region.
The peak fluxes of these regions are given in Table
\ref{table-activity}, as well as the limiting (3$\sigma$) flux 
in the clumps without detections.  The properties of all the detections 
(outside or inside the clumps)
are reported in Table \ref{table-vla}.  The sources are
named based on the IRDC in which they were found (though they are not
necessarily associated with the cloud) and in order of decreasing
brightness (VLA1, VLA2, etc.).  All the Gaussian-type sources
were fit with 2-D elliptical Gaussians, and the
fluxes reported are from the fits.  The $\sigma$ reported is the r.m.s. of
the fit residuals.  For the sources with non-Gaussian morphology, the
fluxes were measured using a circular aperture and an outer annulus of equal area.
The mean of the flux in the outer annulus is assumed to be the background
level and is subtracted from the source flux, while the standard deviation in the
outer annulus is our quoted $\sigma$.
Some of these detections are more evolved H~II regions, peripheral to the IRDC. 
Others are point sources unassociated with any GLIMPSE emission.  These 
sources may be unresolved H~II regions that happen to be in our field of
view but unassociated with the mid-IR, 
or they may be other types of Galactic or extragalactic radio continuum sources.
We restrict our analysis to the four radio continuum sources
associated with the millimeter clump peaks,
as these are almost certainly UCH~II regions in the IRDC.

\begin{figure*}
\centering
\includegraphics{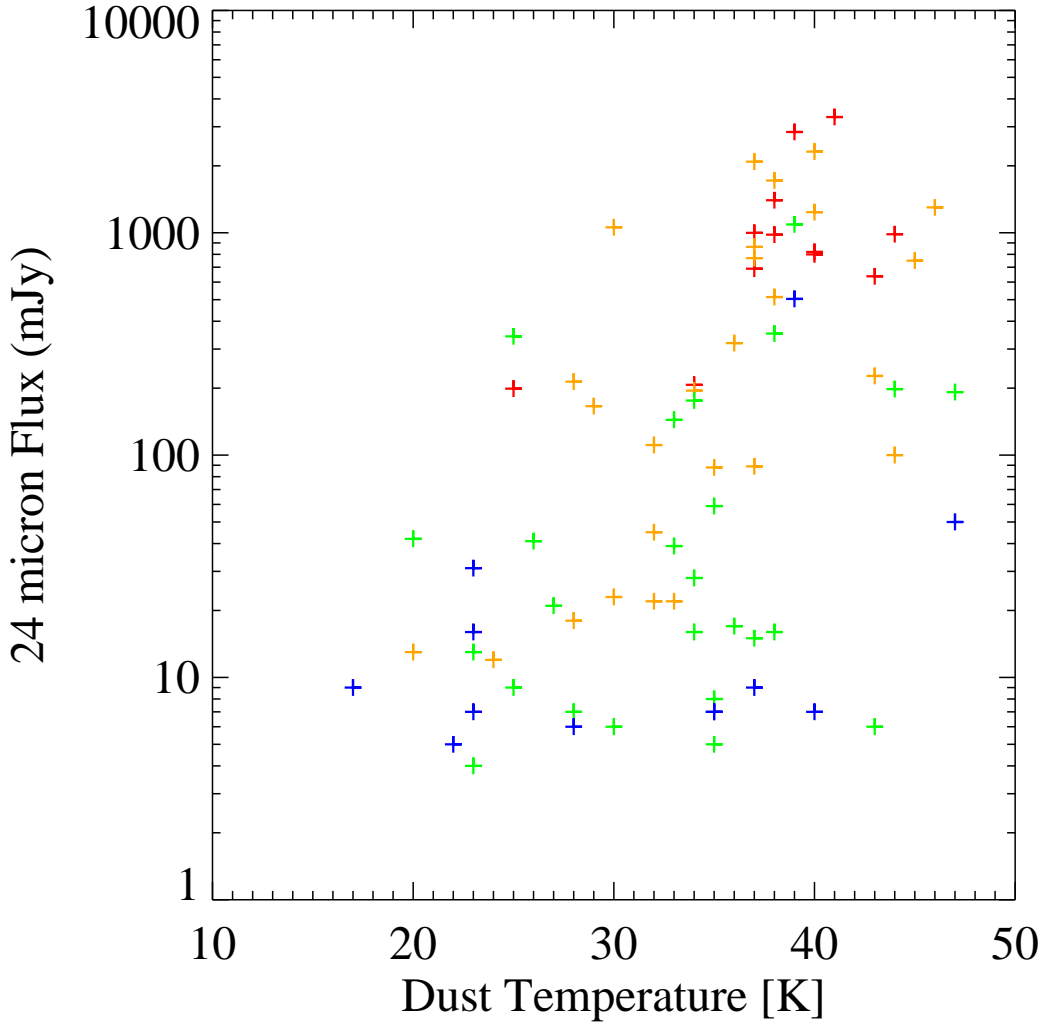}
\caption{24~\micron~flux \citep[as measured by][]{cha09} vs. dust
  temperature \citep[from][]{rat10} for the red: ``red'' clumps,
  orange: clumps with three star formation tracers, green: two star
  formation tracers, and blue: one star formation tracer.  Note the
  that ``red'' clumps are associated with brighter 24~\micron~flux and that
  in general, the more tracers of star formation activity, the warmer the clump.}
\label{fig:24flux}
\end{figure*}

The UCH~II region radio continuum flux is directly related to
the number of Lyman continuum photons (Q) of the ionizing star
if the region is optically thin.  Through comparison
with the 20 cm and 6 cm radio continuum data from the 
Multi-Array Galactic Plane Imaging Survey \citep[MAGPIS;][]{whi05}
we find that G034.43+00.24:~GLM2 (VLA1) is
slightly resolved ($D\sim0.14$ pc) and optically thin at 3.6 cm. 
G034.43+00.24:~GLM1 (VLA3) is a marginal detection in both our survey and the 6 cm survey by 
\citet{she04}.  The measured flux densities at 3.6 and 6 cm are not consistent with
the source being an UCH~II region.  However, the fluxes reported are 
highly uncertain, and systematic errors 
could be large enough to account for this inconsistency if it is in fact an
UCH~II region.  Since an optically thick UCH~II region is the most
likely explanation for a radio continuum source 
associated with star formation in an IRDC clump, we assume that this is the
explanation for the radio continuum source.  We use the 3.6 cm flux to
calculate the Lyman continuum flux of an unresolved, optically thick, UCH~II
region, but acknowledge that this assumption may be incorrect.
For G024.33+00.11:~GLM1 (VLA2) and G023.60+00.00:~GLM1 (VLA4)
the 6 cm upper limits are consistent with the source being an optically thick
UCH~II region at 3.6 cm.

In the optically thin case (G034.43+00.24: GLM2 (VLA1)), 
we fit a thermal bremsstrahlung curve to the
3.6, 6, and 20 cm points.  We find an emission measure EM = 4.4
$\times$ 10$^{6}$ cm$^{-6}$ pc, a turnover frequency $\nu(\tau=1)$ = 1.3 GHz, and Q
= 6.1
$\times$ 10$^{45}$ Lyman continuum photons per second.  This translates to
a B0.5 star \citep{vac96}, in agreement with \citet{she04}.

For the unresolved, optically thick sources, we first derive
a source radius
\begin{equation}
r = \Bigg[\frac{S_{\nu}c^{2}}{2\nu^{2}kT_{e}}4D^{2}\Bigg]^{1/2}
\end{equation}
where S$_{\nu}$ is the radio continuum flux, $\nu$ is the band-center
frequency, T$_{e}$ is the ionized gas temperature, and D is the
distance.  We assume a T$_{e}$ = 8000 K ionized gas, typical of an H~II
region. We conservatively
assume that the observed frequency is the turnoff frequency (where $\tau$ 
=1).  This gives us a lower limit for the emission measure (EM), electron
density (n$_{e}$), and number of Lyman continuum photons (Q).
If our observing frequency is the turnover frequency ($\nu$)
then $\tau$ = 1 in our expression for the emission measure 
\citep{woo89}
\begin{equation}
EM ({\rm cm^{-6}}~{\rm pc)} = \frac{\tau}{8.235\times10^{-2}
\alpha(\nu,T_{e})T_{e}^{-1.35}\nu^{-2.1}}.
\end{equation}
The lower limit for the number density of electrons, which we take to be
equal the number density of ions, is
\begin{equation}
n_{e} = \sqrt{\frac{EM}{2r}}.
\end{equation}
The number of Lyman continuum photons is given by the Stromgren sphere
equation, 
\begin{equation}
Q = \frac{4}{3}\pi r^{3}\alpha_{B}n_{e}^{2}
\end{equation}
where $\alpha_{B}$ is the recombination rate coefficient, 3.1 $\times$
10$^{-13}$ cm$^{3}$ s$^{-1}$ for T$_{e}$ = 8000 K.  The Stromgren sphere
equation assumes that the H~II region is spherical, in equilibrium, that
there is one electron per ion, and that each energetic photon ionizes an
atom or molecule.  Since the observed structure is clumpy, the Stromgren
sphere approximation is probably not a very good one.  
Using this method for the two (possibly three) optically
thick, unresolved UCH~II regions yields a log(Q) of 46.14, 46.55, and 46.15
and stellar types \citep{vac96} of B0, O9.5, and B0 
for G023.60+00.00:~GLM1 (VLA4), G024.33+00.11:~GLM1 (VLA2), and G034.43+00.24:~GLM1 (VLA3)
respectively.

%\clearpage
%\begin{rotate}
\begin{figure*}
  \centering
  \includegraphics[scale=0.45]{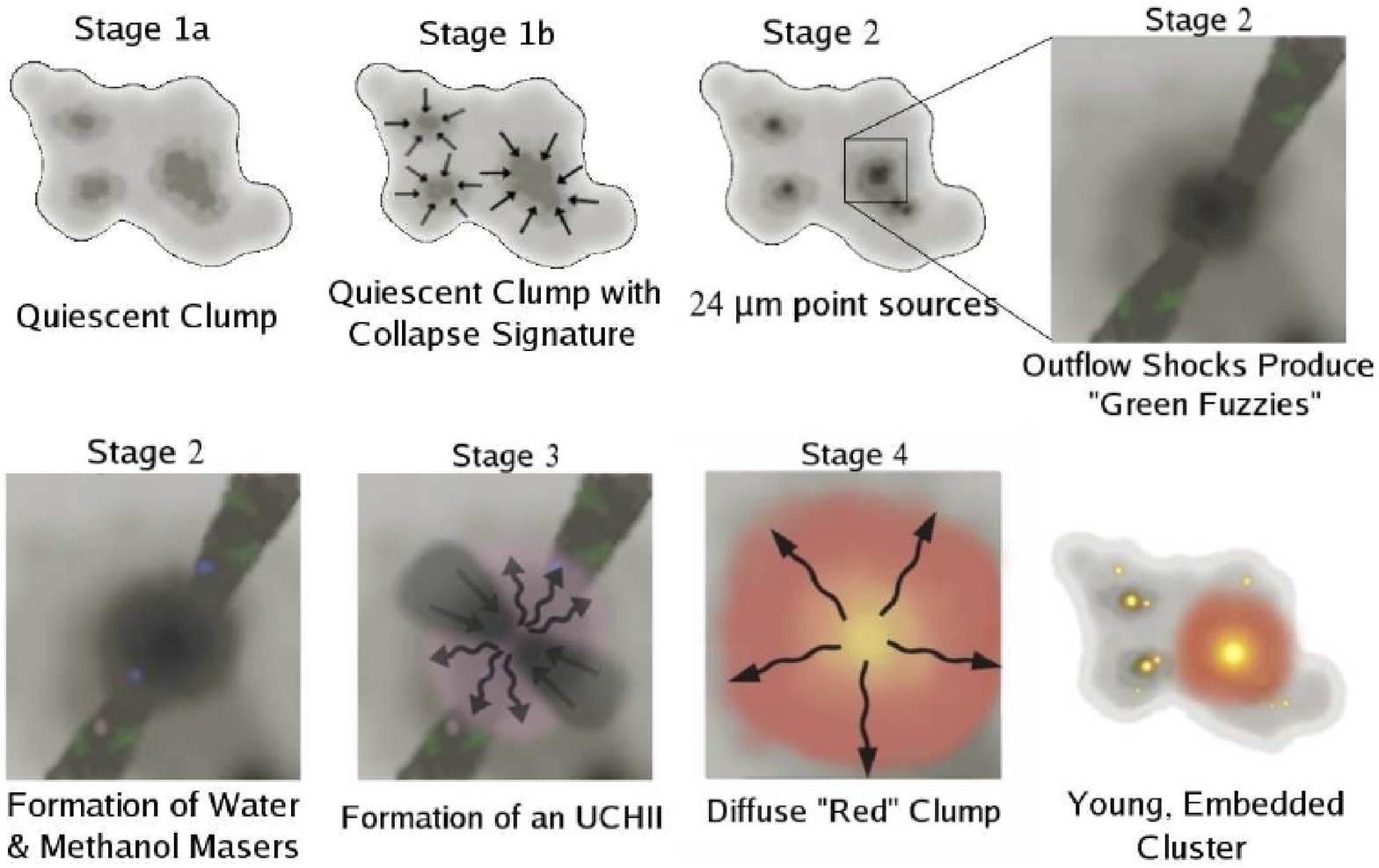}
  \caption{A cartoon depiction of the discussed evolutionary sequence.  See
  \S \ref{sec:evseq} for details.}
  \label{fig:cartoon}
\end{figure*}
%\end{rotate}

%\clearpage
%\begin{rotate}

%\clearpage
\subsection{Clump Activity}
\label{sec:activity}
A major driving question in the study of IRDCs is the evolutionary 
sequence of the clumps and clouds: identifying the different 
evolutionary stages and their relative lifetimes.  This requires an
understanding of the star formation activity.  We emply four different
star formation tracers: 1) an embedded 24 \micron ~point source, 
2) ``green fuzzies,'' regions of enhanced, extended 4.5 \micron~emission, 
indicative of shocks and outflows, 3) H$_{2}$O and Class I CH$_{3}$OH 
maser emission, and finally, 4) Ultra-Compact (UC) H~II regions, 
an unambiguous indication that a massive star that has ``turned on.''
Tracers of star formation activity in these clumps are
summarized in Table \ref{table-activity}.  

Each of these tracers is sensitive to somewhat different environments.
A 24 \micron~point source traces material accreting onto forming
stars, the gravitational contraction of a young stellar object (YSO), or even the
nuclear energy emitted once the star has ``turned on.''  
``Green fuzzies'' \cite[also called Extended Green Objects, EGOs]{cha09, cyg08} 
are thought to arise from shock-excited spectral line emission
\citep{mar04, nor04} due to outflows from young stars.
Many of the \hcop~and \n2hp~spectra (see Figures
\ref{fig:irdc6_allspectra} to \ref{fig:irdc43_allspectra}) show asymmetric
line profiles commonly associated with outflows, many of which
(e.g. G034.43+00.24 GLM3, Figure \ref{fig:irdc43_allspectra}) are
associated with ``green fuzzies,'' further implicating ``green fuzzies'' 
as positive outflow tracers.
H$_{2}$O (22.23 GHz) and CH$_{3}$OH (Class I 24.96 GHz) masers are 
well-known signposts of star formation activity, likely formed 
in shocks and outflows.  Finally, we include the presence of 3.6 cm radio
continuum emission which when associated
with a millimeter peak or mid-IR emission is a clear indicator
of an UCH~II region.  

Together, these tracers are sensitive to warm, embedded dust indicative of 
accretion, shocks and outflows, and thermal bremsstrahlung from 
newly formed H~II regions.  While these are all excellent tracers
of star formation, none are perfect, so we would not expect to see
every tracer in every actively star-forming region.  Viewing angle or dust
obscuration could play a significant role in a non-detection.  We assign the
designation ``active'' to clumps that exhibit three or four signs of
active star formation (``green fuzzy,'' 24 \micron~point source, 
UCH~II region, or maser emission).  This ensures
that each ``active'' clump has either an H~II region (unambiguously 
star-forming) or at least two outflow tracers (maser emission or ``green
fuzzy'') and a 24 \micron~point source.  We assign the designation
``quiescent'' to clumps that exhibit no signs of active star
formation.  We reserve the title ``intermediate'' for clumps 
which exhibit one or two signs of active star formation.  Intermediate
clumps, therefore, may have only shock/outflow signatures or a 24 \micron~
point source.  These categorizations represent a crude separation
between the quiet, massive-starless clumps and the active, 
star-forming clumps, not an evolutionary sequence.  
Evolutionary sequences are discussed in \S \ref{sec:evseq}.  

The detection of several UCH~II regions embedded in 
IRDC clumps unambiguously shows that some IRDCs are forming
massive stars.  We note that all the clumps associated with
UCH~II regions have significantly brighter ($\gtrsim$ 1 Jy) 24
\micron~flux.  All but one of the UCH~II regions are associated with a
CH$_{3}$OH maser and two of the four have a ``green fuzzy.''  The lack of
``green fuzzies'' and a CH$_{3}$OH maser in some sources may indicate 
that the outflow stage has ceased, or could simply be related to an
unfavorable viewing angle.  We also note that the brighest UCH~II regions 
(G034.43+00.24: GLM2 and near the edge of G028.37+00.07: GLM4) are 
associated with ``diffuse red clumps,'' regions of extended, enhanced
8~\micron~emission.  The ionizing radiation from a young H~II region will
excite PAH features in the 8~\micron~band.  As the H~II region grows it
will begin to clear some of the surrounding dust and gas and reveal itself
as a ``diffuse red clump.''  As the H~II region continues to evolve, there
will be PAH destruction by the strong UV flux and it will no longer have
enhanced 8~\micron~emission.  The clumps which contain UCH~II regions, but 
are not classified as ``red'' are either in an earlier evolutionary phase 
or are perhaps still obscured by the dust at 8~\micron. There are no 
``diffuse red clumps'' in our sample
which are not associated with UCH~II regions, so these are potentially very
useful tracers of young H~II regions.  

\citet{cha09} discussed the use of ``red clumps'' (regions of enhanced
8~\micron~emission) as tracers of embedded 
H~II regions.  Since an enhancement at 8~\micron~could presumably be 
caused by any significant UV flux
and most high column density clumps are nearly
optically thick at 8~\micron, we suggest that only the more diffuse ``red
clumps'' are necessarily associated with H~II regions.  Only when the
forming star has cleared out some dust obscuration and is no
longer associated with a millimeter \textit{peak} can it be identified as
a ``diffuse red clump.'' 
Other ``red clumps'' could be embedded UCH~II regions that have
not cleared out their envelopes, or they could be PAH nebulae excited by
main sequence or evolved AB stars.

Figure \ref{fig:24flux} shows the 24~\micron~flux \citep[as measured
by][]{cha09} vs. the dust temperature \citep[measured through SED fits 
in the sample of 100 clumps by][]{rat10}.
Notice that the ``red'' clumps are associated
with brighter 24~\micron~flux and that in general, the more star formation
tracers, the warmer the clump is.  This general trend indicates that these
tracers combined are successfully measuring star formation activity.
However, the large scatter may be due to
viewing angle biases, mis-assigned star formation tracers, or the
inadequacy of single-temperature SED fits.

\begin{figure*} 
  \centering
  \subfigure{
    \label{fig:sftracerhist}
    \includegraphics[width=0.46\textwidth]{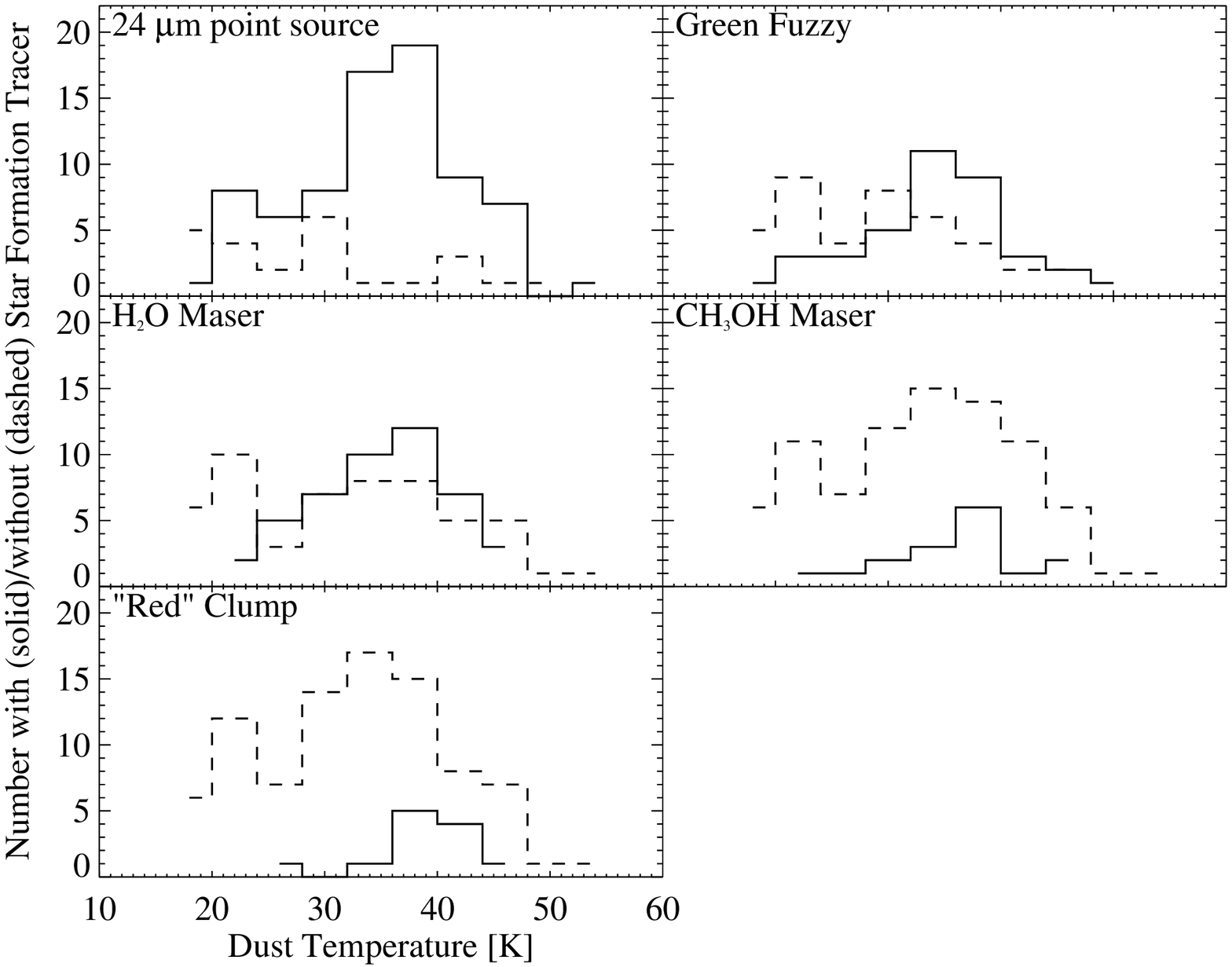}}
%  \vspace{-0.2in}
  \subfigure{
    \label{fig:stageshist}
    \includegraphics[width=0.46\textwidth]{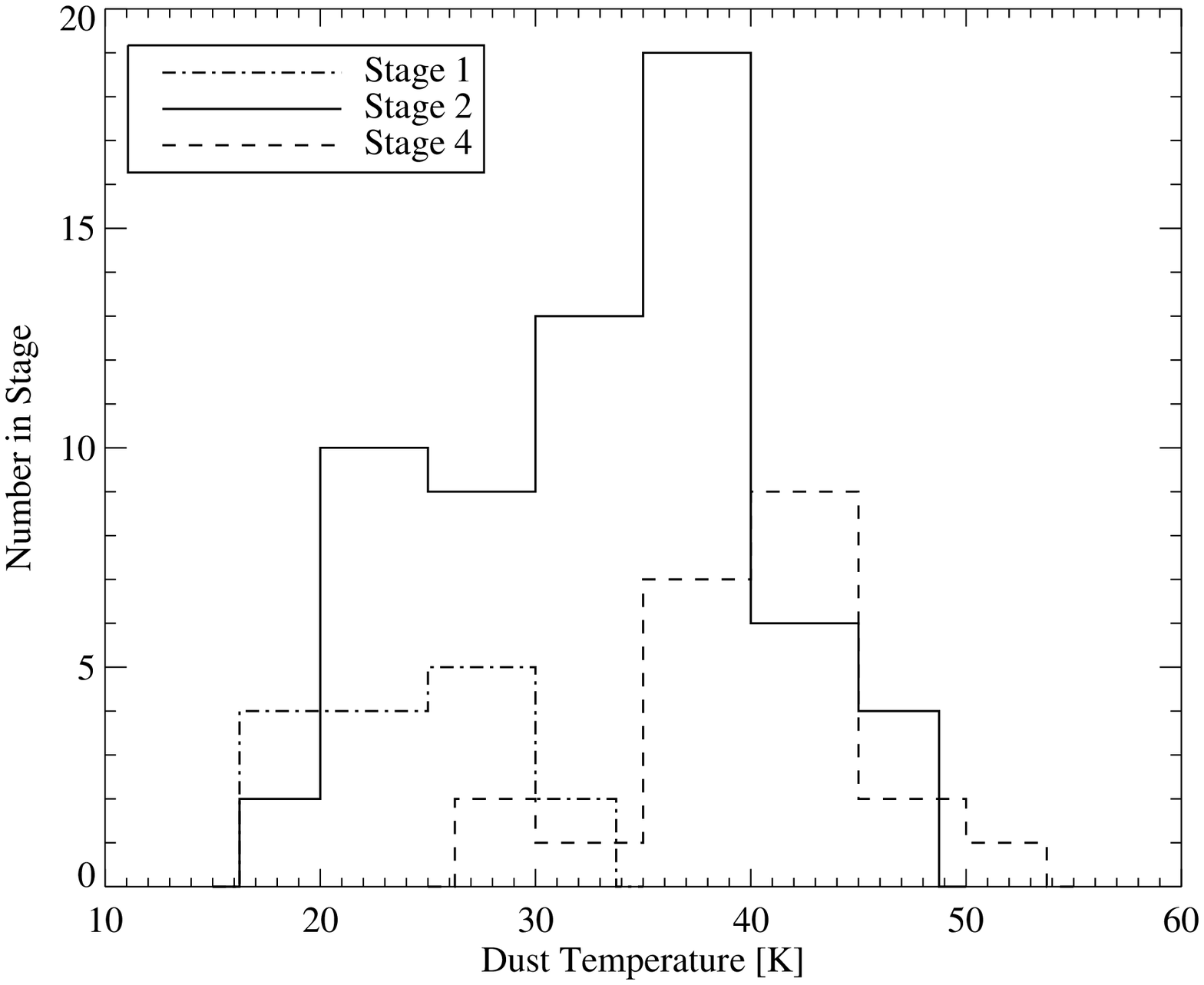}}\\
\caption{Left: For each star formation activity tracer used, we plot a dust
  temperature \citep[as measured by][]{rat10} histogram
  of the clumps with (solid line) and without (dashed line) that star
  formation tracer \citep[as measured by][]{cha09} from the sample of
  \citet{rat10}.  Note that all the star formation tracers exist over a
  wide range of temperatures, except for ``red'' clumps which are found
  primarily at higher temperatures.  The top four panels are Stage 2
  sources while the ``red'' clump panel are Stage 4 sources (see
  Section \ref{sec:evseq}).  Right: A dust temperature histogram of
  possible evolutionary stages in the sample of 100 clumps
  from \citet{rat10}.  This sample is likely incomplete at the 
  low-temperature end, as many of the coldest sources would also be faint
  and below the detection threshhold for the SED fits.}
\label{fig:temphist}
\end{figure*}

\section{Disscussion of an Evolutionary Sequence for IRDC Clumps}
\label{sec:evseq}
Based on a sample of 190 IRDC clumps, \citet{cha09} 
proposed an evolutionary sequence for IRDC clumps which begins
with a quiescent clump (in their definition, this means no
``green fuzzy'' or 24 \micron~point source), transitions into
an active clump (the clump contains a ``green fuzzy'' 
and a 24 \micron~point source), and finally, becomes
 a ``red'' (enhanced 8~\micron~emission) clump.
Our radio continuum data support this sequence, especially 
that the final 
evolutionary stage is a ``red'' clump, in which the H~II region
is exciting the PAH spectral feature in the 8~\micron~band.
However, we suggest that the term ``red'' clump be reserved for more
diffuse, evolved ``red'' clumps, not associated with a millimeter peak.  This
prevents any confusion with embedded B or A stars which have UV luminosity
signficant enough to excite the PAH emission in the 8
\micron~band, but not enough to produce an UCH~II region.  Some
clumps with H~II regions are ``red'' and some are not, which may indicate
different environments or evolutionary stages.  Clumps
with UCH~II regions tend to have a significantly brighter 
24~\micron~source flux ($\gtrsim$ 1 Jy). 

Presented in Figure \ref{fig:cartoon} is a cartoon of star formation
tracers and how they may possibly relate to evolutionary stages.  
Stage 1 is a quiescent clump; a cold, millimeter peak with no signs of
active star formation.  Stage 1 can be split into two sub-categories: a)
a quiescent clump which has not yet begun to collapse and b) a
quiescent clump that shows collapse signatures, but no signs of active star
formation.
The next stage is a hotter, denser clump that shows signs of
active star formation.  Stage 2 is a clump with at least 1 
sign of active star formation, a 24~\micron~point source, ``green fuzzy,''
or a H$_{2}$O or CH$_{3}$OH maser.  We cannot infer any sequential nature
between these observational signatures at this time.  However, the authors
hypothesize that, in analogy with low-mass star
forming regions, an outflow (``green fuzzy'') and 24~\micron~point source
will occur first (the one seen first depends on viewing angle), 
followed by an H$_{2}$O maser, and then a CH$_{3}$OH maser as the 
accreting protostars gain mass.  

The third stage in this sequence is the 
ignition of an UCH~II region (see Figure \ref{fig:cartoon}).  A Stage 3
source will have a strong ($\gtrsim$ 1 Jy) 24~\micron~point source and
possibly maser emission and outflow indicators.  Modern
theory \citep[e.g.][]{mck07} suggests that massive stars reach the
main sequence while still accreting, because the accretion time scale is
greater than the Kelvin-Helmholtz timescale.  Thus, a main-sequence B star may
continue to accrete on the main sequence, eventually becoming an O star.
A Stage 3 source, therefore, may remain compact and continue accreting
for some time.  The fourth stage in this evolutionary sequence 
is the expansion of
the H~II region, producing a diffuse ``red'' clump (see Figure
\ref{fig:cartoon}).  However, this is not a necessary condition for massive
star formation, as in the densest cores, the H~II region may be quenched 
by the remaining envelope \citep{ket03} for some time and a 
``diffuse red clump'' will not form.  For UCH~II regions not trapped by
their surrounding envelope, the UCH~II region will expand as the 
radiation field increases or instabilities develop.  This
expansion evacuates a dust cavity, creating
a diffuse enhancement at 8~\micron, that is spatially distinct from any
millimeter continuum peaks.  In Table \ref{table-activity} we assign each
clump to these coarse evolutionary stages.

One potential test of the sequential nature of star formation tracers is
the clump temperature.  We suggest that as a clump evolves, its
temperature monotonically increases with time.  In a statistical sense,
warmer objects should be more evolved than colder objects.  We use the dust
temperatures derived by \citet{rat10} in a sample of 100 clumps and the
star formation tracers as catalogued by \citet{cha09} to investigate how
dust temperatures change with star formation tracers in Figure
\ref{fig:temphist}.  All of the individual star formation tracers exist
over a surprisingly wide range of temperatures, except for ``red'' clumps
which are found primarily at higher temperatures. Figure
\ref{fig:24flux} shows that the more star formation tracers a clump has,
the warmer it is on average, however, no sequential nature can be inferred
between the individual star formation tracers in Stage 2.  The lower panel
of Figure \ref{fig:temphist} shows that the broad categorizations between
Stage 1 (quiescent), Stage 2 (intermediate/active), and Stage 4 (``diffuse
red clump'') clumps are well-separated by temperature, and likely
represent a true evolutionary sequence (we note that many
Stage 1 clumps were too faint to reliably fit an SED, so the true 
histogram is likely even further to the colder side of the histogram).
These data support the coarse separation of Stage 1, Stage 2, and Stage 4
clumps as an evolutionary sequence, however, no further sequential nature
of star formation tracers can be inferred.

When discussing evolutionary sequences, there is always the added 
uncertainty of whether we are truly seeing an evolutionary sequence or just
different environments.  We point out that it is likely that not all Stage
1 (quiescent) clumps will form Stage 2 (active) clumps, however, we are
suggesting that all Stage 2 clumps form from Stage 1 clumps and so on.
Some quiescent
clumps may be forming low/intermediate mass stars which will remain 
undetected by our star formation tracers.  Alternatively, some quiescent
clumps may be transient density enhancements which will eventually
disperse.  This makes the question of lifetimes more complicated.  The lack
of a sequence between the observed star formation tracers is a bit
surprising, however, the fact that more star formation tracers indicates a
higher temperature indicates that viewing angle/obscuration may be playing
a large role in the observed scatter.  Perhaps only when a clump is more
evolved and somewhat less obscured by dust can one detect many of these
star formation tracers.  Higher sensitivity and better spatial resolution 
observations, as well as a better control of systematics may reveal
some sequential nature between the observed star formation tracers.
We note that the overlap of the Stage 2 and Stage 4 sources in Figure
\ref{fig:stageshist} hint at some number of UCH~II regions undected as
``red'' clumps or some number of ``red'' clumps incorrectly associated with
UCH~II regions.

%\clearpage
\section{Conclusion}
\label{sec:conclusion}
We examined a sample of 17 clumps
within 8 IRDCs using
existing infrared, millimeter, and radio data, as well as 
new VLA radio continuum and HHT dense gas spectral data.
Our main conclusions are: 
\begin{itemize}
\item 8~\micron~extinction mass and BGPS 1.1 mm mass are
  complementary tracers of mass in IRDCs, except for the most active clumps
  (notably those containing UCH~II regions), for which both mass tracers
  suffer biases.
\item The measured virial masses of IRDC clumps are uniformly larger than
  the dust continuum masses on the scale of $\sim$ 1 pc. 
\item We do not detect a chemical differentiation of 
  \hcop~and \n2hp~between active and quiescent clumps on the scale of $\sim$
  1 pc.
  However, both \hcop~and \n2hp~are brighter in active clumps, due to an
increase in temperature and/or density.
\item We report the identification of four UCH~II regions embedded
  within IRDC clumps.  We find that UCH~II regions are associated with 
  bright ($\gtrsim$ 1 Jy) 24~\micron~point sources, and the brightest 
  UCH~II regions are associated with ``diffuse red clumps'' (an extended
  enhancement at 8~\micron).
\item We discuss an evolutionary sequence for cluster-forming clumps and
  find that the sequential separation of the broad evolutionary stages, 
  quiescent, active, and 
  ``diffuse red clumps,'' is supported by dust temperatures, however, no
  sequential nature can be inferred between the individual star formation
  tracers. 
\end{itemize}

\acknowledgments

We are thankful to our referee, Sean Carey, for a thorough and insightful
review, which has significantly improved the quality of this manuscript.
We are grateful to Edward Chambers, Jeremy Darling, Erica Ellingson,
Susanna Finn, Michael Shull, and Irena Stojimirovi\'c for useful
comments and discussion.  This work has made use of the GLIMPSE and MIPSGAL
surveys, and we thank those teams for their help and support.  We would like
to thank the staff at the HHT and VLA for their assistance.  The National
Radio Astronomy Observatory is a facility of the National Science
Foundation 
operated under cooperative agreement by Associated Universities, Inc. 
The BGPS project is supported
by the National Science Foundation through NSF
grant AST-0708403.  We are
grateful to Jill Rathborne for sharing her MAMBO 1.2 mm data.  
This publication makes use of molecular line data from the Boston
University-FCRAO Galactic Ring Survey (GRS). The GRS is a joint project 
of Boston University and Five College Radio Astronomy Observatory, 
funded by the National Science Foundation under grants AST-9800334, 
AST-0098562, AST-0100793, AST-0228993, \& AST-0507657.  This work has
made use of ds9, Aplpy (aplpy.sourceforge.net), and the Goddard Space
Flight Center's IDL Astronomy Library.
C.B. is supported by the National Science
Foundation (NSF) through the Graduate Research Fellowship Program (GRFP).

{\it Facilities:} \facility{HHT}, \facility{VLA}, 
\facility{Spitzer (IRAC)}, \facility{Spitzer (MIPS)},
\facility{CSO (Bolocam)}

\bibliography{references1}{}

\begin{thebibliography}{56}
\expandafter\ifx\csname natexlab\endcsname\relax\def\natexlab#1{#1}\fi

\bibitem[{{Aguirre} {et~al.}(2010){Aguirre}, {Ginsburg}, {Dunham}, {Drosback},
  {Bally}, {Battersby}, {Bradley}, {Cyganowski}, {Dowell}, {Evans II}, {Glenn},
  {Harvey}, {Rosolowsky}, {Stringfellow}, {Walawender}, \& {Williams}}]{agu10}
{Aguirre}, J.~A., {Ginsburg}, A.~G., {Dunham}, M.~K., {Drosback}, M.~D.,
  {Bally}, J., {Battersby}, C., {Bradley}, E., {Cyganowski}, C., {Dowell}, D.,
  {Evans II}, N.~J., {Glenn}, J., {Harvey}, P.~M., {Rosolowsky}, E.~W.,
  {Stringfellow}, G., {Walawender}, J., \& {Williams}, J.~P. 2010, submitted

\bibitem[{{Benjamin} {et~al.}(2003){Benjamin}, {Churchwell}, {Babler}, {Bania},
  {Clemens}, {Cohen}, {Dickey}, {Indebetouw}, {Jackson}, {Kobulnicky},
  {Lazarian}, {Marston}, {Mathis}, {Meade}, {Seager}, {Stolovy}, {Watson},
  {Whitney}, {Wolff}, \& {Wolfire}}]{ben03}
{Benjamin}, R.~A., {Churchwell}, E., {Babler}, B.~L., {Bania}, T.~M.,
  {Clemens}, D.~P., {Cohen}, M., {Dickey}, J.~M., {Indebetouw}, R., {Jackson},
  J.~M., {Kobulnicky}, H.~A., {Lazarian}, A., {Marston}, A.~P., {Mathis},
  J.~S., {Meade}, M.~R., {Seager}, S., {Stolovy}, S.~R., {Watson}, C.,
  {Whitney}, B.~A., {Wolff}, M.~J., \& {Wolfire}, M.~G. 2003, \pasp, 115, 953

\bibitem[{{Beuther} {et~al.}(2005){Beuther}, {Sridharan}, \& {Saito}}]{beu05}
{Beuther}, H., {Sridharan}, T.~K., \& {Saito}, M. 2005, \apjl, 634, L185

\bibitem[{{Butler} \& {Tan}(2009)}]{but09}
{Butler}, M.~J., \& {Tan}, J.~C. 2009, \apj, 696, 484

\bibitem[{{Carey} {et~al.}(1998){Carey}, {Clark}, {Egan}, {Price}, {Shipman},
  \& {Kuchar}}]{car98}
{Carey}, S.~J., {Clark}, F.~O., {Egan}, M.~P., {Price}, S.~D., {Shipman},
  R.~F., \& {Kuchar}, T.~A. 1998, \apj, 508, 721

\bibitem[{{Carey} {et~al.}(2000){Carey}, {Feldman}, {Redman}, {Egan},
  {MacLeod}, \& {Price}}]{car00}
{Carey}, S.~J., {Feldman}, P.~A., {Redman}, R.~O., {Egan}, M.~P., {MacLeod},
  J.~M., \& {Price}, S.~D. 2000, \apjl, 543, L157

\bibitem[{{Carey} {et~al.}(2009){Carey}, {Noriega-Crespo}, {Mizuno}, {Shenoy},
  {Paladini}, {Kraemer}, {Price}, {Flagey}, {Ryan}, {Ingalls}, {Kuchar},
  {Pinheiro Gon{\c c}alves}, {Indebetouw}, {Billot}, {Marleau}, {Padgett},
  {Rebull}, {Bressert}, {Ali}, {Molinari}, {Martin}, {Berriman}, {Boulanger},
  {Latter}, {Miville-Deschenes}, {Shipman}, \& {Testi}}]{car09}
{Carey}, S.~J., {Noriega-Crespo}, A., {Mizuno}, D.~R., {Shenoy}, S.,
  {Paladini}, R., {Kraemer}, K.~E., {Price}, S.~D., {Flagey}, N., {Ryan}, E.,
  {Ingalls}, J.~G., {Kuchar}, T.~A., {Pinheiro Gon{\c c}alves}, D.,
  {Indebetouw}, R., {Billot}, N., {Marleau}, F.~R., {Padgett}, D.~L., {Rebull},
  L.~M., {Bressert}, E., {Ali}, B., {Molinari}, S., {Martin}, P.~G.,
  {Berriman}, G.~B., {Boulanger}, F., {Latter}, W.~B., {Miville-Deschenes},
  M.~A., {Shipman}, R., \& {Testi}, L. 2009, \pasp, 121, 76

\bibitem[{{Chambers} {et~al.}(2009){Chambers}, {Jackson}, {Rathborne}, \&
  {Simon}}]{cha09}
{Chambers}, E.~T., {Jackson}, J.~M., {Rathborne}, J.~M., \& {Simon}, R. 2009,
  \apjs, 181, 360

\bibitem[{{Cyganowski} {et~al.}(2008){Cyganowski}, {Whitney}, {Holden},
  {Braden}, {Brogan}, {Churchwell}, {Indebetouw}, {Watson}, {Babler},
  {Benjamin}, {Gomez}, {Meade}, {Povich}, {Robitaille}, \& {Watson}}]{cyg08}
{Cyganowski}, C.~J., {Whitney}, B.~A., {Holden}, E., {Braden}, E., {Brogan},
  C.~L., {Churchwell}, E., {Indebetouw}, R., {Watson}, D.~F., {Babler}, B.~L.,
  {Benjamin}, R., {Gomez}, M., {Meade}, M.~R., {Povich}, M.~S., {Robitaille},
  T.~P., \& {Watson}, C. 2008, \aj, 136, 2391

\bibitem[{{Daniel} {et~al.}(2007){Daniel}, {Cernicharo}, {Roueff}, {Gerin}, \&
  {Dubernet}}]{dan07}
{Daniel}, F., {Cernicharo}, J., {Roueff}, E., {Gerin}, M., \& {Dubernet}, M.~L.
  2007, \apj, 667, 980

\bibitem[{{de Wit} {et~al.}(2005){de Wit}, {Testi}, {Palla}, \&
  {Zinnecker}}]{dew05}
{de Wit}, W.~J., {Testi}, L., {Palla}, F., \& {Zinnecker}, H. 2005, \aap, 437,
  247

\bibitem[{{Du} \& {Yang}(2008)}]{duy08}
{Du}, F., \& {Yang}, J. 2008, \apj, 686, 384

\bibitem[{{Dunham} {et~al.}(2010){Dunham}, {Rosolowsky}, {Evans II}, {Aguirre},
  {Bally}, {Battersby}, {Bradley}, {Dowell}, {Drosback}, {Ginsburg}, {Glenn},
  {Harvey}, {Merello}, {Schlingman}, {Shirley}, {Stringfellow}, {Walawender},
  \& {Williams}}]{dun10}
{Dunham}, M.~K., {Rosolowsky}, E., {Evans II}, N.~J., C.~C., {Aguirre}, J.~A.,
  {Bally}, J., {Battersby}, C., {Bradley}, E., {Dowell}, D., {Drosback}, M.~D.,
  {Ginsburg}, A.~G., {Glenn}, J., {Harvey}, P.~M., {Merello}, M., {Schlingman},
  W., {Shirley}, Y.~L., {Stringfellow}, G., {Walawender}, J., \& {Williams},
  J.~P. 2010, submitted

\bibitem[{{Egan} {et~al.}(1998){Egan}, {Shipman}, {Price}, {Carey}, {Clark}, \&
  {Cohen}}]{ega98}
{Egan}, M.~P., {Shipman}, R.~F., {Price}, S.~D., {Carey}, S.~J., {Clark},
  F.~O., \& {Cohen}, M. 1998, \apjl, 494, L199+

\bibitem[{{Enoch} {et~al.}(2006){Enoch}, {Young}, {Glenn}, {Evans}, {Golwala},
  {Sargent}, {Harvey}, {Aguirre}, {Goldin}, {Haig}, {Huard}, {Lange},
  {Laurent}, {Maloney}, {Mauskopf}, {Rossinot}, \& {Sayers}}]{eno06}
{Enoch}, M.~L., {Young}, K.~E., {Glenn}, J., {Evans}, II, N.~J., {Golwala}, S.,
  {Sargent}, A.~I., {Harvey}, P., {Aguirre}, J., {Goldin}, A., {Haig}, D.,
  {Huard}, T.~L., {Lange}, A., {Laurent}, G., {Maloney}, P., {Mauskopf}, P.,
  {Rossinot}, P., \& {Sayers}, J. 2006, \apj, 638, 293

\bibitem[{{Glenn} {et~al.}(2003){Glenn}, {Ade}, {Amarie}, {Bock}, {Edgington},
  {Goldin}, {Golwala}, {Haig}, {Lange}, {Laurent}, {Mauskopf}, {Yun}, \&
  {Nguyen}}]{gle03}
{Glenn}, J., {Ade}, P.~A.~R., {Amarie}, M., {Bock}, J.~J., {Edgington}, S.~F.,
  {Goldin}, A., {Golwala}, S., {Haig}, D., {Lange}, A.~E., {Laurent}, G.,
  {Mauskopf}, P.~D., {Yun}, M., \& {Nguyen}, H. 2003, in Society of
  Photo-Optical Instrumentation Engineers (SPIE) Conference Series, Vol. 4855,
  Society of Photo-Optical Instrumentation Engineers (SPIE) Conference Series,
  ed. T.~G. {Phillips} \& J.~{Zmuidzinas}, 30--40

\bibitem[{{Heitsch} {et~al.}(2008){Heitsch}, {Hartmann}, {Slyz}, {Devriendt},
  \& {Burkert}}]{hei08}
{Heitsch}, F., {Hartmann}, L.~W., {Slyz}, A.~D., {Devriendt}, J.~E.~G., \&
  {Burkert}, A. 2008, \apj, 674, 316

\bibitem[{{Jackson} {et~al.}(2006){Jackson}, {Rathborne}, {Shah}, {Simon},
  {Bania}, {Clemens}, {Chambers}, {Johnson}, {Dormody}, {Lavoie}, \&
  {Heyer}}]{jac06}
{Jackson}, J.~M., {Rathborne}, J.~M., {Shah}, R.~Y., {Simon}, R., {Bania},
  T.~M., {Clemens}, D.~P., {Chambers}, E.~T., {Johnson}, A.~M., {Dormody}, M.,
  {Lavoie}, R., \& {Heyer}, M.~H. 2006, \apjs, 163, 145

\bibitem[{{J{\o}rgensen} {et~al.}(2004){J{\o}rgensen}, {Sch{\"o}ier}, \& {van
  Dishoeck}}]{jor04}
{J{\o}rgensen}, J.~K., {Sch{\"o}ier}, F.~L., \& {van Dishoeck}, E.~F. 2004,
  \aap, 416, 603

\bibitem[{{Kauffmann} {et~al.}(2008){Kauffmann}, {Bertoldi}, {Bourke}, {Evans},
  \& {Lee}}]{kau08}
{Kauffmann}, J., {Bertoldi}, F., {Bourke}, T.~L., {Evans}, II, N.~J., \& {Lee},
  C.~W. 2008, \aap, 487, 993

\bibitem[{{Keto}(2003)}]{ket03}
{Keto}, E. 2003, \apj, 599, 1196

\bibitem[{{Lada} \& {Lada}(2003)}]{lad03}
{Lada}, C.~J., \& {Lada}, E.~A. 2003, \araa, 41, 57

\bibitem[{{Lucas} \& {Liszt}(1998)}]{luc98}
{Lucas}, R., \& {Liszt}, H. 1998, \aap, 337, 246

\bibitem[{{Mangum} {et~al.}(2007){Mangum}, {Emerson}, \& {Greisen}}]{man07}
{Mangum}, J.~G., {Emerson}, D.~T., \& {Greisen}, E.~W. 2007, \aap, 474, 679

\bibitem[{{Marston} {et~al.}(2004){Marston}, {Reach}, {Noriega-Crespo}, {Rho},
  {Smith}, {Melnick}, {Fazio}, {Rieke}, {Carey}, {Rebull}, {Muzerolle},
  {Egami}, {Watson}, {Pipher}, {Latter}, \& {Stapelfeldt}}]{mar04}
{Marston}, A.~P., {Reach}, W.~T., {Noriega-Crespo}, A., {Rho}, J., {Smith},
  H.~A., {Melnick}, G., {Fazio}, G., {Rieke}, G., {Carey}, S., {Rebull}, L.,
  {Muzerolle}, J., {Egami}, E., {Watson}, D.~M., {Pipher}, J.~L., {Latter},
  W.~B., \& {Stapelfeldt}, K. 2004, \apjs, 154, 333

\bibitem[{{McKee} \& {Ostriker}(2007)}]{mck07}
{McKee}, C.~F., \& {Ostriker}, E.~C. 2007, \araa, 45, 565

\bibitem[{{McKee} \& {Tan}(2003)}]{mck03}
{McKee}, C.~F., \& {Tan}, J.~C. 2003, \apj, 585, 850

\bibitem[{{McKee} \& {Williams}(1997)}]{mck97}
{McKee}, C.~F., \& {Williams}, J.~P. 1997, \apj, 476, 144

\bibitem[{{Mueller} {et~al.}(2002){Mueller}, {Shirley}, {Evans}, \&
  {Jacobson}}]{mue02}
{Mueller}, K.~E., {Shirley}, Y.~L., {Evans}, II, N.~J., \& {Jacobson}, H.~R.
  2002, \apjs, 143, 469

\bibitem[{{Noriega-Crespo} {et~al.}(2004){Noriega-Crespo}, {Morris}, {Marleau},
  {Carey}, {Boogert}, {van Dishoeck}, {Evans}, {Keene}, {Muzerolle},
  {Stapelfeldt}, {Pontoppidan}, {Lowrance}, {Allen}, \& {Bourke}}]{nor04}
{Noriega-Crespo}, A., {Morris}, P., {Marleau}, F.~R., {Carey}, S., {Boogert},
  A., {van Dishoeck}, E., {Evans}, II, N.~J., {Keene}, J., {Muzerolle}, J.,
  {Stapelfeldt}, K., {Pontoppidan}, K., {Lowrance}, P., {Allen}, L., \&
  {Bourke}, T.~L. 2004, \apjs, 154, 352

\bibitem[{{Omont} {et~al.}(2003){Omont}, {Gilmore}, {Alard}, {Aracil},
  {August}, {Baliyan}, {Beaulieu}, {B{\'e}gon}, {Bertou}, {Blommaert},
  {Borsenberger}, {Burgdorf}, {Caillaud}, {Cesarsky}, {Chitre}, {Copet}, {de
  Batz}, {Egan}, {Egret}, {Epchtein}, {Felli}, {Fouqu{\'e}}, {Ganesh},
  {Genzel}, {Glass}, {Gredel}, {Groenewegen}, {Guglielmo}, {Habing},
  {Hennebelle}, {Jiang}, {Joshi}, {Kimeswenger}, {Messineo},
  {Miville-Desch{\^e}nes}, {Moneti}, {Morris}, {Ojha}, {Ortiz}, {Ott},
  {Parthasarathy}, {P{\'e}rault}, {Price}, {Robin}, {Schultheis}, {Schuller},
  {Simon}, {Soive}, {Testi}, {Teyssier}, {Tiph{\`e}ne}, {Unavane}, {van Loon},
  \& {Wyse}}]{omo03}
{Omont}, A., {Gilmore}, G.~F., {Alard}, C., {Aracil}, B., {August}, T.,
  {Baliyan}, K., {Beaulieu}, S., {B{\'e}gon}, S., {Bertou}, X., {Blommaert},
  J.~A.~D.~L., {Borsenberger}, J., {Burgdorf}, M., {Caillaud}, B., {Cesarsky},
  C., {Chitre}, A., {Copet}, E., {de Batz}, B., {Egan}, M.~P., {Egret}, D.,
  {Epchtein}, N., {Felli}, M., {Fouqu{\'e}}, P., {Ganesh}, S., {Genzel}, R.,
  {Glass}, I.~S., {Gredel}, R., {Groenewegen}, M.~A.~T., {Guglielmo}, F.,
  {Habing}, H.~J., {Hennebelle}, P., {Jiang}, B., {Joshi}, U.~C.,
  {Kimeswenger}, S., {Messineo}, M., {Miville-Desch{\^e}nes}, M.~A., {Moneti},
  A., {Morris}, M., {Ojha}, D.~K., {Ortiz}, R., {Ott}, S., {Parthasarathy}, M.,
  {P{\'e}rault}, M., {Price}, S.~D., {Robin}, A.~C., {Schultheis}, M.,
  {Schuller}, F., {Simon}, G., {Soive}, A., {Testi}, L., {Teyssier}, D.,
  {Tiph{\`e}ne}, D., {Unavane}, M., {van Loon}, J.~T., \& {Wyse}, R. 2003,
  \aap, 403, 975

\bibitem[{{Ossenkopf} \& {Henning}(1994)}]{oss94}
{Ossenkopf}, V., \& {Henning}, T. 1994, \aap, 291, 943

\bibitem[{{Perault} {et~al.}(1996){Perault}, {Omont}, {Simon}, {Seguin},
  {Ojha}, {Blommaert}, {Felli}, {Gilmore}, {Guglielmo}, {Habing}, {Price},
  {Robin}, {de Batz}, {Cesarsky}, {Elbaz}, {Epchtein}, {Fouque}, {Guest},
  {Levine}, {Pollock}, {Prusti}, {Siebenmorgen}, {Testi}, \& {Tiphene}}]{per96}
{Perault}, M., {Omont}, A., {Simon}, G., {Seguin}, P., {Ojha}, D., {Blommaert},
  J., {Felli}, M., {Gilmore}, G., {Guglielmo}, F., {Habing}, H., {Price}, S.,
  {Robin}, A., {de Batz}, B., {Cesarsky}, C., {Elbaz}, D., {Epchtein}, N.,
  {Fouque}, P., {Guest}, S., {Levine}, D., {Pollock}, A., {Prusti}, T.,
  {Siebenmorgen}, R., {Testi}, L., \& {Tiphene}, D. 1996, \aap, 315, L165

\bibitem[{{Peretto} \& {Fuller}(2009)}]{per09}
{Peretto}, N., \& {Fuller}, G.~A. 2009, \aap, 505, 405

\bibitem[{{Ragan} {et~al.}(2009){Ragan}, {Bergin}, \& {Gutermuth}}]{rag09}
{Ragan}, S.~E., {Bergin}, E.~A., \& {Gutermuth}, R.~A. 2009, \apj, 698, 324

\bibitem[{{Rathborne} {et~al.}(2010){Rathborne}, {Jackson}, {Chambers},
  {Stojimirovic}, {Simon}, {Shipman}, \& {Frieswijk}}]{rat10}
{Rathborne}, J.~M., {Jackson}, J.~M., {Chambers}, E.~T., {Stojimirovic}, I.,
  {Simon}, R., {Shipman}, R., \& {Frieswijk}, W. 2010, ArXiv e-prints

\bibitem[{{Rathborne} {et~al.}(2006){Rathborne}, {Jackson}, \& {Simon}}]{rat06}
{Rathborne}, J.~M., {Jackson}, J.~M., \& {Simon}, R. 2006, \apj, 641, 389

\bibitem[{{Rathborne} {et~al.}(2008){Rathborne}, {Jackson}, {Zhang}, \&
  {Simon}}]{rat08}
{Rathborne}, J.~M., {Jackson}, J.~M., {Zhang}, Q., \& {Simon}, R. 2008, \apj,
  689, 1141

\bibitem[{{Rathborne} {et~al.}(2007){Rathborne}, {Simon}, \& {Jackson}}]{rat07}
{Rathborne}, J.~M., {Simon}, R., \& {Jackson}, J.~M. 2007, \apj, 662, 1082

\bibitem[{{Reach} {et~al.}(2006){Reach}, {Rho}, {Tappe}, {Pannuti}, {Brogan},
  {Churchwell}, {Meade}, {Babler}, {Indebetouw}, \& {Whitney}}]{rea06}
{Reach}, W.~T., {Rho}, J., {Tappe}, A., {Pannuti}, T.~G., {Brogan}, C.~L.,
  {Churchwell}, E.~B., {Meade}, M.~R., {Babler}, B., {Indebetouw}, R., \&
  {Whitney}, B.~A. 2006, \aj, 131, 1479

\bibitem[{{Redman} {et~al.}(2003){Redman}, {Feldman}, {Wyrowski},
  {C{\^o}t{\'e}}, {Carey}, \& {Egan}}]{red03}
{Redman}, R.~O., {Feldman}, P.~A., {Wyrowski}, F., {C{\^o}t{\'e}}, S., {Carey},
  S.~J., \& {Egan}, M.~P. 2003, \apj, 586, 1127

\bibitem[{{Reid} \& {Ho}(1985)}]{rei85}
{Reid}, M.~J., \& {Ho}, P.~T.~P. 1985, \apjl, 288, L17

\bibitem[{{Reid} {et~al.}(2009){Reid}, {Menten}, {Zheng}, {Brunthaler},
  {Moscadelli}, {Xu}, {Zhang}, {Sato}, {Honma}, {Hirota}, {Hachisuka}, {Choi},
  {Moellenbrock}, \& {Bartkiewicz}}]{re09}
{Reid}, M.~J., {Menten}, K.~M., {Zheng}, X.~W., {Brunthaler}, A., {Moscadelli},
  L., {Xu}, Y., {Zhang}, B., {Sato}, M., {Honma}, M., {Hirota}, T.,
  {Hachisuka}, K., {Choi}, Y.~K., {Moellenbrock}, G.~A., \& {Bartkiewicz}, A.
  2009, ArXiv e-prints

\bibitem[{{Rosolowsky} {et~al.}(2010){Rosolowsky}, {Dunham}, {Ginsburg},
  {Bradley}, {Aguirre}, {Bally}, {Battersby}, {Cyganowski}, {Dowell},
  {Drosback}, {Evans}, {Glenn}, {Harvey}, {Stringfellow}, {Walawender}, \&
  {Williams}}]{ros10}
{Rosolowsky}, E., {Dunham}, M.~K., {Ginsburg}, A., {Bradley}, E.~T., {Aguirre},
  J., {Bally}, J., {Battersby}, C., {Cyganowski}, C., {Dowell}, D., {Drosback},
  M., {Evans}, N.~J., {Glenn}, J., {Harvey}, P., {Stringfellow}, G.~S.,
  {Walawender}, J., \& {Williams}, J.~P. 2010, \apjs, 188, 123

\bibitem[{{Shepherd} {et~al.}(2004){Shepherd}, {N{\"u}rnberger}, \&
  {Bronfman}}]{she04}
{Shepherd}, D.~S., {N{\"u}rnberger}, D.~E.~A., \& {Bronfman}, L. 2004, \apj,
  602, 850

\bibitem[{{Shirley} {et~al.}(2003){Shirley}, {Evans}, {Young}, {Knez}, \&
  {Jaffe}}]{shi03}
{Shirley}, Y.~L., {Evans}, II, N.~J., {Young}, K.~E., {Knez}, C., \& {Jaffe},
  D.~T. 2003, \apjs, 149, 375

\bibitem[{{Simon} {et~al.}(2006{\natexlab{a}}){Simon}, {Jackson}, {Rathborne},
  \& {Chambers}}]{sim06}
{Simon}, R., {Jackson}, J.~M., {Rathborne}, J.~M., \& {Chambers}, E.~T.
  2006{\natexlab{a}}, \apj, 639, 227

\bibitem[{{Simon} {et~al.}(2006{\natexlab{b}}){Simon}, {Rathborne}, {Shah},
  {Jackson}, \& {Chambers}}]{sim06b}
{Simon}, R., {Rathborne}, J.~M., {Shah}, R.~Y., {Jackson}, J.~M., \&
  {Chambers}, E.~T. 2006{\natexlab{b}}, \apj, 653, 1325

\bibitem[{{Tafalla} {et~al.}(2002){Tafalla}, {Myers}, {Caselli}, {Walmsley}, \&
  {Comito}}]{taf02}
{Tafalla}, M., {Myers}, P.~C., {Caselli}, P., {Walmsley}, C.~M., \& {Comito},
  C. 2002, \apj, 569, 815

\bibitem[{{Vacca} {et~al.}(1996){Vacca}, {Garmany}, \& {Shull}}]{vac96}
{Vacca}, W.~D., {Garmany}, C.~D., \& {Shull}, J.~M. 1996, \apj, 460, 914

\bibitem[{{Wang} {et~al.}(2008){Wang}, {Zhang}, {Pillai}, {Wyrowski}, \&
  {Wu}}]{wan08}
{Wang}, Y., {Zhang}, Q., {Pillai}, T., {Wyrowski}, F., \& {Wu}, Y. 2008, \apjl,
  672, L33

\bibitem[{{White} {et~al.}(2005){White}, {Becker}, \& {Helfand}}]{whi05}
{White}, R.~L., {Becker}, R.~H., \& {Helfand}, D.~J. 2005, \aj, 130, 586

\bibitem[{{Wood} \& {Churchwell}(1989)}]{woo89}
{Wood}, D.~O.~S., \& {Churchwell}, E. 1989, \apjs, 69, 831

\bibitem[{{Wyrowski} {et~al.}(2000){Wyrowski}, {Carey}, {Egan}, {Feldman}, \&
  {Redman}}]{wyr00}
{Wyrowski}, F., {Carey}, S.~J., {Egan}, M.~P., {Feldman}, P.~A., \& {Redman},
  R.~O. 2000, in Bulletin of the American Astronomical Society, Vol. 197,
  Bulletin of the American Astronomical Society, 515--+

\bibitem[{{Zhang} {et~al.}(2009){Zhang}, {Wang}, {Pillai}, \&
  {Rathborne}}]{zha09}
{Zhang}, Q., {Wang}, Y., {Pillai}, T., \& {Rathborne}, J. 2009, \apj, 696, 268

\bibitem[{{Zinnecker} \& {Yorke}(2007)}]{zin07}
{Zinnecker}, H., \& {Yorke}, H.~W. 2007, \araa, 45, 481

\end{thebibliography}

%\clearpage
%\newpage
\begin{figure*} 
  \centering
  \subfigure{
    \label{fig:irdc6_glm3_iram}
    \includegraphics[width=.46\textwidth]{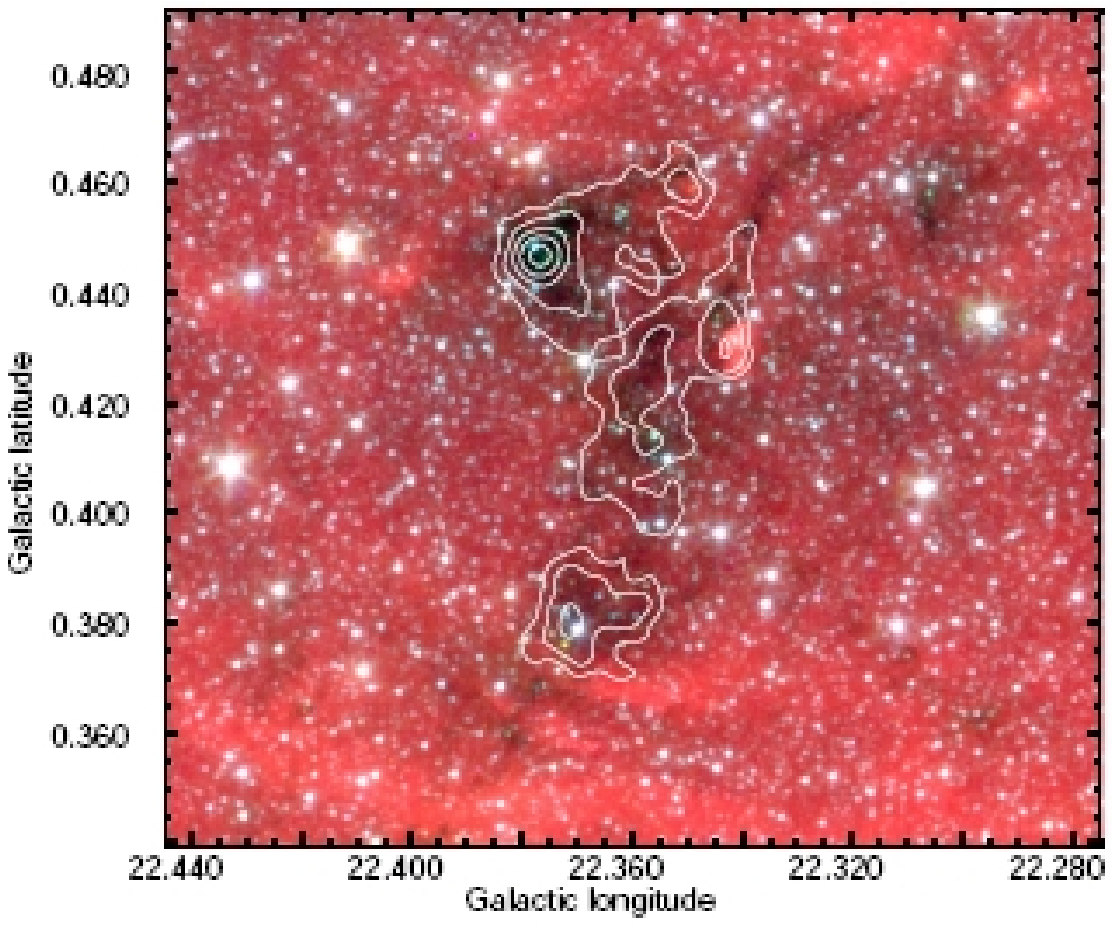}}
  \subfigure{
    \label{fig:irdc6_mips_bgps}
    \includegraphics[width=.46\textwidth]{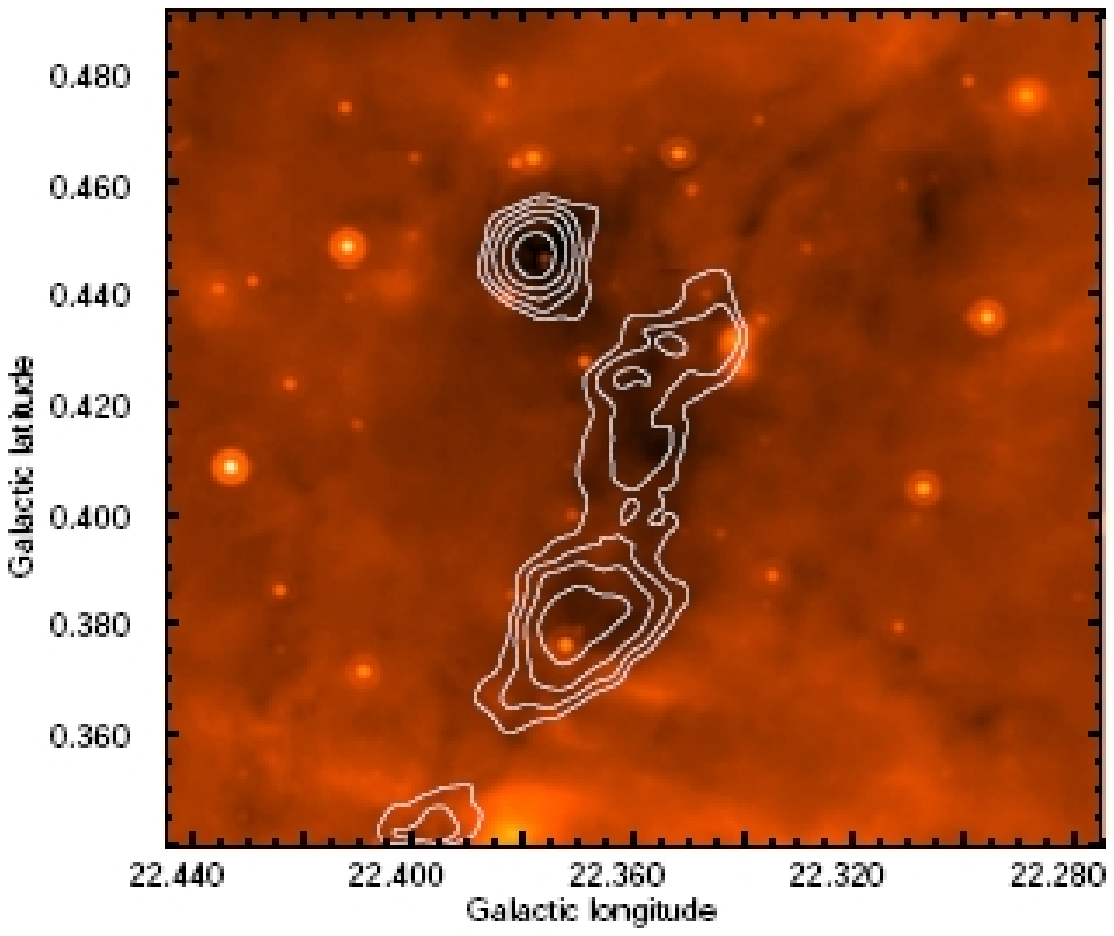}}\\
\caption{G022.35+00.41: Left: GLIMPSE three-color image, red is 8~\micron, green is
    4.5 \micron~and blue is 3.6 \micron~with MAMBO 1.2 mm contours 
    overlaid.  The contours are on a log scale from 15 to 410 mJy beam$^{-1}$.
    Right: MIPSGAL 24 \micron~image with BGPS 1.1 mm contours.  The contours
    are on a log scale from 0.12 to 0.7 Jy beam$^{-1}$}
\label{fig:irdc6_glm_mips}
\end{figure*}

%\newpage
\begin{figure*}
  \centering
  \subfigure{
    \label{fig:irdc6_apertures}
    \includegraphics[width=.46\textwidth]{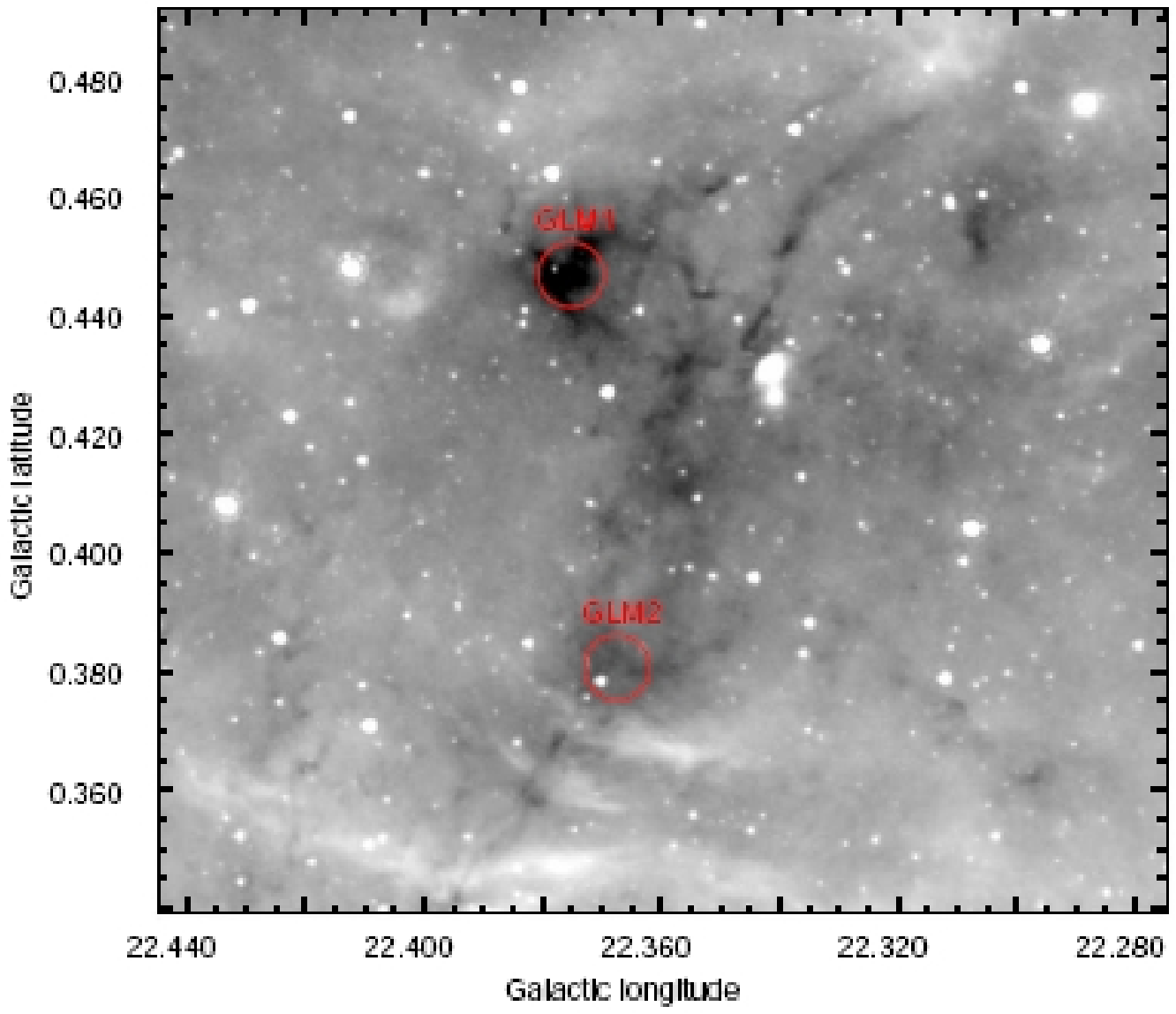}}
%  \vspace{-0.2in}
  \subfigure{
    \label{fig:irdc6_allspectra}
    \includegraphics[width=.46\textwidth]{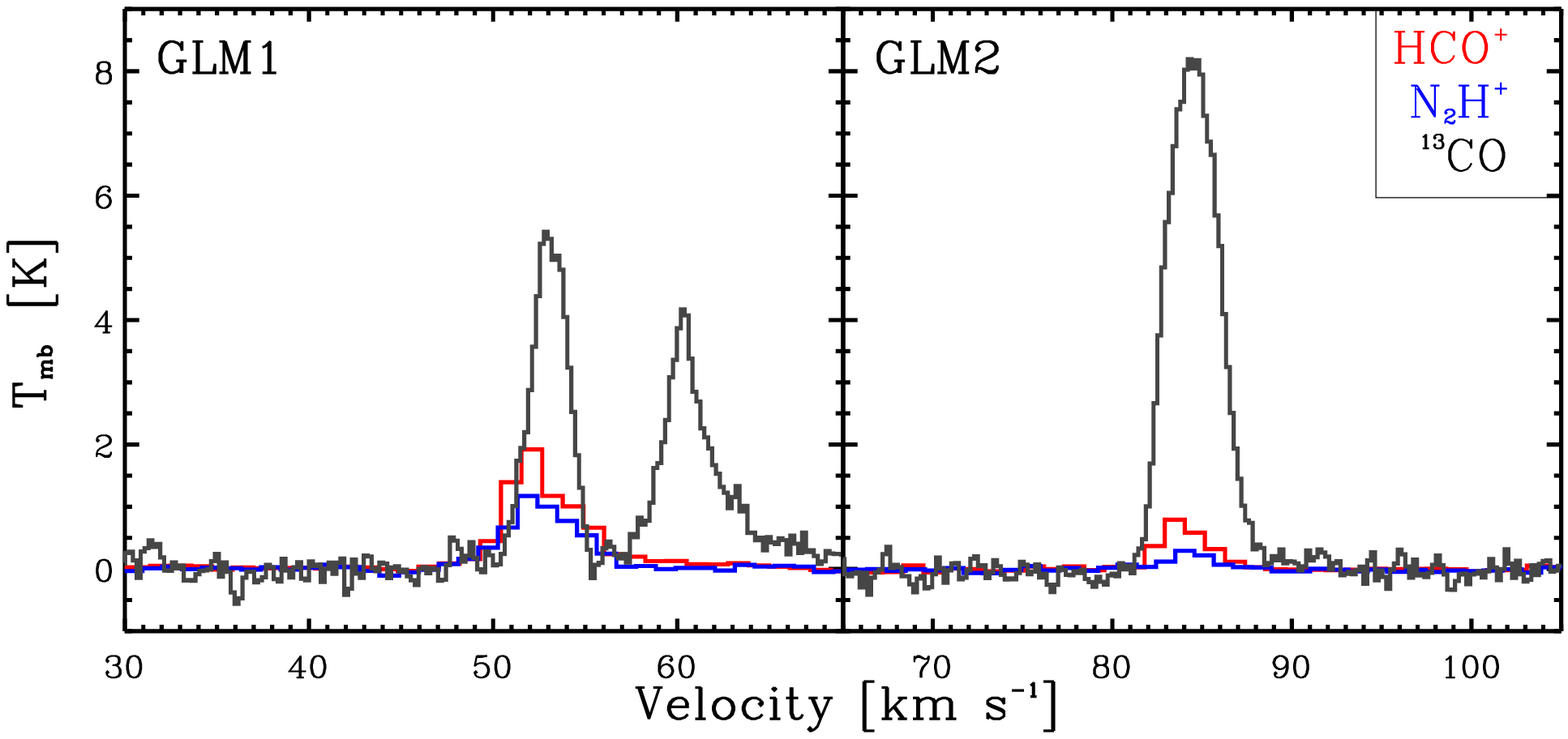}}\\
  \caption{G022.35+00.41: Left: GLIMPSE 8 $\mu$m 
    overplotted with the BGPS beam-sized apertures that were
    used to determine clump masses.  Right:  \hcop, \n2hp~and \13CO
    spectra in clumps GLM1 (Stage 2) and GLM2 (Stage 1). }
  \label{fig:irdc6_aps_spec}
  \end{figure*}

%\clearpage
\begin{figure*} 
  \centering
  \subfigure{
    \label{fig:irdc33_glm3_iram}
    \includegraphics[width=.46\textwidth]{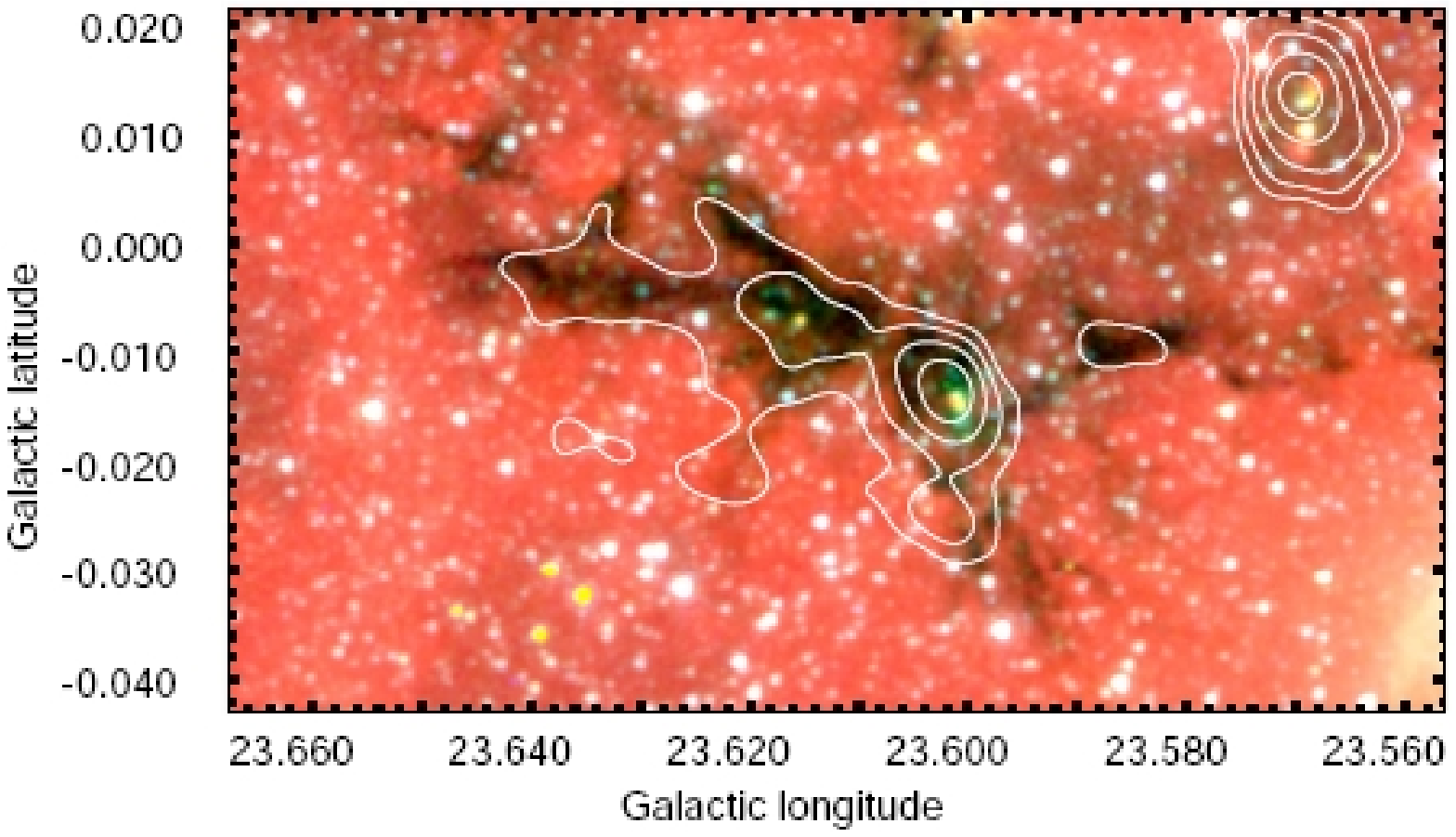}}
%  \vspace{-0.2in}
  \subfigure{
    \label{fig:irdc33_mips_bgps}
    \includegraphics[width=.46\textwidth]{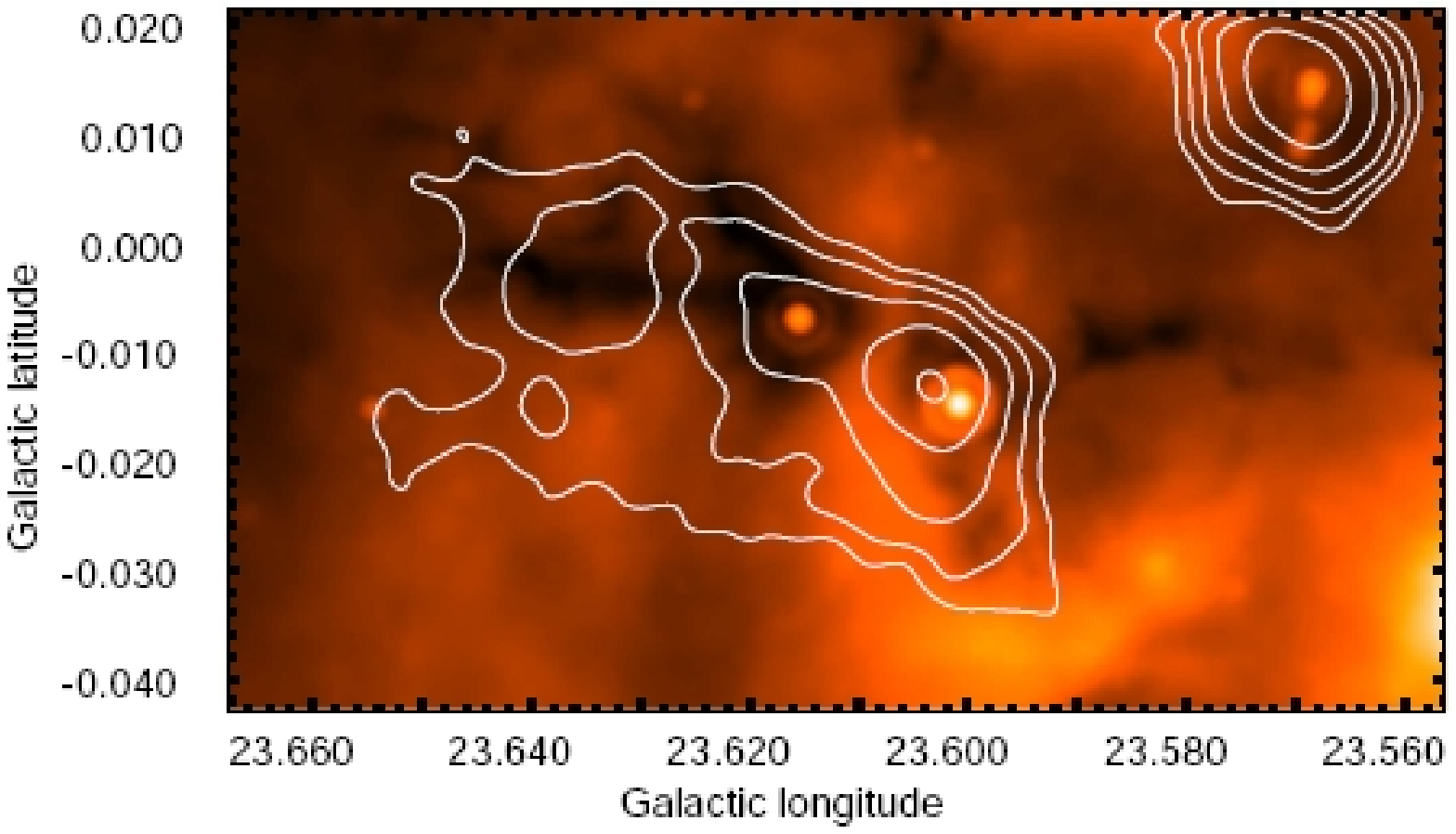}}\\
\caption{G023.60+00.00: Left: GLIMPSE three-color image, red is 8~\micron, green is
    4.5 \micron~and blue is 3.6 \micron~with MAMBO 1.2 mm contours 
    overlaid.  The contours are on a log scale from 30 to 440 mJy beam$^{-1}$.
    Right: MIPSGAL 24 \micron~image with BGPS 1.1 mm contours.  The contours
    are on a log scale from 0.12 to 1.1 Jy beam$^{-1}$. }
\label{fig:irdc33_glm_mips}
\end{figure*}
%\newpage
\begin{figure*}
  \centering
 \subfigure{
    \label{fig:irdc33_bgps_hcop}
    \includegraphics[width=.46\textwidth]{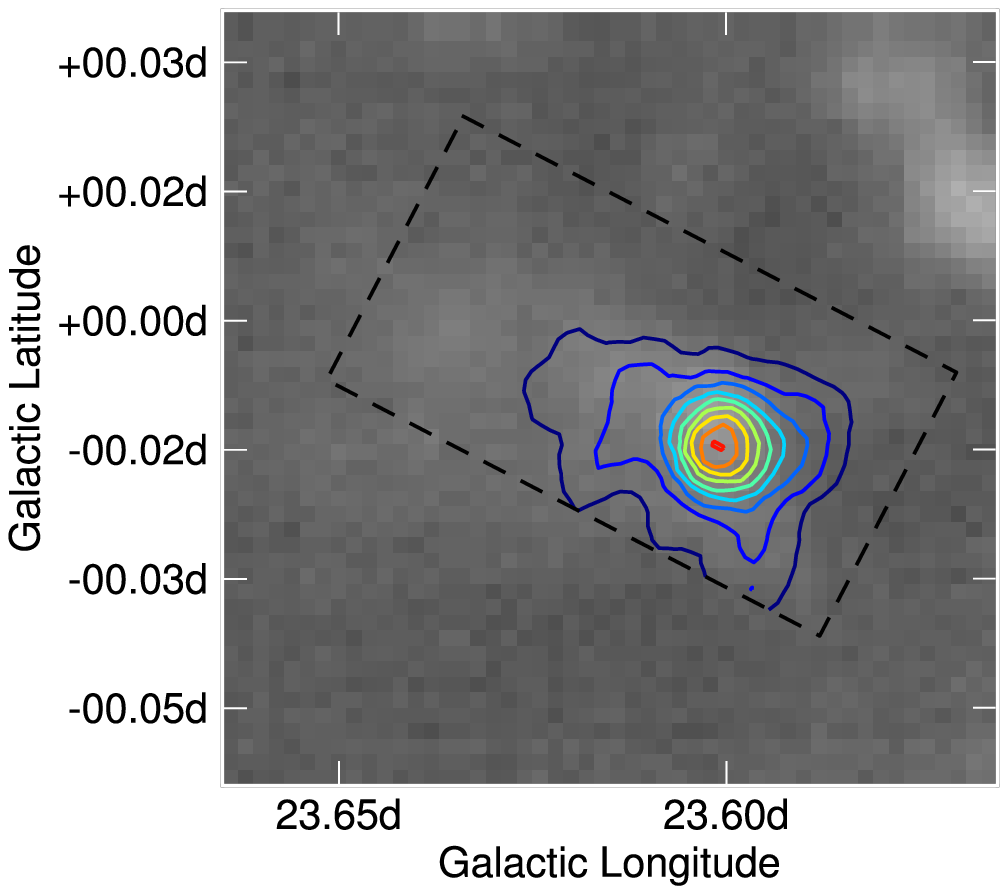}}
  \subfigure{
    \label{fig:irdc33_bgps_n2hp}
    \includegraphics[width=.46\textwidth]{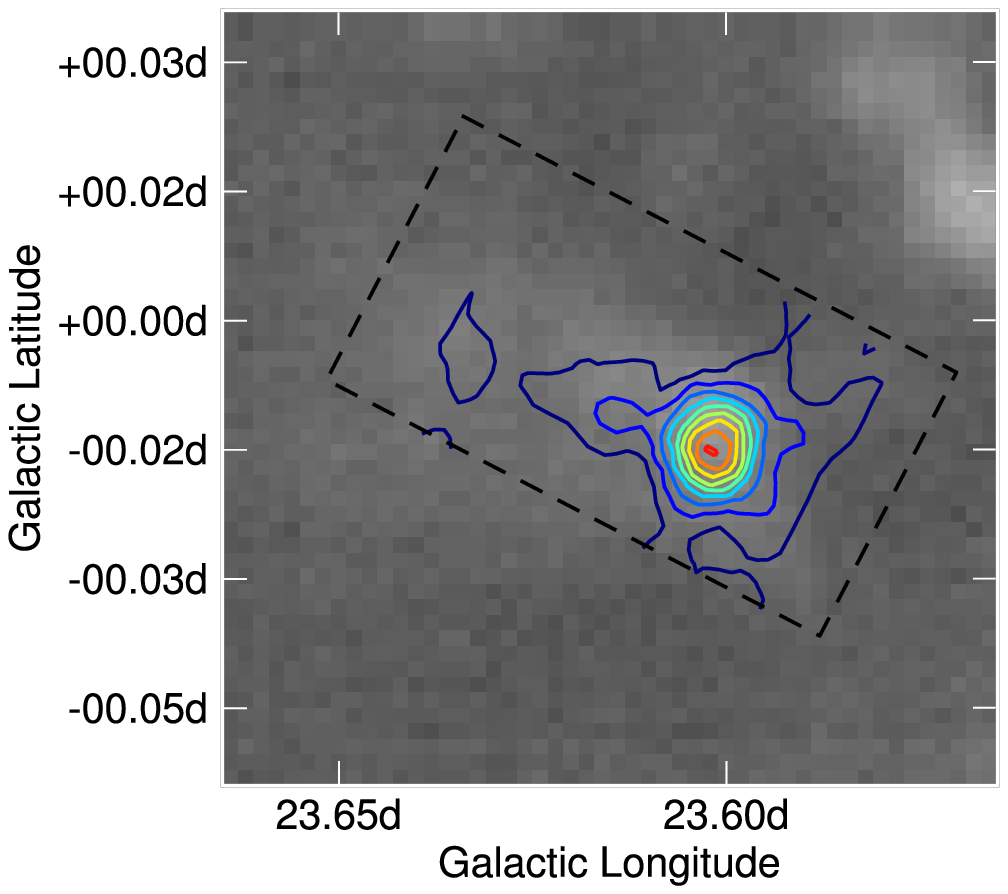}}
 \subfigure{
    \label{fig:irdc33_apertures}
    \includegraphics[scale=0.6]{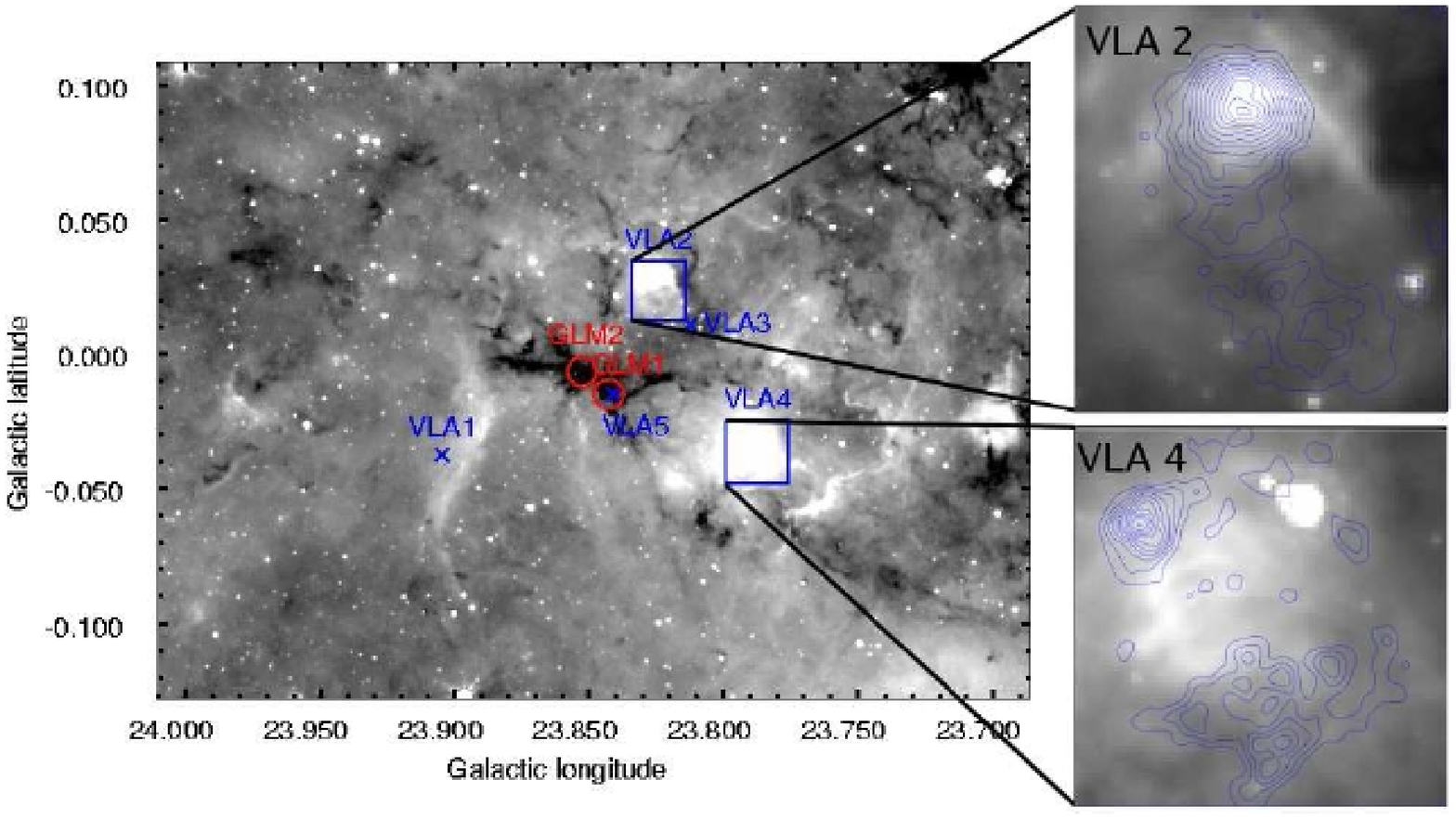}}\\
  \subfigure{
    \label{fig:irdc33_allspectra}
    \includegraphics[scale=0.5]{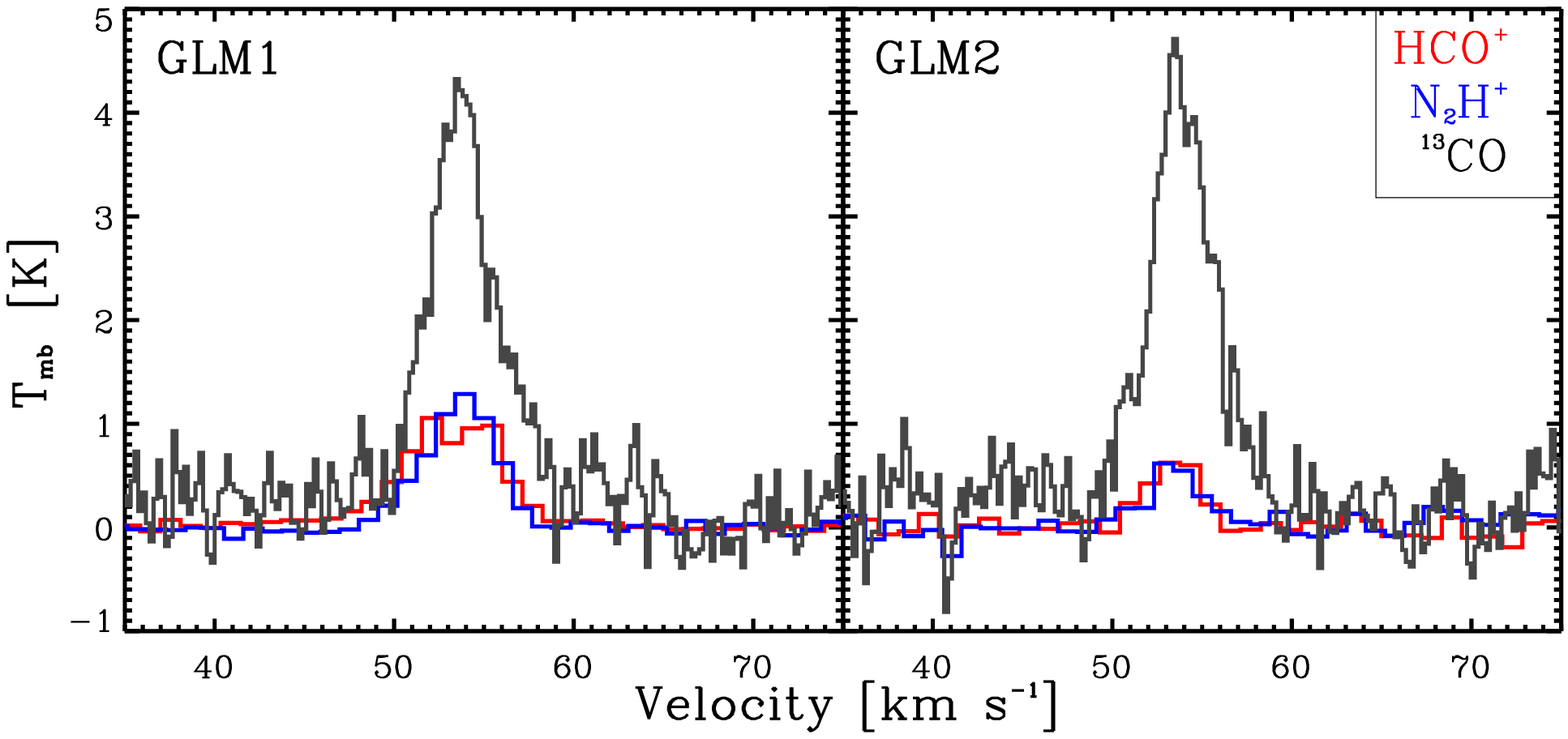}}\\
  \caption{G023.60+00.00: Top Left: BGPS 1.1 mm continuum dust emission overlaid with linear \hcop~
    contours from 1 to 10 K \kms.  Top right: BGPS 1.1 mm continuum
    dust emission overlaid with linear \n2hp~contours from 1.2 to 
    8.2 K \kms Middle: GLIMPSE 8 $\mu$m 
    overplotted with red BGPS beam-sized apertures that were
    used to determine clump masses.  The blue X's are VLA 3.6 cm point
    sources, and extended 3.6 cm sources are depicted as blue contours on
    the adjacent box.  Right:  \hcop, \n2hp~and \13CO
    spectra in clumps GLM1 (Stage 3) and GLM2 (Stage 2).}
  \label{fig:irdc33_hcop_n2hp}
\end{figure*}

%\clearpage
\appendix
\section{Individual clumps and clouds}
\label{sec:appendix}
This appendix explores the properties of each cloud in more detail.
Presented are GLIMPSE three-color images of each cloud with MAMBO 1.2 mm
contours and MIPSGAL 24~\micron~images with BGPS 1.1 mm contours.  For
clouds where \hcop~and \n2hp~maps were made, they are presented as contours
on a BGPS 1.1 mm greyscale image of the cloud.  Also presented are the
position and sizes of the apertures used for clump masses, the location of
VLA 3.6 cm point sources, and where resolved, contours of 3.6 cm emission 
on top of a GLIMPSE 8~\micron~image.  We also present \13CO, \hcop, and
\n2hp spectra at each clump position.

\subsection{G022.35+00.41}
G022.35+00.41 (see Figures \ref{fig:irdc6_glm_mips} and \ref{fig:irdc6_aps_spec}) 
is actually composed of two clumps along the line-of-sight, one
at 3.6 kpc and the other at 4.8 kpc.  At least one of the two
clumps is truly an active star-forming region.  G022.35+00.41:~GLM1 shows a bright
millimeter peak, masers, a ``green fuzzy," and a 24 \micron~point source.  
Interferometric measurements at 1 and 3 mm on the 
IRAM Plateau de Bure Interferometer by \citet{rat07} show that 
GLM1 contains 2 cores, less than 0.026 pc each.
The second clump (and associated filament) is quiescent and has diffuse millimeter
emission.  While the superposition of two clouds along the line-of-sight
lessens the intrinsic column of either, this example shows us that in the
confused inner Galaxy, it is not terribly uncommon to have two spatially
coincident dense clumps.  

\subsection{G023.60+00.00}
G023.60+00.00 (see Figures \ref{fig:irdc33_glm_mips} and \ref{fig:irdc33_hcop_n2hp})
is a particularly interesting example with regard to the comparison
of 8~\micron~extinction and dust emission.  The extinction masses of
the two clumps are very close (100 and 120 \Msun), however, the BGPS
1.1 mm masses differ by almost a factor of two (140 and 80 \Msun).  GLM1 is an
active clump, so it is likely warmer and denser than GLM2, an intermediate
clump.  Also, there is a bright
H~II region to the southwest corner of the image shown in Figure
\ref{fig:irdc33_glm_mips}, which increases the foregound 8
\micron~emission.  Also, the \hcop~and \n2hp~maps show 
 only diffuse emission around GLM2 and 
a bright peak toward GLM1
(see Figures \ref{fig:irdc33_bgps_hcop}
and \ref{fig:irdc33_bgps_n2hp}), tracing the dense, hot gas associated with
an active clump.

At 2.9 $\times$ 1.8 pc (major $\times$ minor axes of the IRDC ellipse, see
Figure\ref{fig:irdc33extmass}), 
G023.60+00.00 is on the smaller side for an IRDC, and shows a filamentary
morphology.  It is near two evolved H~II regions and a complicated
web of mid-infrared bubbles, filamentary IRDCs, and Photo-Dissociation
Regions (PDRs; see Figure
\ref{fig:irdc33_apertures}).  The location of G023.60+00.00 is at 3.6 kpc, right
amidst the active star formation in the molecular ring \citep{jac06}.

%\clearpage
\begin{figure*} 
  \centering
  \subfigure{
    \label{fig:irdc51_glm3_iram}
    \includegraphics[width=.46\textwidth]{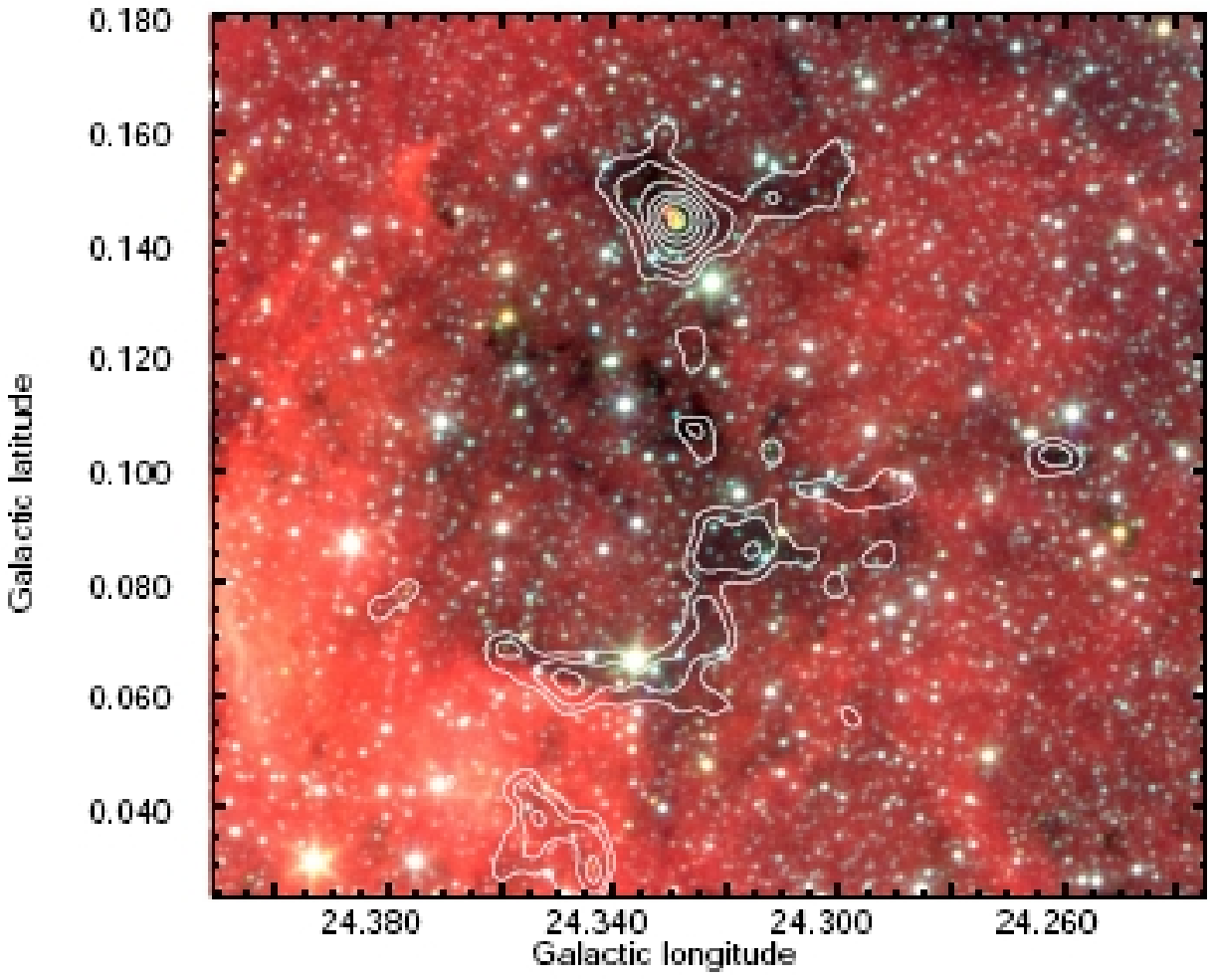}}
%  \vspace{-0.2in}
  \subfigure{
    \label{fig:irdc51_mips_bgps}
    \includegraphics[width=.46\textwidth]{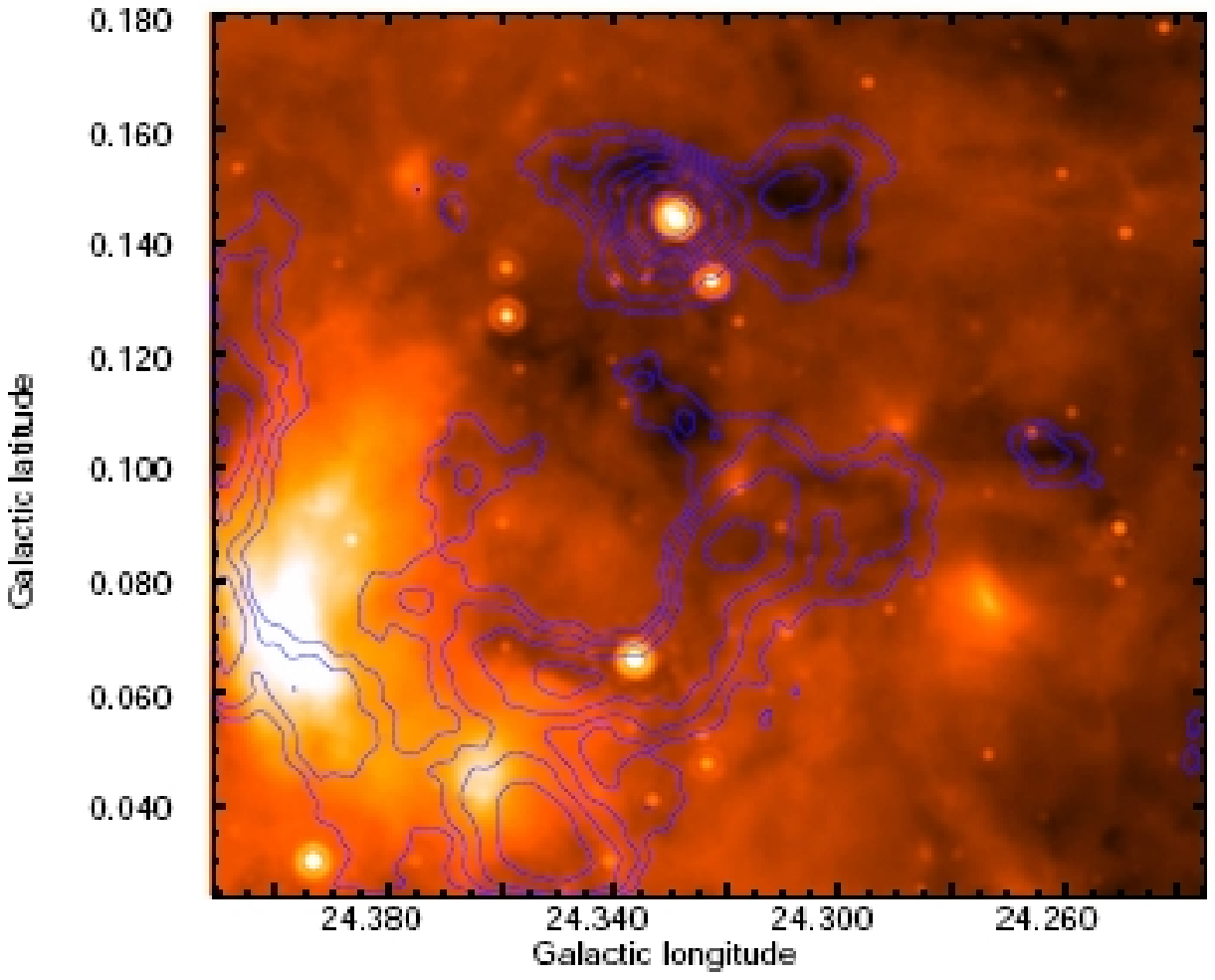}}\\
  \caption{G024.33+00.11: Left: GLIMPSE three-color image, red is 8~\micron, green is
    4.5 \micron~and blue is 3.6 \micron~with MAMBO 1.2 mm contours 
    overlaid.  The contours are on a log scale from 30 to 1280 mJy beam$^{-1}$.
    Right: MIPSGAL 24 \micron~image with BGPS 1.1 mm contours.  The contours
    are on a log scale from 0.12 to 2.5 Jy beam$^{-1}$.}
  \label{fig:irdc51_glm_mips}
\end{figure*}
%\newpage
\begin{figure*}
  \centering
  \subfigure{
    \label{fig:irdc51_apertures}
    \includegraphics[width=.46\textwidth]{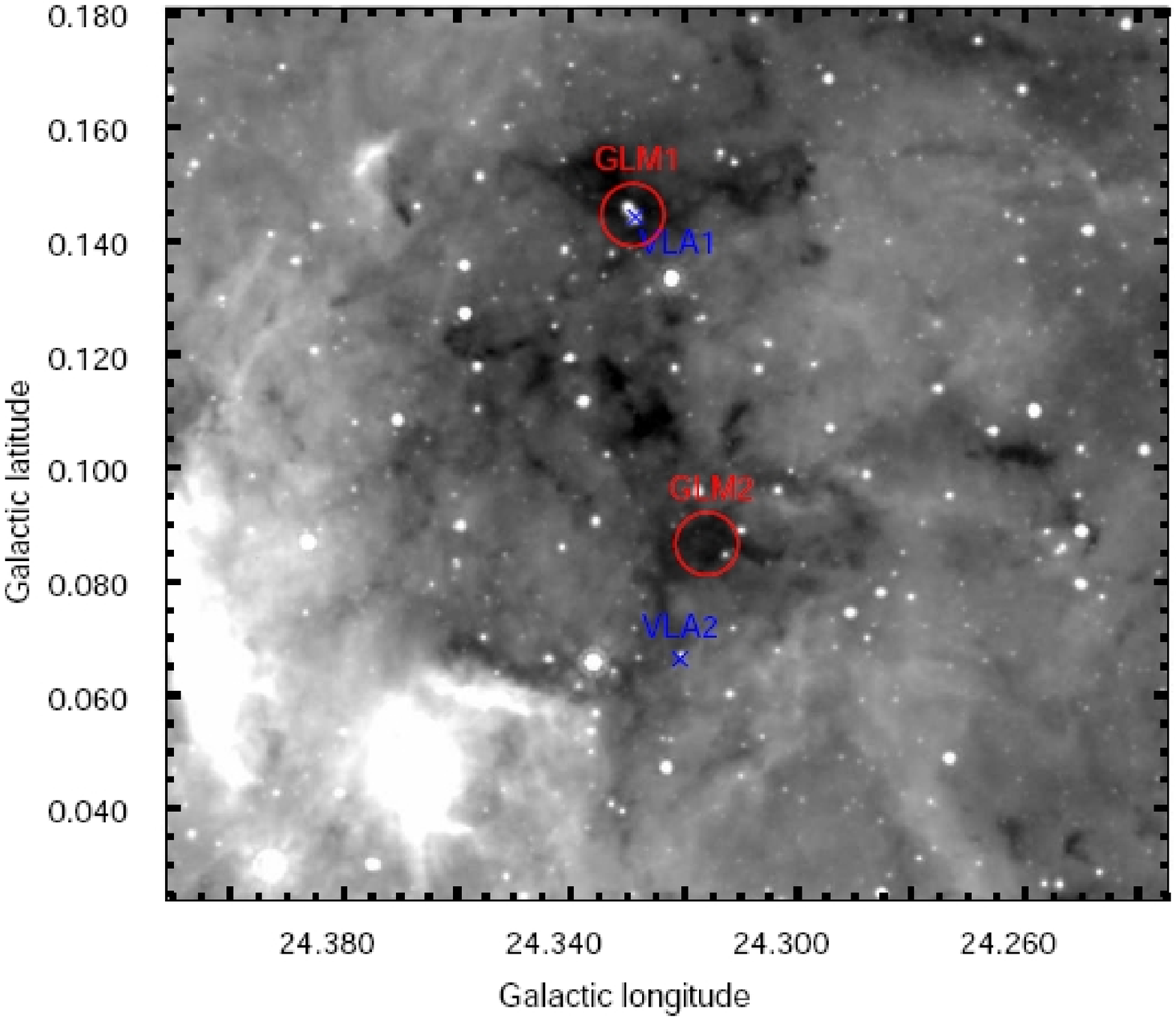}}
%  \vspace{-0.2in}
  \subfigure{
    \label{fig:irdc51_allspectra}
    \includegraphics[width=.46\textwidth]{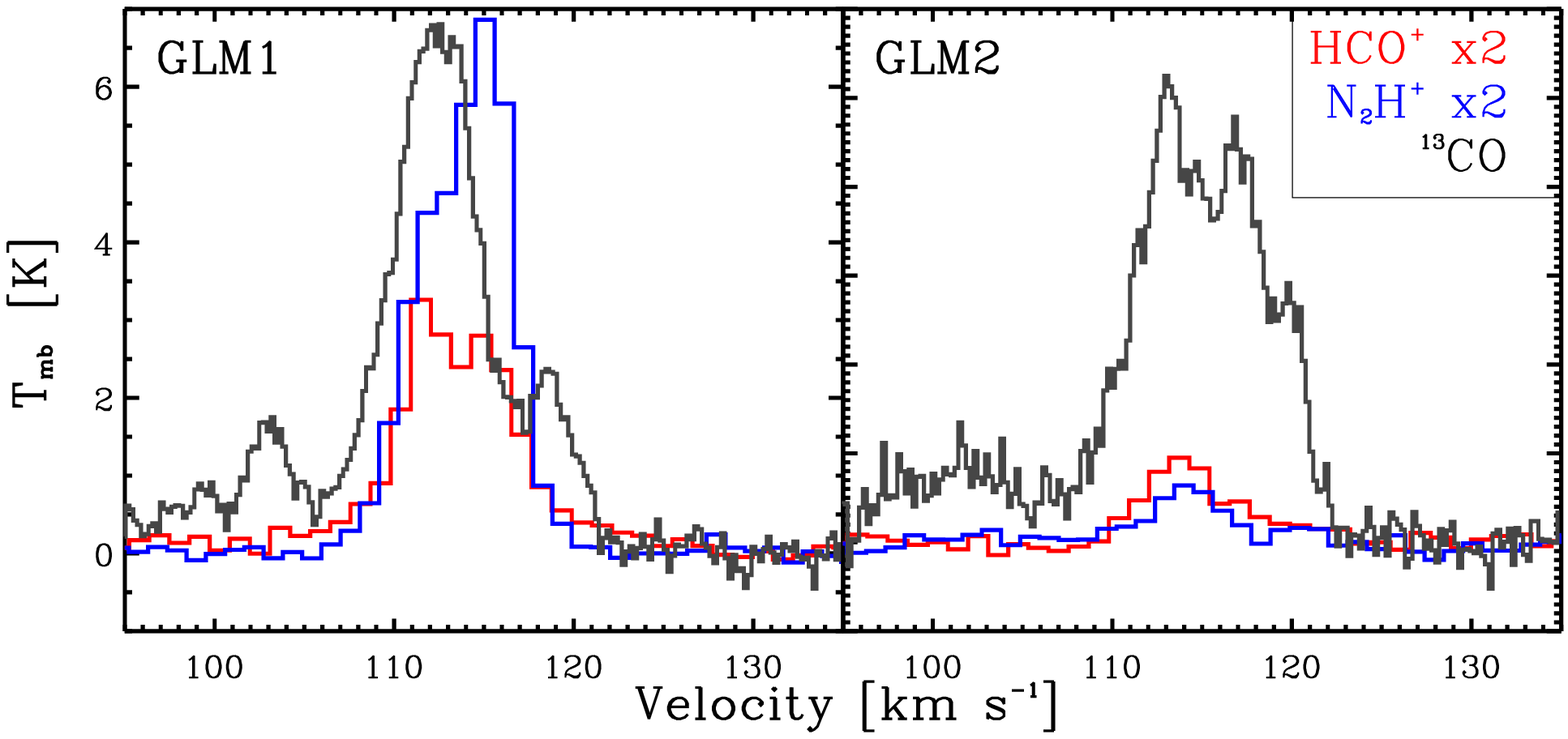}}\\
  \caption{G024.33+00.11: Left: GLIMPSE 8 $\mu$m 
    overplotted with red BGPS beam-sized apertures that were
    used to determine clump masses.  The blue X's are VLA 3.6 cm point
    sources.  Right:  \hcop, \n2hp~and \13CO
    spectra in clumps GLM1 (Stage 3) and GLM2 (Stage 1).  The \hcop~and \n2hp~spectra have been
    multiplied by a factor of two, in order to see them more clearly
    relative to the \13CO.}
  \label{fig:irdc51_aps_spec}
\end{figure*}

\begin{figure*} 
  \centering
  \subfigure{
    \label{fig:irdc9_glm3_iram}
    \includegraphics[width=.46\textwidth]{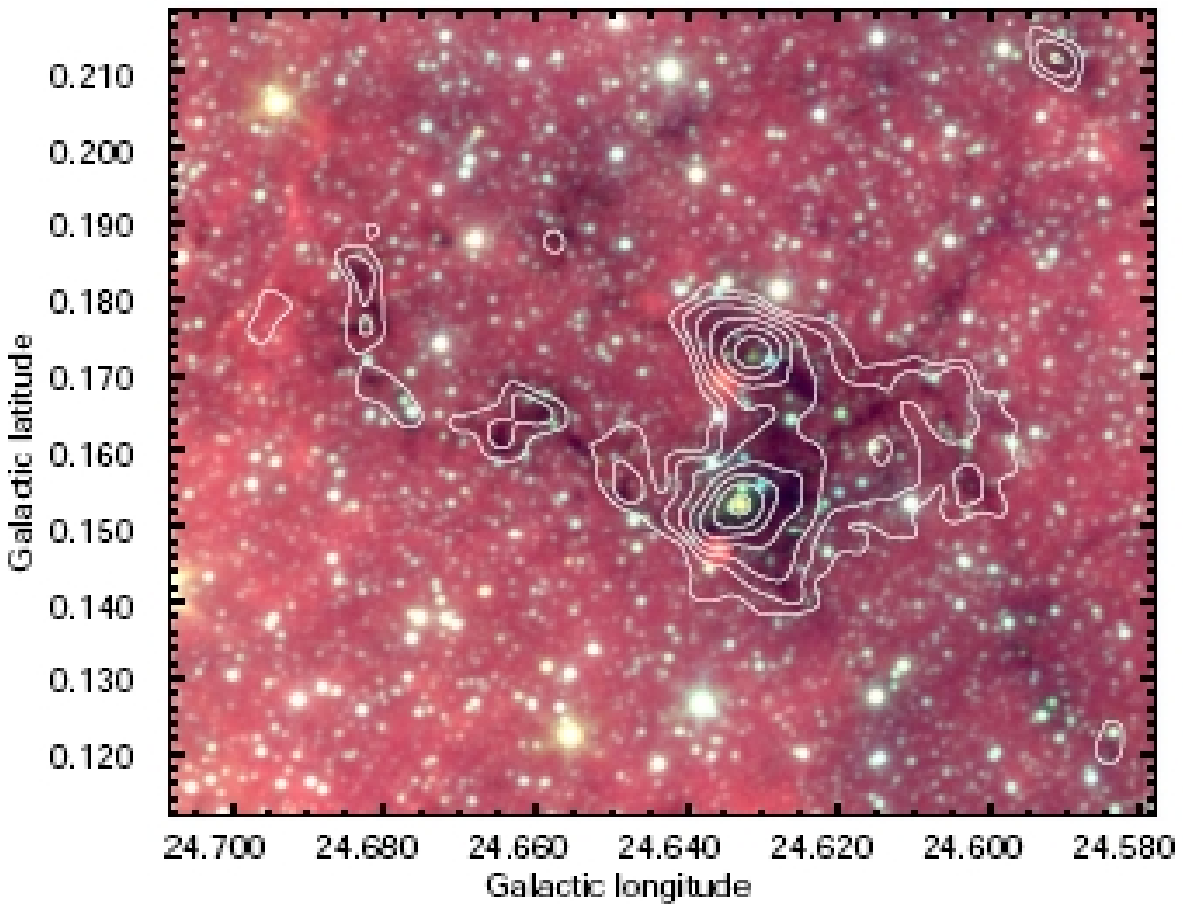}}
%  \vspace{-0.2in}
  \subfigure{
    \label{fig:irdc9_mips_bgps}
    \includegraphics[width=.46\textwidth]{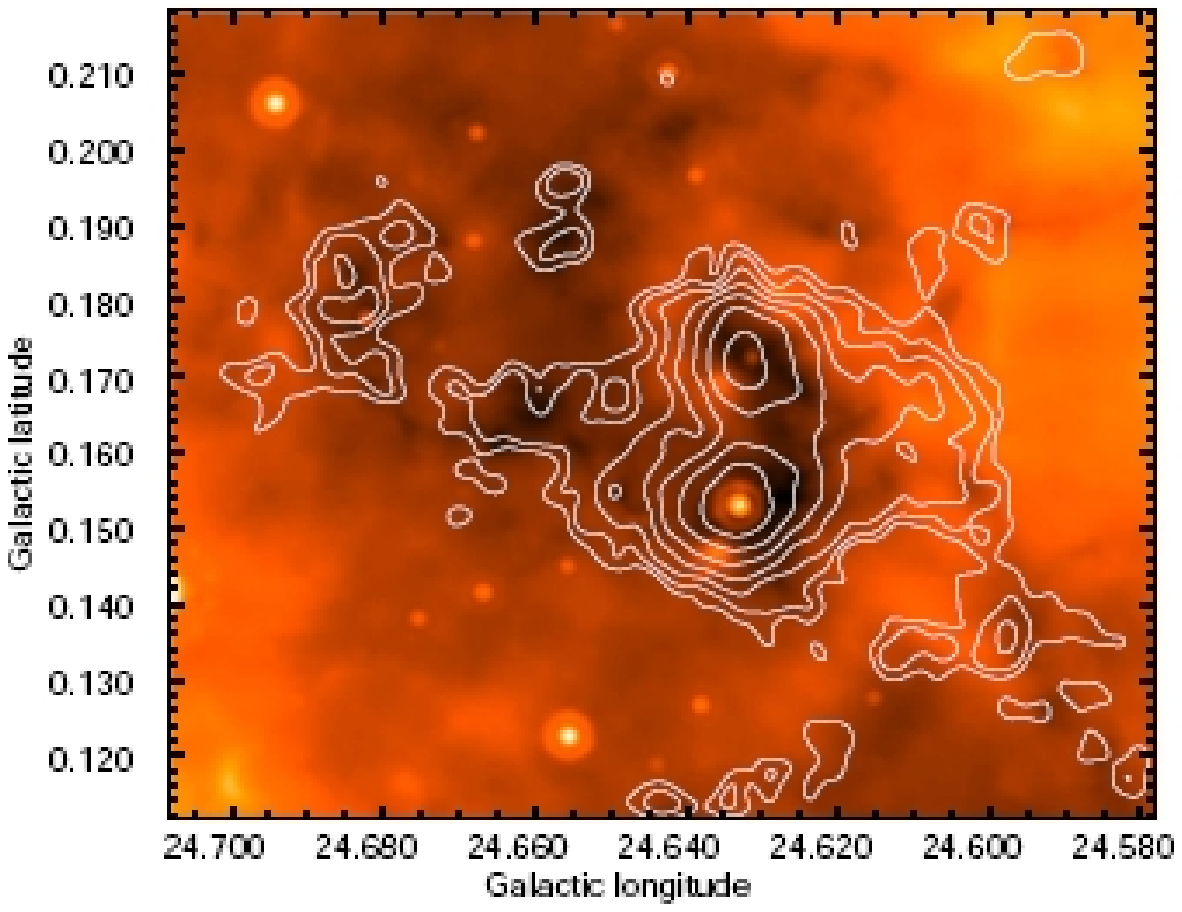}}\\
  \caption{G024.60+00.08: Left: GLIMPSE three-color image, red is 8~\micron, green is
    4.5 \micron~and blue is 3.6 \micron~with MAMBO 1.2 mm contours 
    overlaid.  The contours are on a log scale from 15 to 360 mJy beam$^{-1}$.
    Right: MIPSGAL 24 \micron~image with BGPS 1.1 mm contours.  The contours
    are on a log scale from 0.12 to 0.8 Jy beam$^{-1}$. }
  \label{fig:irdc9_glm_mips}
\end{figure*}
\vspace{-1in}

%\newpage
\begin{figure*}
  \centering
  \subfigure{
    \label{fig:irdc9_1_bgps_hcop}
    \includegraphics[width=.46\textwidth]{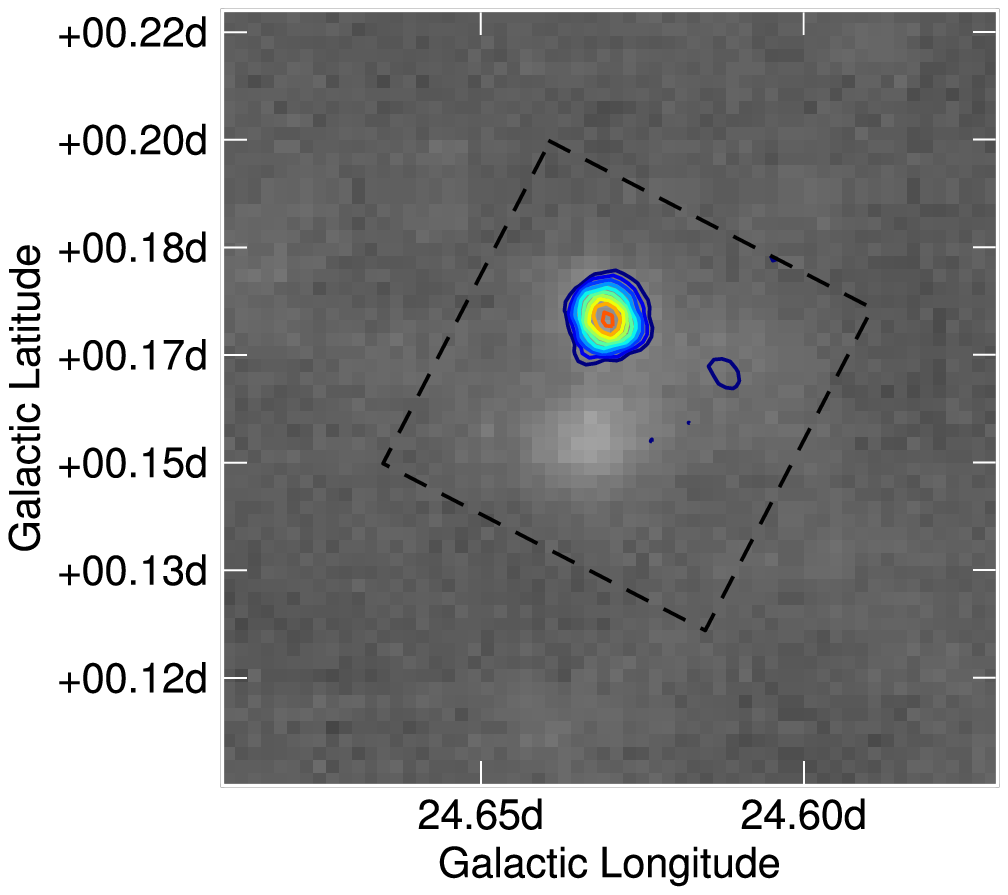}}
  \subfigure{
    \label{fig:irdc9_1_bgps_n2hp}
    \includegraphics[width=.46\textwidth]{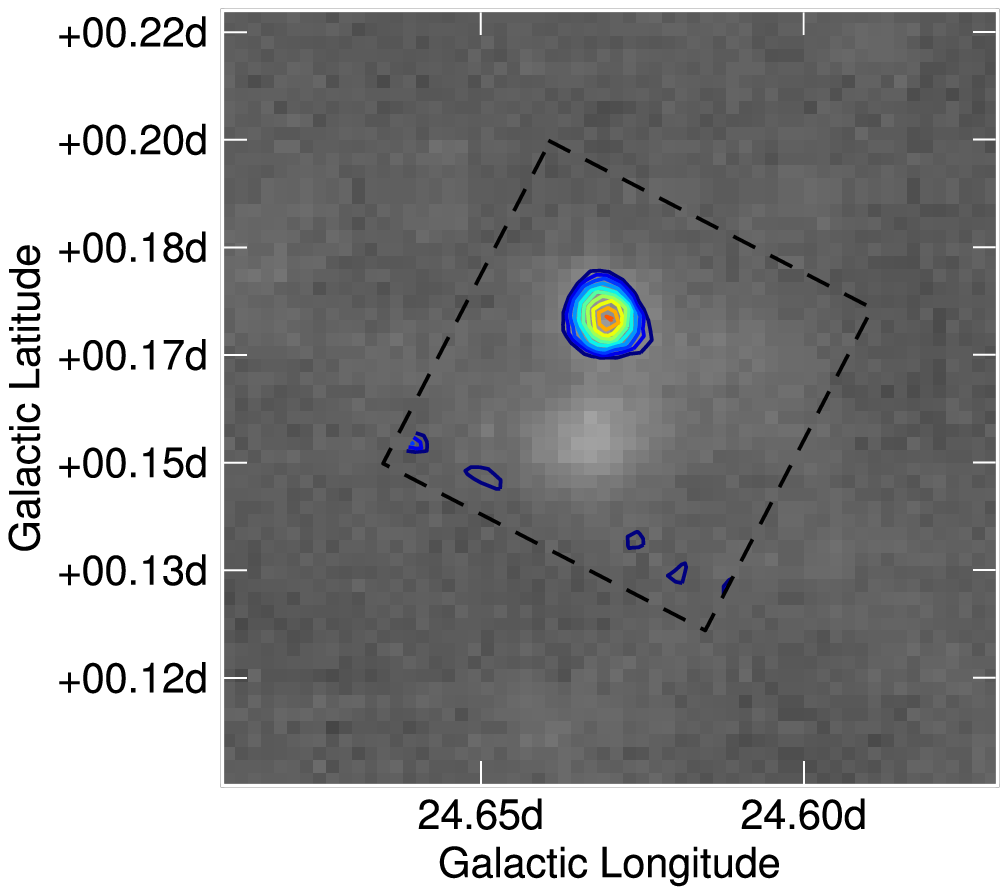}}\\
\vspace{-1in}
 \subfigure{
    \label{fig:irdc9_2_bgps_hcop}
    \includegraphics[width=.46\textwidth]{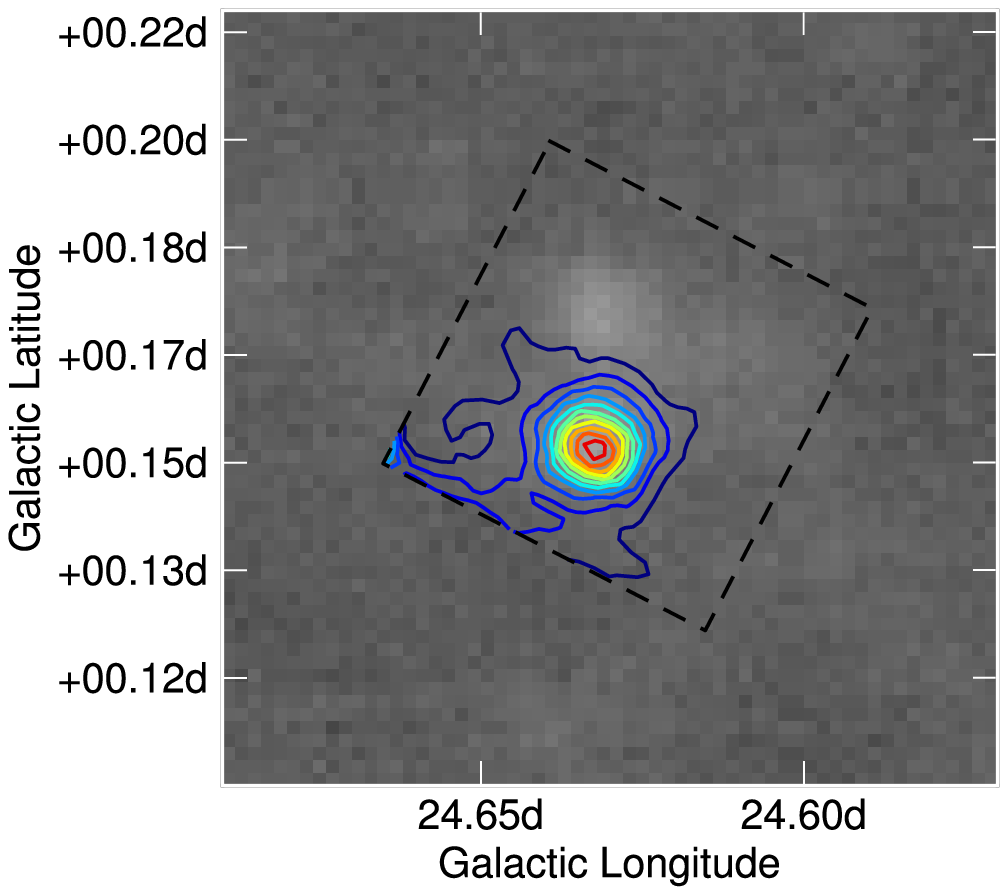}}
  \subfigure{
    \label{fig:irdc9_2_bgps_n2hp}
    \includegraphics[width=.46\textwidth]{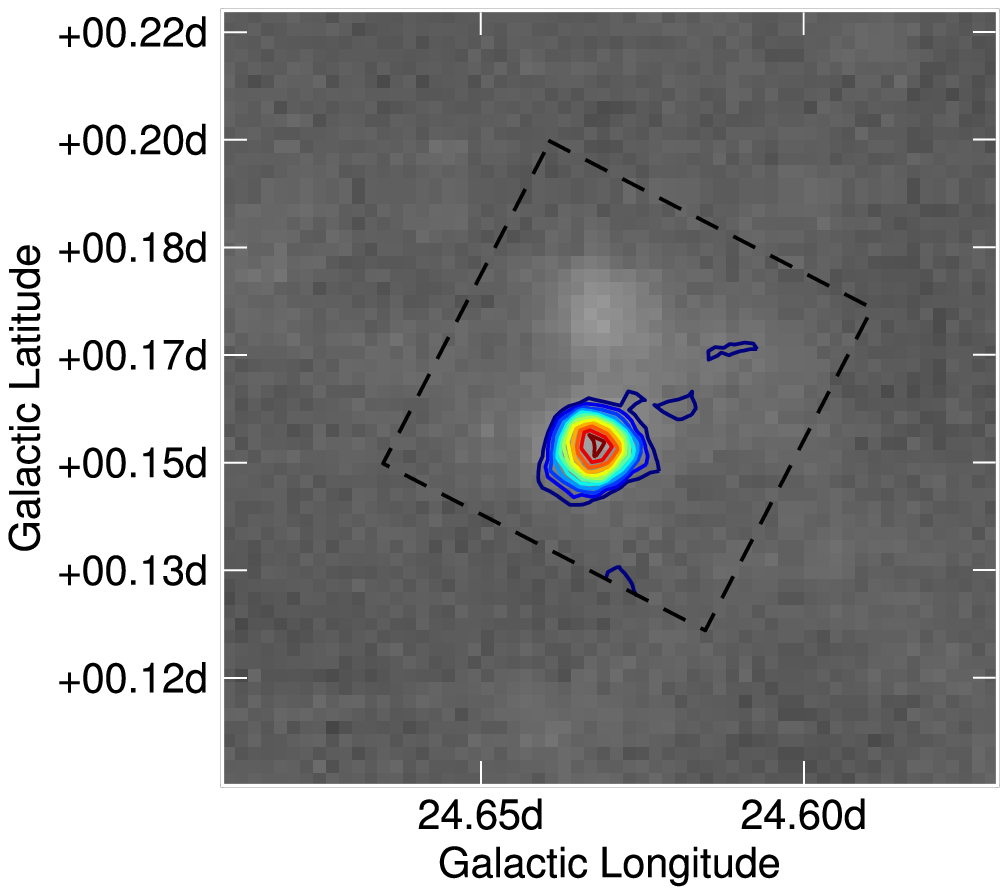}}\\
  \caption{G024.60+00.08: two distinct velocity components/clumps along the
    line-of-sight: Top left: BGPS 1.1 mm continuum dust emission overlaid with linear \hcop~
    contours from 1.2 to 4.4 K \kms from GLM1.  Top right: BGPS 1.1 mm continuum
    dust emission overlaid with linear \n2hp~contours from 1.5 to 
    5 K \kms from GLM1.
    Bottom left: BGPS 1.1 mm continuum dust emission overlaid with linear \hcop~
    contours from 1.8 to 9.8 K \kms from GLM2.  Bottom right: BGPS 1.1 mm continuum
    dust emission overlaid with linear \n2hp~contours from 1.5 to 
    5.5 K \kms from GLM2}
  \label{fig:irdc9_hcop_n2hp}
\end{figure*}
%\newpage
\begin{figure*}
  \centering
 \subfigure{
    \label{fig:irdc9_apertures}
    \includegraphics[width=0.75\textwidth]{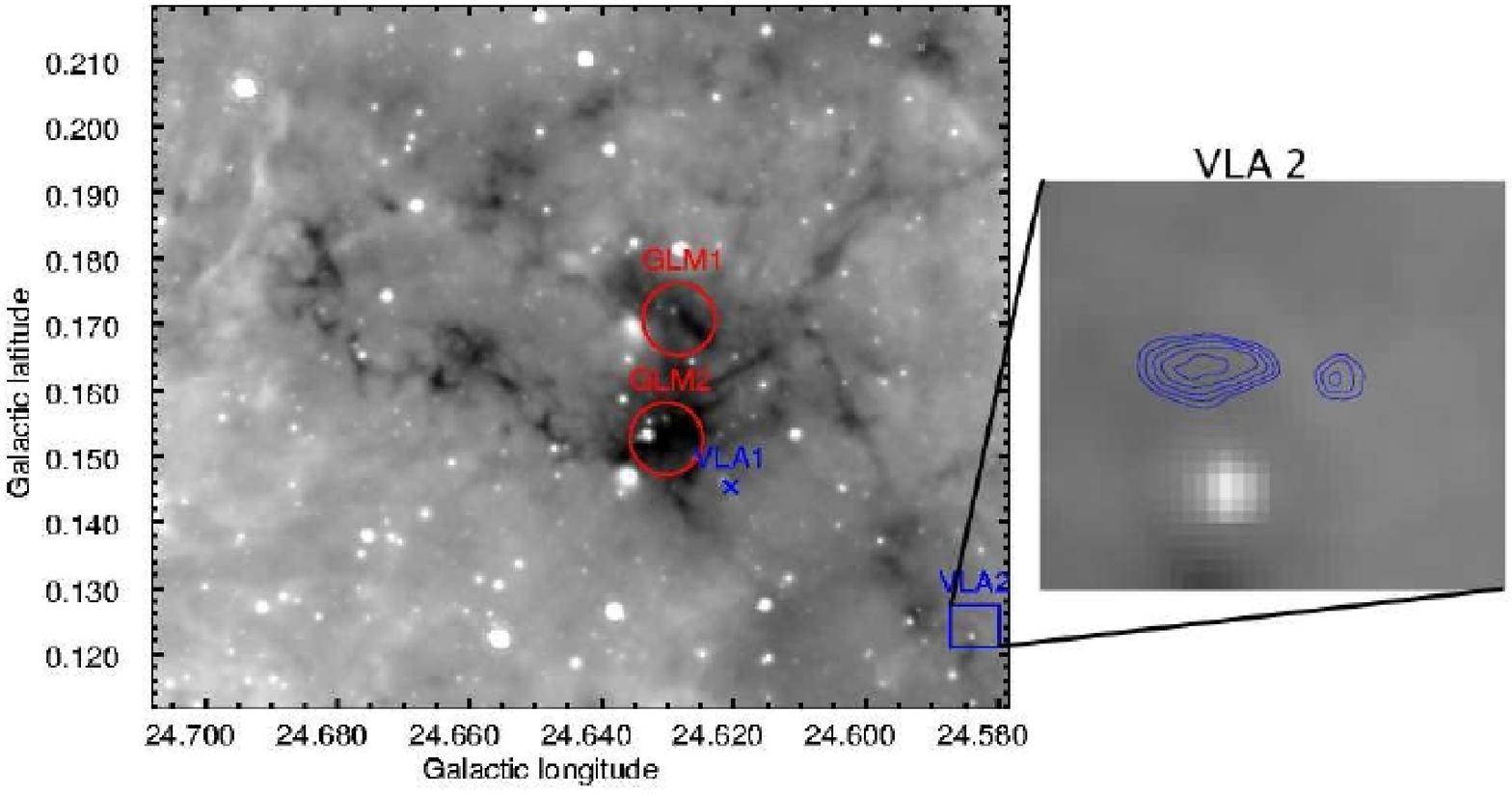}}\\
 %\vspace{-0.2pt}
  \subfigure{
    \label{fig:irdc9_allspectra}
    \includegraphics[width=0.75\textwidth]{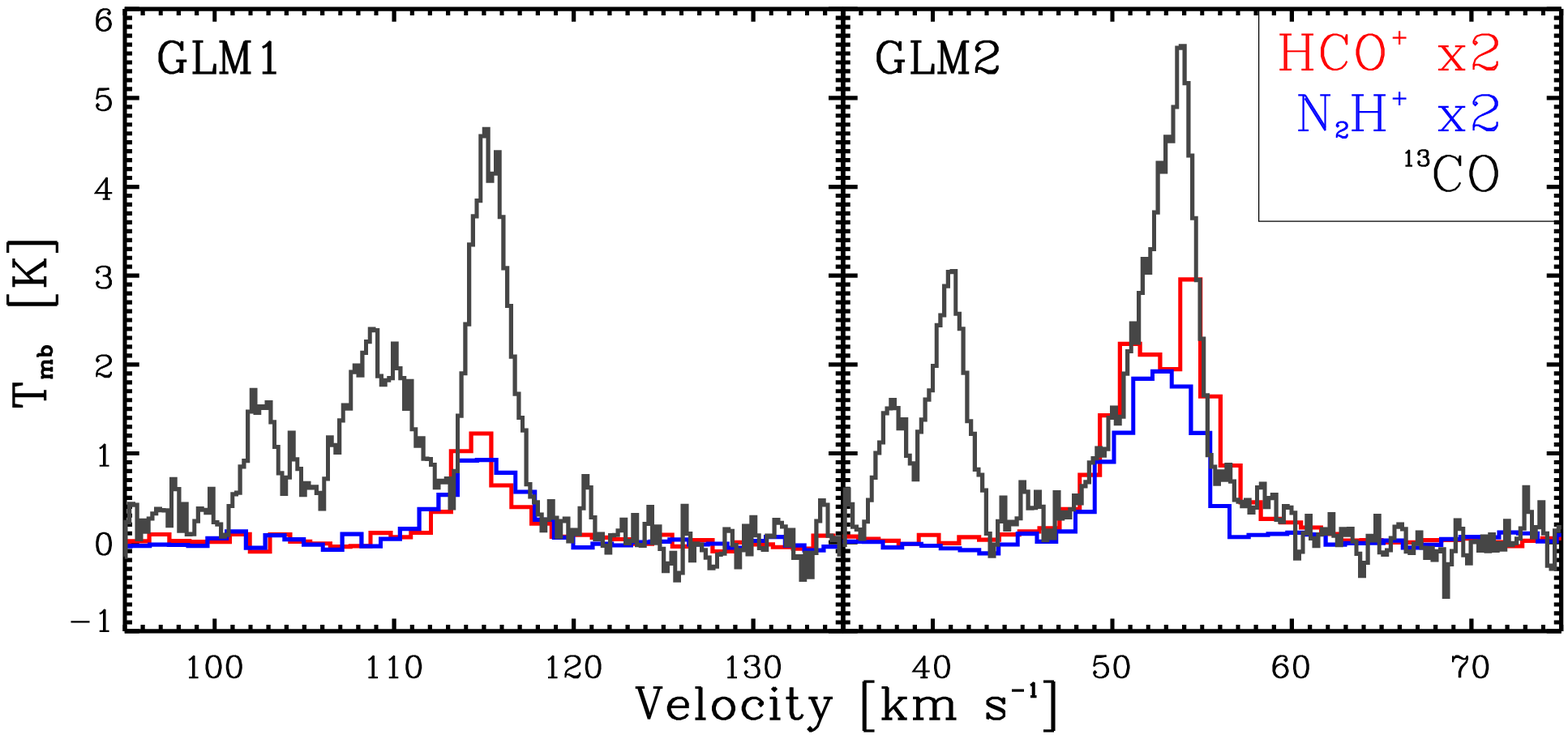}}\\
  \caption{G024.60+00.08: 
    Left: GLIMPSE 8 $\mu$m 
    overplotted with red BGPS beam-sized apertures that were
    used to determine clump masses.  The blue X's are VLA 3.6 cm point
    sources, and extended 3.6 cm sources are depicted as blue contours in
    the adjacent box.  Right:  \hcop, \n2hp~and \13CO
    spectra in clumps GLM1 (Stage 2) and GLM2 (Stage 2).  The \hcop~and \n2hp~spectra have been
    multiplied by a factor of two, in order to see them more clearly
    relative to the \13CO.  Note the distinctly different clump
    velocities}
  \label{fig:irdc9_aps_spec}
\end{figure*}
\clearpage
\newpage

%\clearpage
\subsection{G024.33+00.11}
G024.33+00.11 (see Figures \ref{fig:irdc51_glm_mips} and
\ref{fig:irdc51_aps_spec}) contains a dense, active clump and more diffuse
filamentary structure.  There is a ridge
between dark 8~\micron~extinction and bright 8~\micron~emission 
at the southern edge of the cloud.  At a
distance of 5.9 kpc, this cloud is among the most distant in our sample.  G024.33+00.11
was not mapped in \hcop~and \n2hp, but it would be a good candidate for a
future mapping project, to see if the dense ridge is associated with the
IRDC.  Additionally, portions of the cloud are 
dark at 8~\micron, but show very little millimeter emission.

G024.33+00.11:~GLM1 is a particularly good example of a dense, active
clump.  Interferometric measurements at 1 and 3 mm on the 
IRAM Plateau de Bure Interferometer by \citet{rat07} show that 
G024.33+00.11:~GLM1 has a single core smaller than 0.035 pc.  Figure
\ref{fig:irdc51_allspectra} exemplifies the increase of \hcop~and \n2hp~in
an active clump, and the self-absorption of \13CO~and \hcop~in a
particularly dense environment.  This is in contrast to the weak
\hcop~and \n2hp~detections in G024.33+00.11:~GLM2.

\subsection{G024.60+00.08}
G024.60+00.08 (see Figures \ref{fig:irdc9_glm_mips}, \ref{fig:irdc9_hcop_n2hp}, and
\ref{fig:irdc9_aps_spec}) is the second (of two) examples in our sample
which is
comprised of two unique clumps along the line-of-sight, rather than one
contiguous object.  The two clumps are spatially adjacent with only a
slight overlap of diffuse emission, which is an astonishing
coincidence.  The \hcop~and \n2hp~maps confirm (see Figure
\ref{fig:irdc9_hcop_n2hp}) this chance alignment.  

At 6.0 kpc, G024.60+00.08:~GLM1 is the most distant clump in our sample.  The 8
\micron~extinction does not appear abnormally large, but given the amount
of foreground emission between us and the clump, it has one of the highest
extinction masses in our sample.  G024.60+00.08:~GLM2, on the other hand, is located
in the molecular ring at 3.4 kpc.  This clump shows self-absorption of
\hcop, and an asymmetric line profile in \13CO, indicative of outflows.
These clumps are both categorized as intermediate.
Interferometric measurements at 1 and 3 mm on the 
IRAM Plateau de Bure Interferometer by \citet{rat07} show that 
G024.60+00.08:~GLM1 contains 5 cores, less than 0.041 pc each, while
GLM2 contains 3 cores, smaller than 0.024 pc each.

%\clearpage
\begin{figure*} 
  \centering
  \subfigure{
    \label{fig:irdc2_glm3_iram}
    \includegraphics[width=0.46\textwidth]{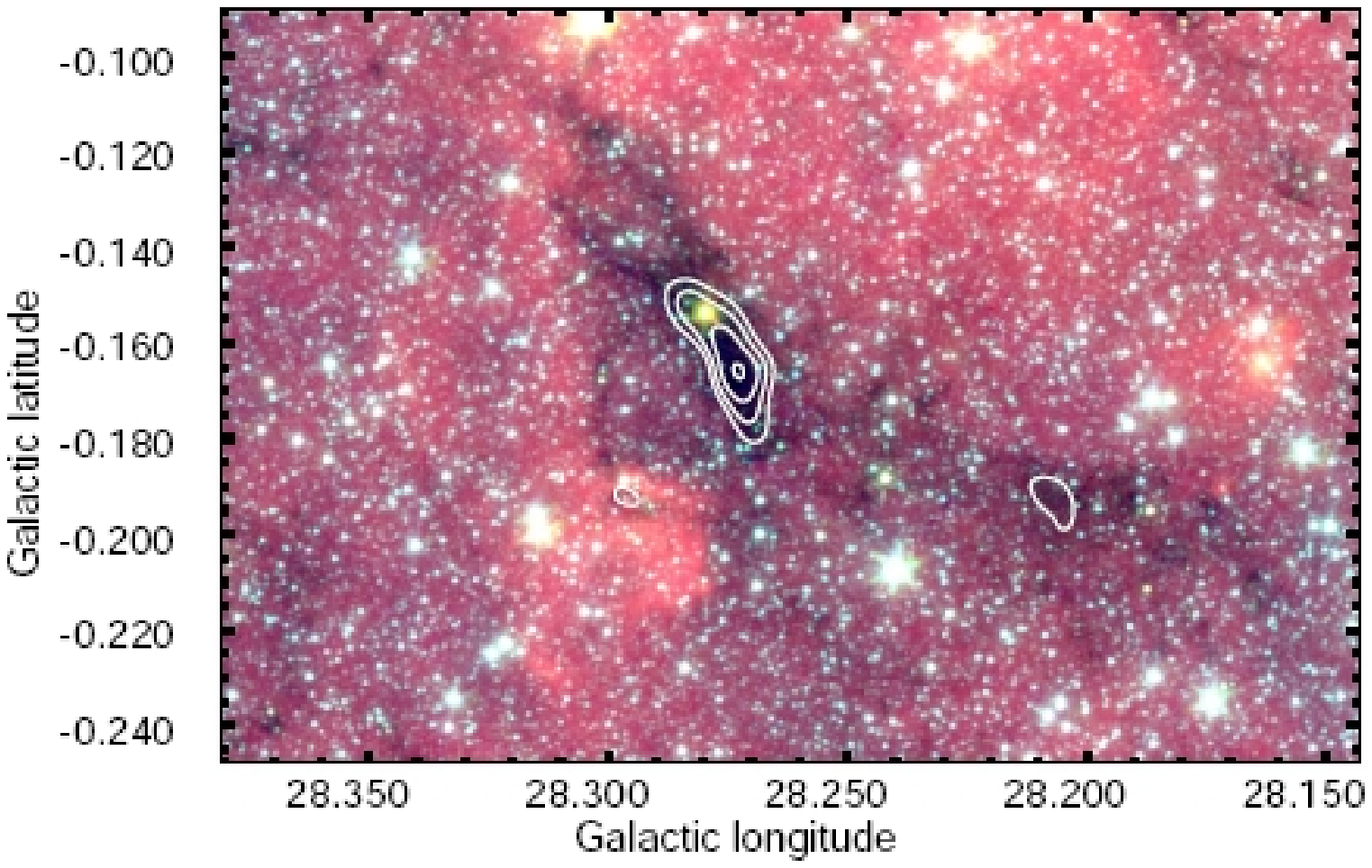}}
  \subfigure{
    \label{fig:irdc2_mips_bgps}
    \includegraphics[width=0.46\textwidth]{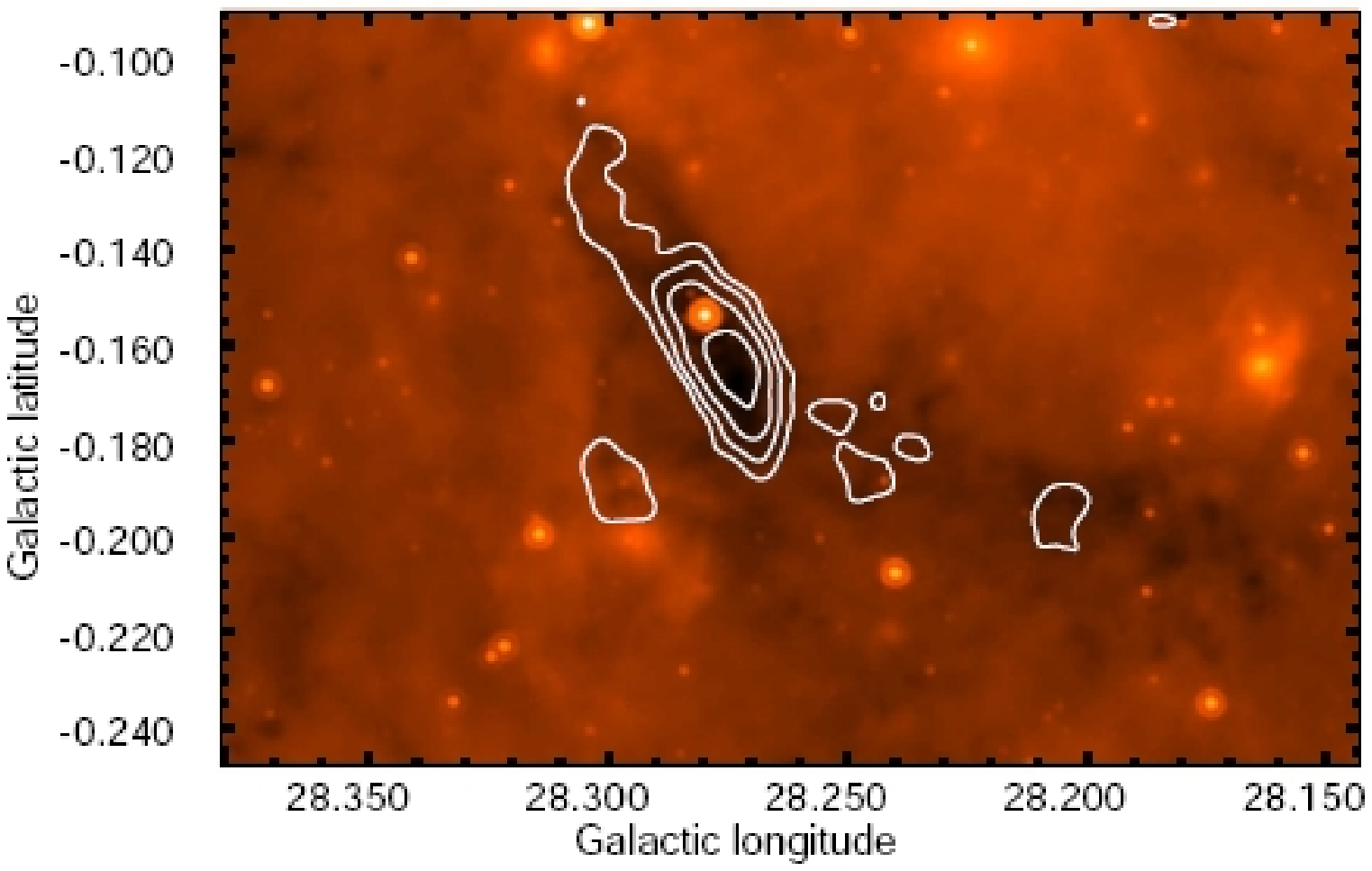}}\\
  \caption{G028.23-00.19: Left: GLIMPSE three-color image, red is 8~\micron, green is
      4.5 \micron~and blue is 3.6 \micron~with MAMBO 1.2 mm contours 
      overlaid.  The contours are on a log scale from 30 to 112 mJy beam$^{-1}$.
      Right: MIPSGAL 24 \micron~image with BGPS 1.1 mm contours.  The contours
      are on a log scale from 0.12 to 0.64 Jy beam$^{-1}$.}
  \label{fig:irdc2_glm_mips}
\end{figure*}
%\newpage
\begin{figure*}
  \centering
  \subfigure{
    \label{fig:irdc2_apertures}
    \includegraphics[width=0.46\textwidth]{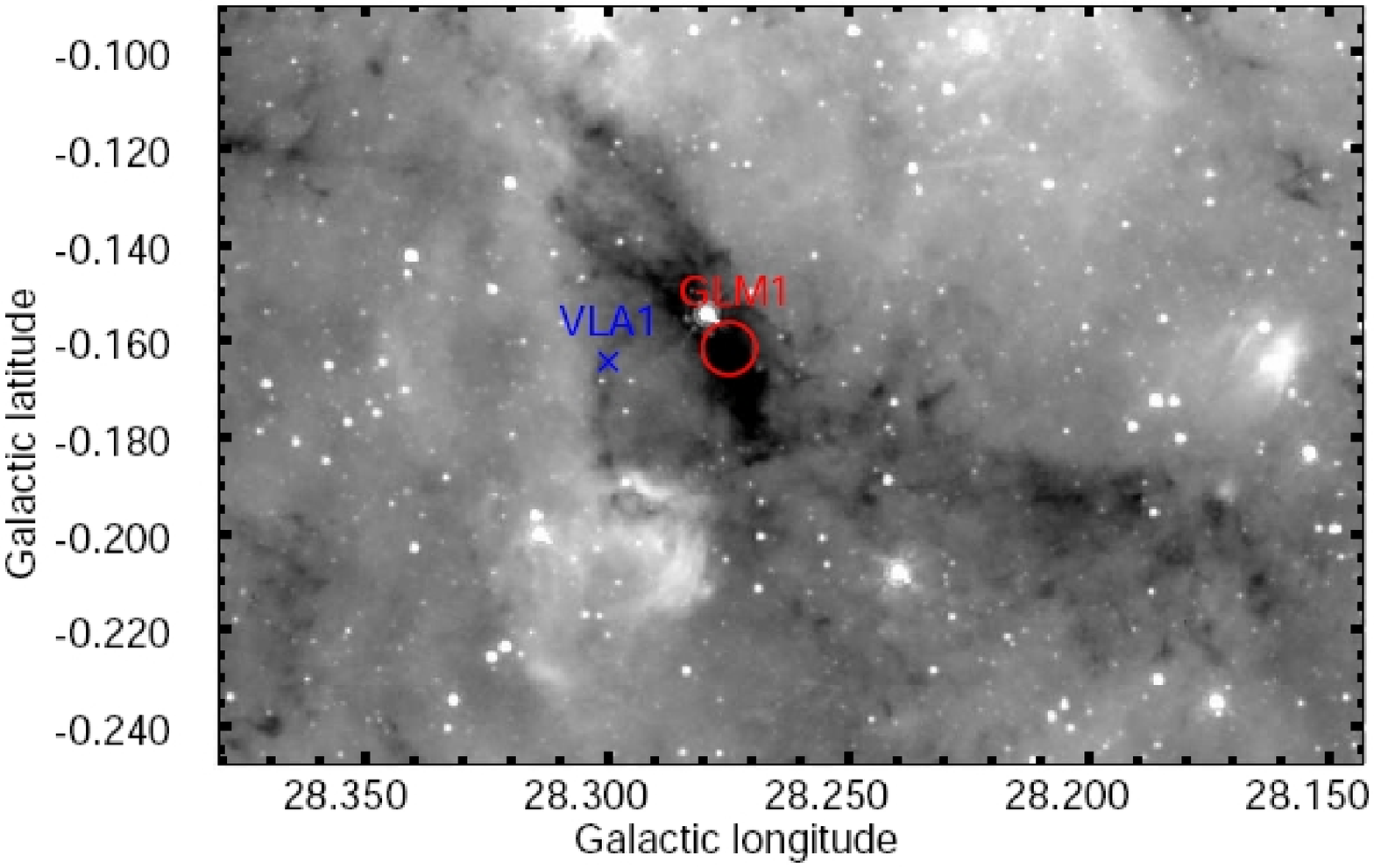}}
  \subfigure{
    \label{fig:irdc2_allspectra}
    \includegraphics[width=0.46\textwidth]{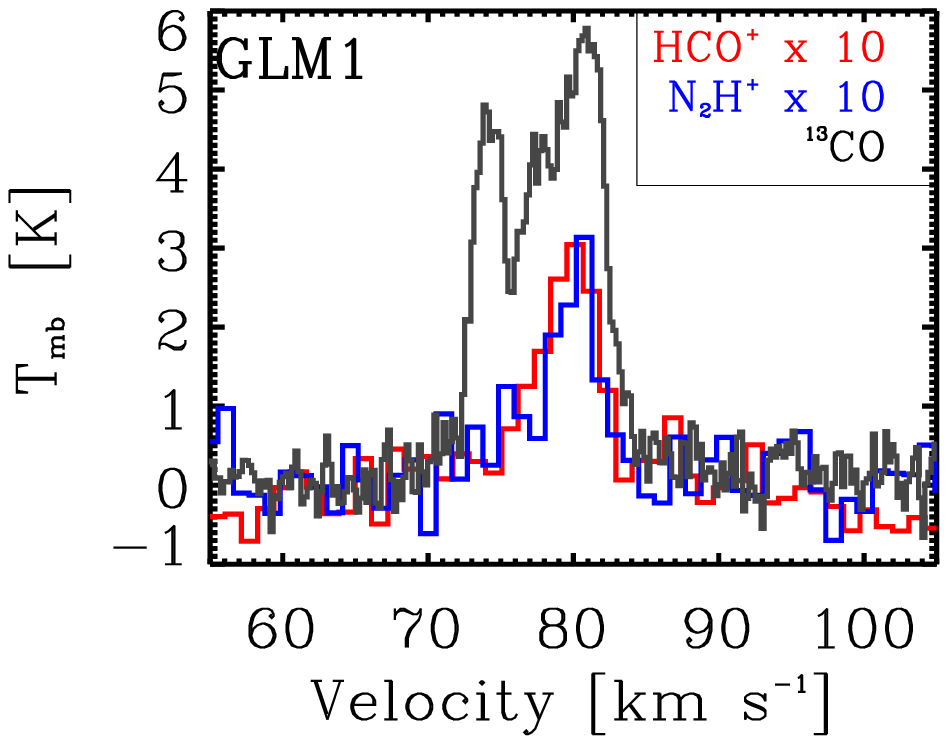}}\\
   \caption{G028.23-00.19: Left: GLIMPSE 8 $\mu$m 
    overplotted with red BGPS beam-sized aperture that was
    used to determine the clump mass.  The blue X's are VLA 3.6 cm point
    sources.  Right:  \hcop, \n2hp~and \13CO
    spectra in clump GLM1 (Stage 1).  The \hcop~and \n2hp~spectra have been
    multiplied by a factor of ten, in order to see them more clearly
    relative to the \13CO.}
  \label{fig:irdc2_aps_spec}
\end{figure*}

%\clearpage
\begin{figure*} 
  \centering
  \subfigure{
    \label{fig:irdc1_glm3_iram}
    \includegraphics[width=0.46\textwidth]{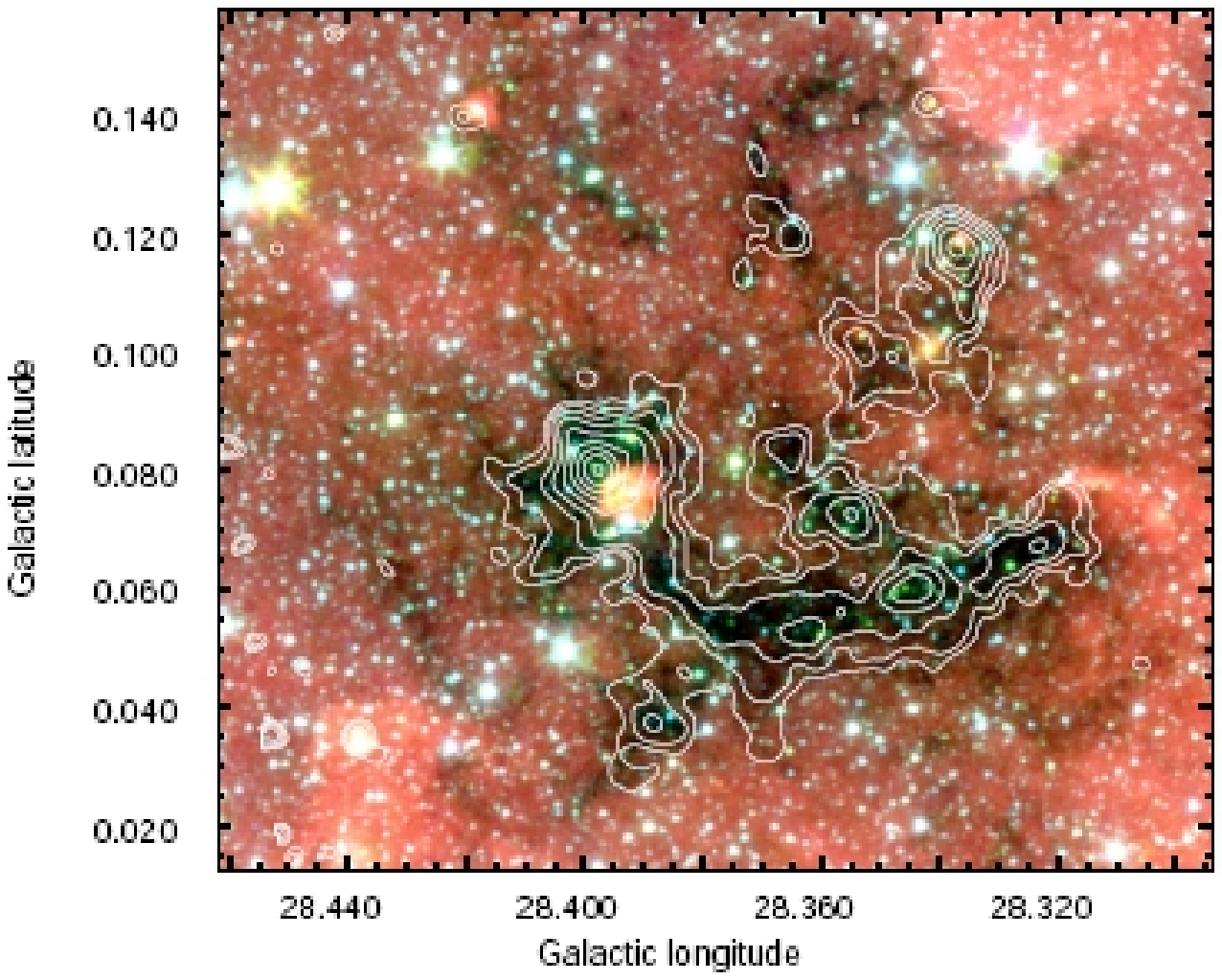}}
%  \vspace{-0.2in}
  \subfigure{
    \label{fig:irdc1_mips_bgps}
    \includegraphics[width=0.46\textwidth]{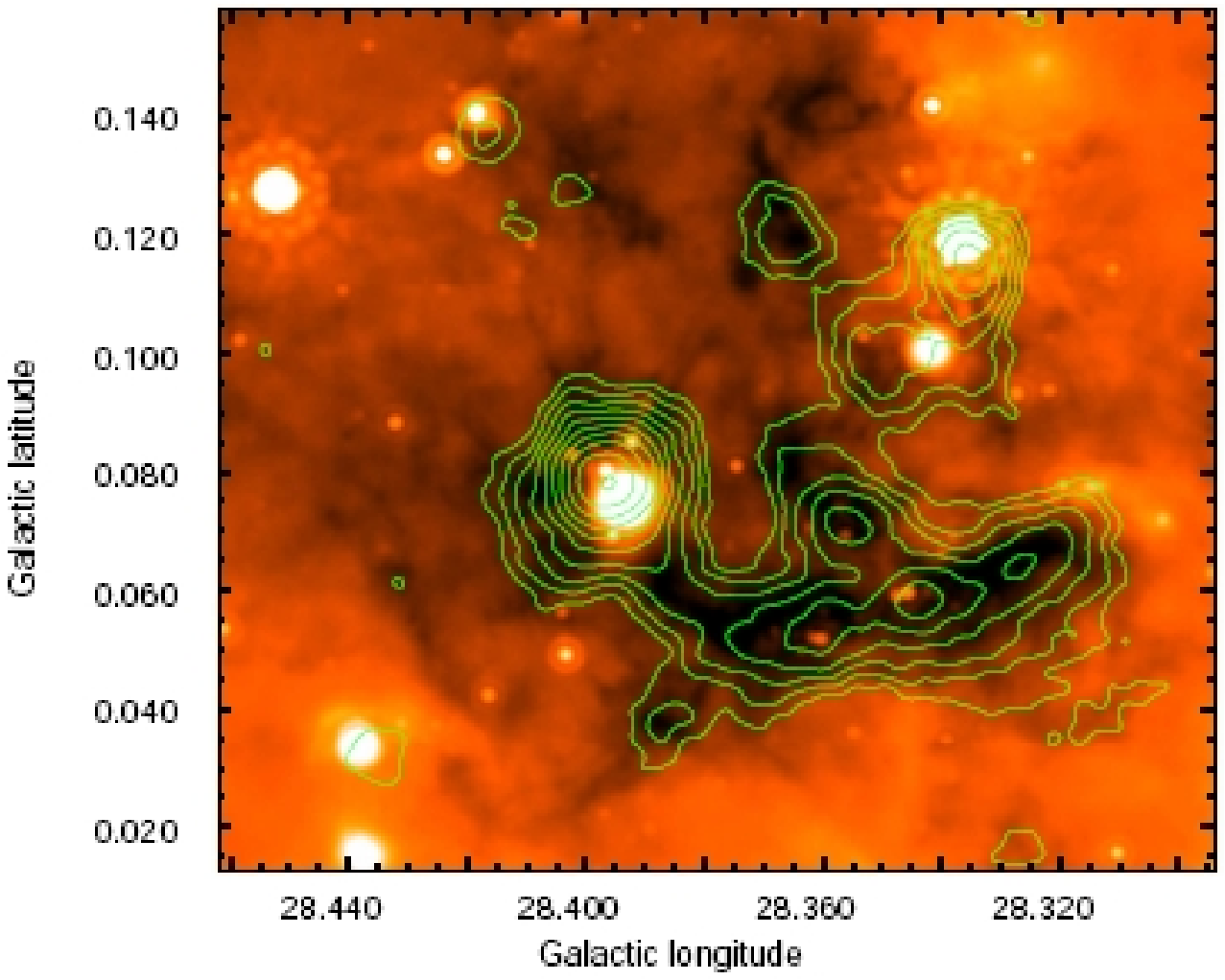}}\\
  \caption{G028.37+00.07: Left: GLIMPSE three-color image, red is 8~\micron, green is
    4.5 \micron~and blue is 3.6 \micron~with MAMBO 1.2 mm contours 
    overlaid.  The contours are on a log scale from 30 to 1340 mJy beam$^{-1}$.
    Right: MIPSGAL 24 \micron~image with BGPS 1.1 mm contours.  The contours
    are on a log scale from 0.2 to 2.2 Jy beam$^{-1}$. }
  \label{fig:irdc1_glm_mips}
\end{figure*}
%\newpage

\begin{figure*}
  \centering
  \subfigure{
    \centering
    \label{fig:irdc1_bgps_hcop}
    \vspace{-60pt}
    \includegraphics[trim=0mm 3mm 0mm 3mm, clip, scale=0.5]{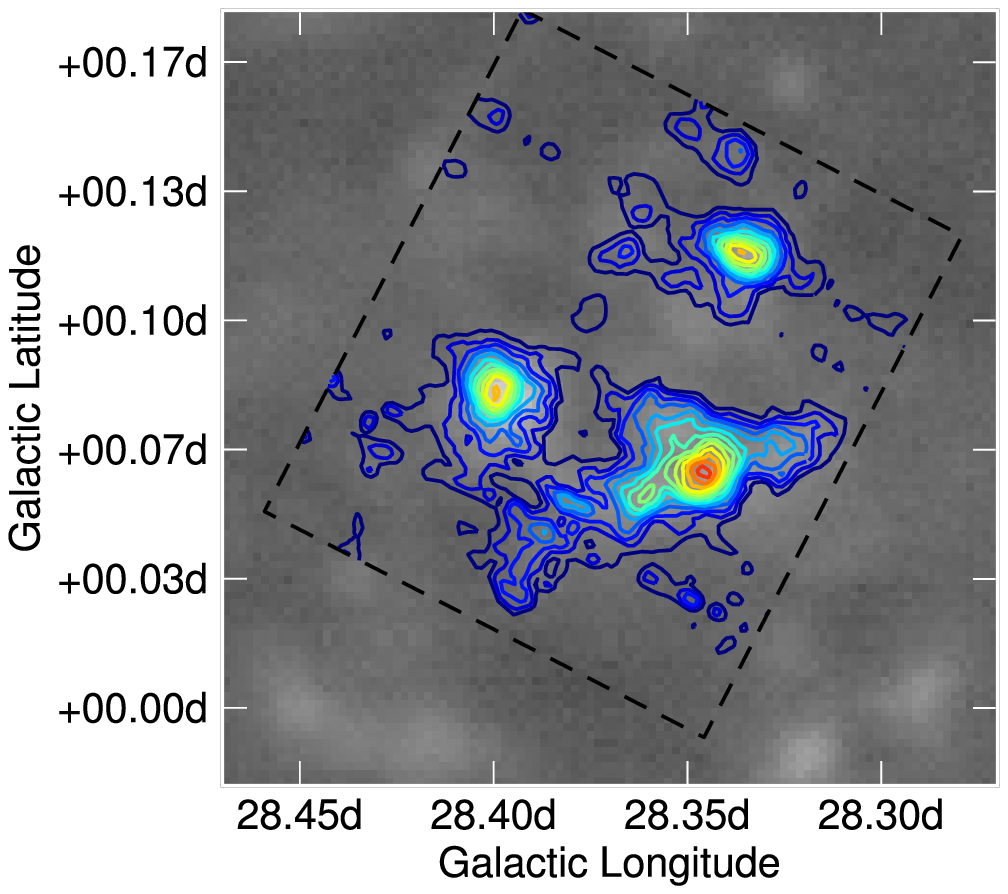}}
  \subfigure{
    \centering
    \label{fig:irdc1_bgps_n2hp}
    \includegraphics[trim=0mm 3mm 0mm 3mm, clip, scale=0.5]{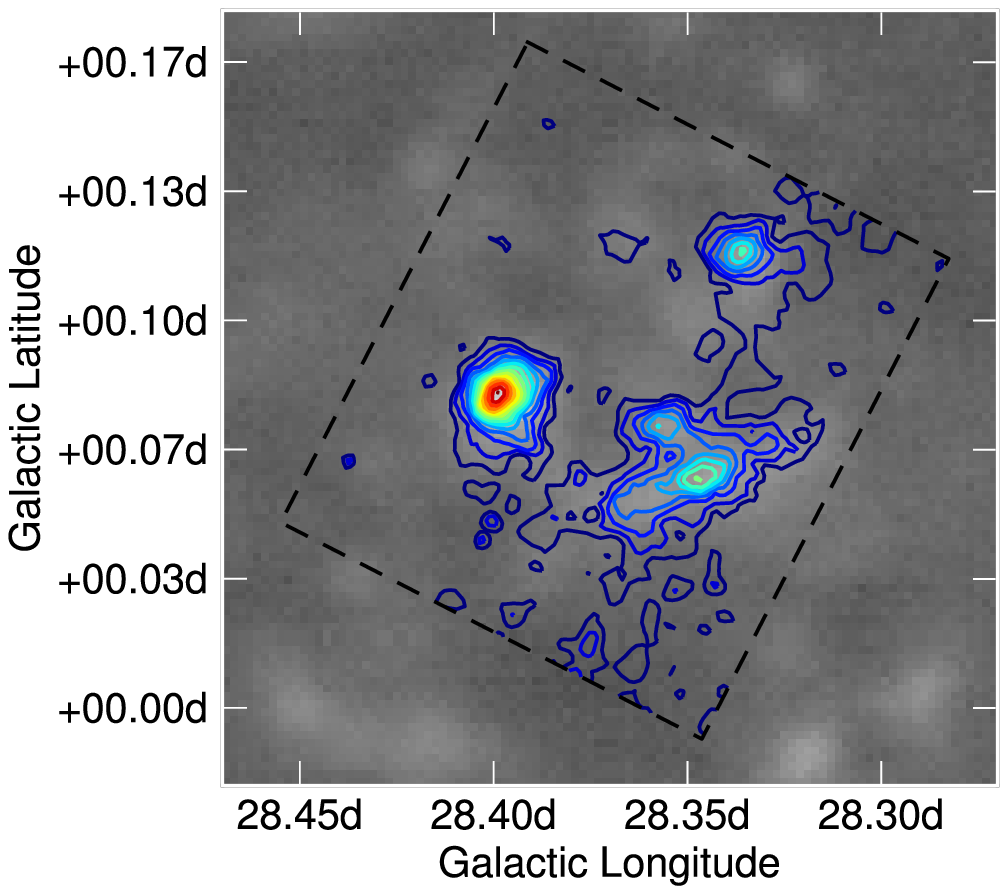}}\\
 \subfigure{
   \centering
    \label{fig:irdc1_apertures}
    \vspace{-60pt}
    \includegraphics[width=.9\textwidth]{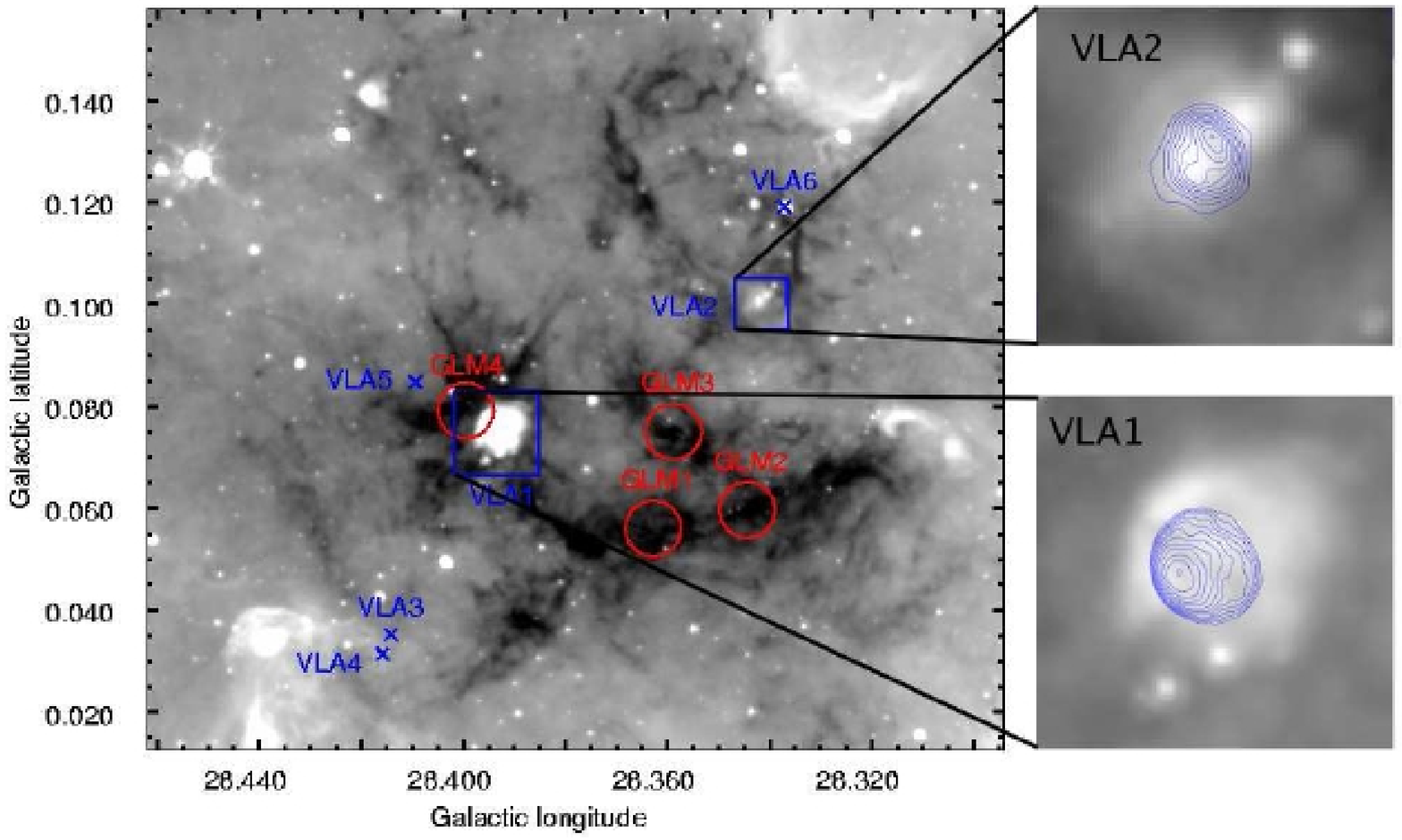}}\\
  \subfigure{
    \centering
    \label{fig:irdc1_allspectra}
    \vspace{-40pt}
    \includegraphics[scale=0.7]{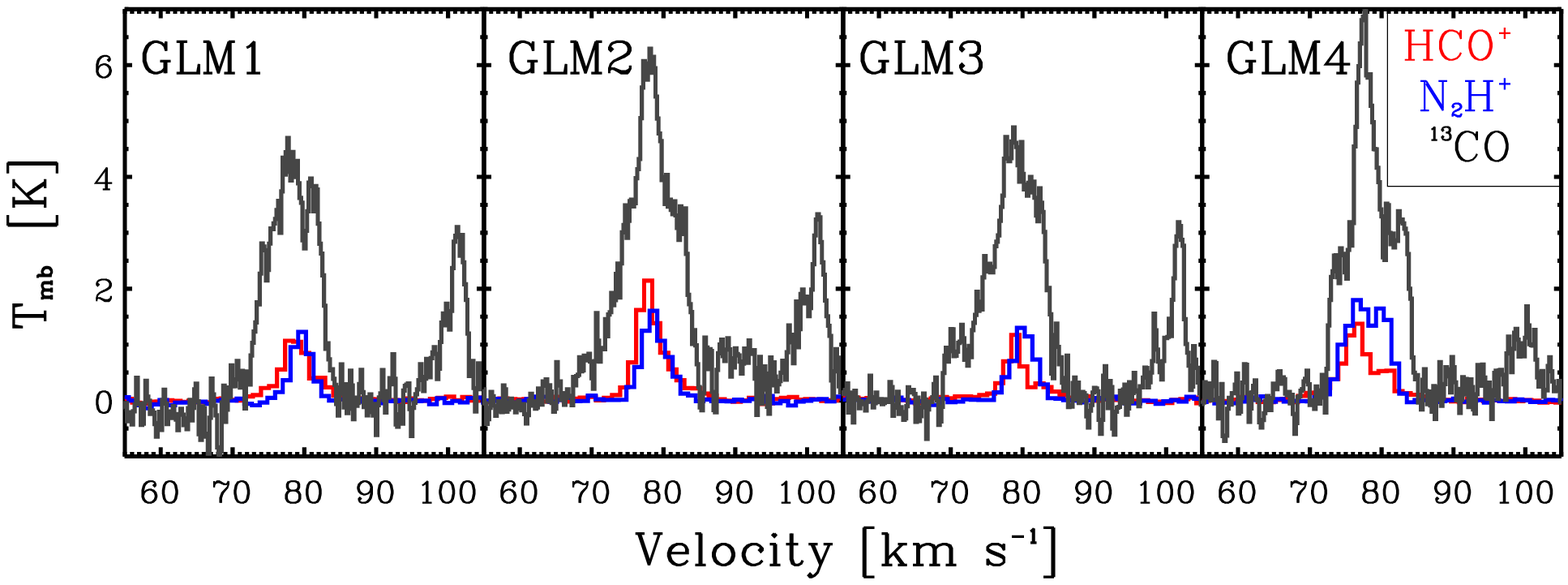}}
\vspace{-5pt}
 \caption{G028.37+00.07: 
    Top left: BGPS 1.1 mm continuum dust emission overlaid with linear \hcop~
    contours from 1.2 to 9 K \kms.  Top right: BGPS 1.1 mm continuum
    dust emission overlaid with linear \n2hp~contours from 0.8 to 
    10 K \kms. 
    Middle: GLIMPSE 8 $\mu$m 
    overplotted with the BGPS beam-sized apertures that were
    used to determine clump masses.  The blue X's are VLA 3.6 cm point
    sources, and extended 3.6 cm sources are depicted as blue contours in
    the adjacent boxes. Right:  \hcop, \n2hp~and \13CO
    spectra in clumps GLM1 (Stage 1), GLM2 (Stage 2), GLM3 (Stage 2), and
    GLM4 (Stage 2, slightly overlapping with Stage 4 ``diffuse red clump'').}
  \label{fig:irdc1_hcop_n2hp}
\end{figure*}

%\clearpage
\subsection{G028.23-00.19}
G028.23-00.19:~GLM1 (see Figures \ref{fig:irdc2_glm_mips} and \ref{fig:irdc2_aps_spec})
is the prime example of a starless IRDC clump.  It is typical of
an IRDC clump in size, mass, 8~\micron~extinction, and in having a
relatively compact millimeter core.  However, this clump has absolutely no
signs of star formation.  The detections of \hcop~and \n2hp~in this source
are incredibly weak.  The dust column is comparable to many other clumps
with bright \hcop~and \n2hp, the primary difference being the lack of
density and a heating source in G028.23-00.19:~GLM1.  The 8~\micron~point source
just north of GLM1 may be an indication of nearby, recent star formation.
Given the compact nature of GLM1, its high column density, and
nearby star formation, it seems likely that it is a quiescent
precursor to massive star formation.

\subsection{G028.37+00.07}
G028.37+00.07 (see Figures \ref{fig:irdc1_glm_mips} and \ref{fig:irdc1_hcop_n2hp})
is an extremely varied cloud.  It was identified by
\citet{sim06} as the darkest (by contrast) cloud in the First Galactic
quadrant.  Due to a particularly bright background around the cloud at 8
\micron~(as discussed in \S \ref{sec:extmasses}) the extinction
column density is over-estimated as compared with the BGPS 1.1 mm column
density.  This cloud is the most massive (by all tracers) in our sample.
G028.37+00.07 is composed of dense filaments and surrounded by an extremely bright
mid-IR background, including nearby evolved H~II regions. 

G028.37+00.07 hosts a prime example of a ``diffuse red clump.''  In Figure
\ref{fig:irdc1_glm3_iram}, we see a ``diffuse red clump,'' spatially
separated from the bright millimter peak.  The H~II region producing this
``red'' clump is asymmetric.  Away from the millimeter clump, the H~II
region has expanded and become more diffuse and toward the millimeter
clump the H~II region remains confined by the high-density.

The structure of \hcop~and \n2hp~in G028.37+00.07 is a great example of the
differing ratio of \hcop/\n2hp.  The
\hcop~and \n2hp~maps show a similar overall structure, but the intensities
of the two tracers vary across different clumps.  In \hcop, GLM2 is
the brightest, and in \n2hp, GLM4 is by far the brightest.  As pointed out
previously, however, this is not due to chemical differentiation.  Rather,
the \hcop~and even the \n2hp~become optically thick (and self-absorbed) in
the densest clump, GLM4.  The self-absorption of \hcop~makes \n2hp~appear
brighter by comparison.  

\citet{wan08} and \citet{zha09} observed G028.37+00.07: GLM2 and GLM4 (referred 
to as P1 and P2) with the VLA and SMA, respectively.  In
GLM2, they find 5 cores along the filament with line widths of about 1.2
\kms and temperature of about 15 K.  
In GLM4, they find 2 cores with line widths of about 4.3 \kms,
temperatures of about 30 K, and a rich molecular spectra.

%\clearpage
\begin{figure*} 
  \centering
  \subfigure{
    \label{fig:irdc30_glm3_iram}
    \includegraphics[width=0.46\textwidth]{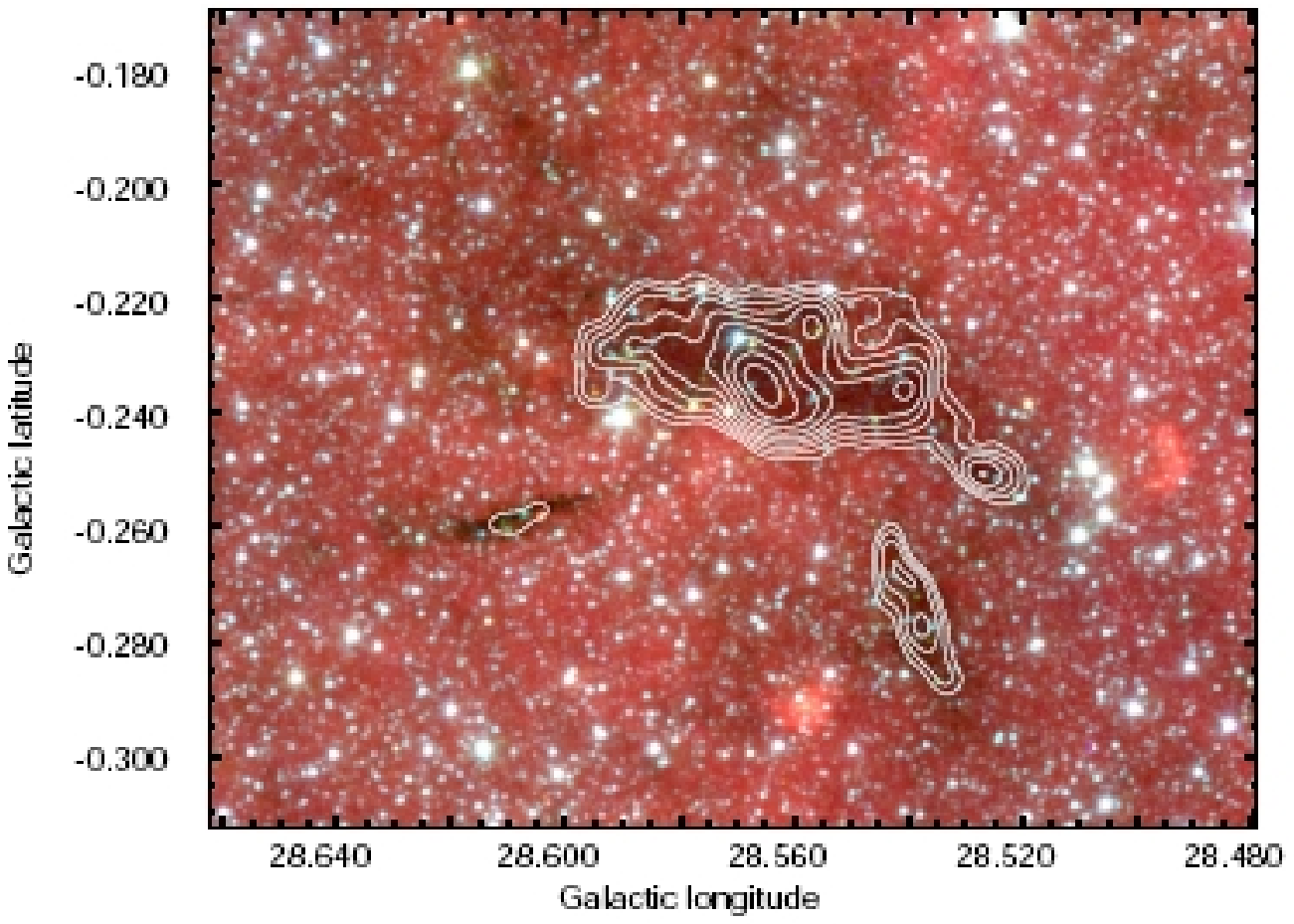}}
  \subfigure{
    \label{fig:irdc30_mips_bgps}
    \includegraphics[width=0.46\textwidth]{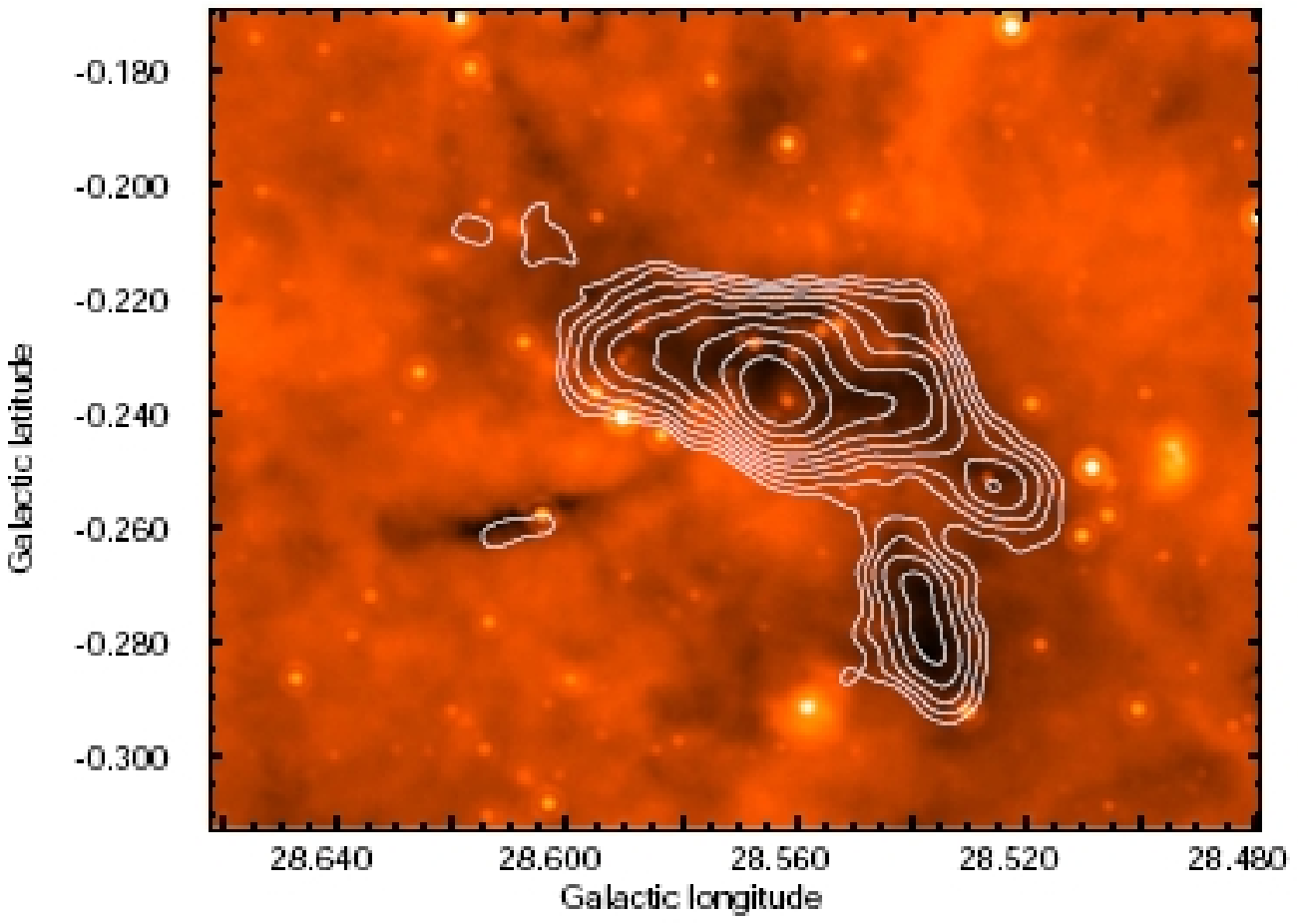}}\\
  \caption{G028.53-00.25: Left: GLIMPSE three-color image, red is 8~\micron, green is
    4.5 \micron~and blue is 3.6 \micron~with MAMBO 1.2 mm contours 
    overlaid.  The contours are on a log scale from 30 to 320 mJy beam$^{-1}$.
    Right: MIPSGAL 24 \micron~image with BGPS 1.1 mm contours.  The contours
    are on a log scale from 0.12 to 1.3 Jy beam$^{-1}$. }
  \label{fig:irdc30_glm_mips}
\end{figure*}
%\newpage
\begin{figure*}
  \centering
  \subfigure{
    \label{fig:irdc30_bgps_hcop}
   \includegraphics[width=.46\textwidth]{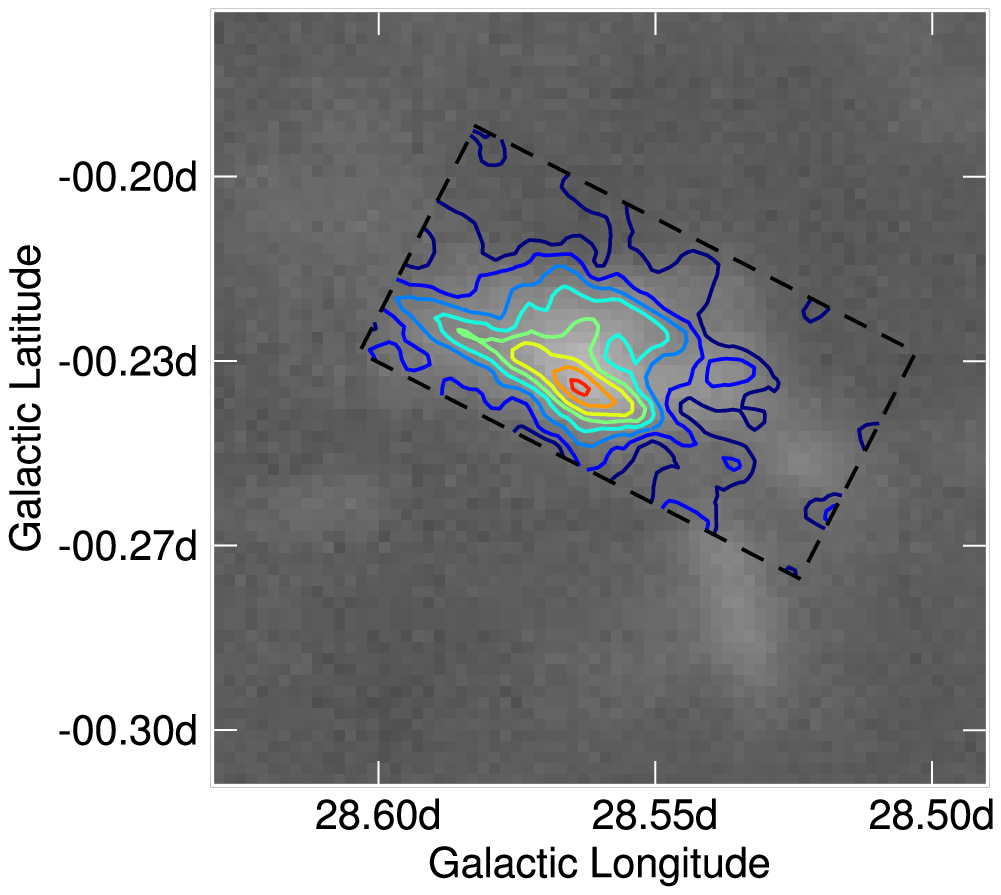}}
  \subfigure{
    \label{fig:irdc30_bgps_n2hp}
   \includegraphics[width=.46\textwidth]{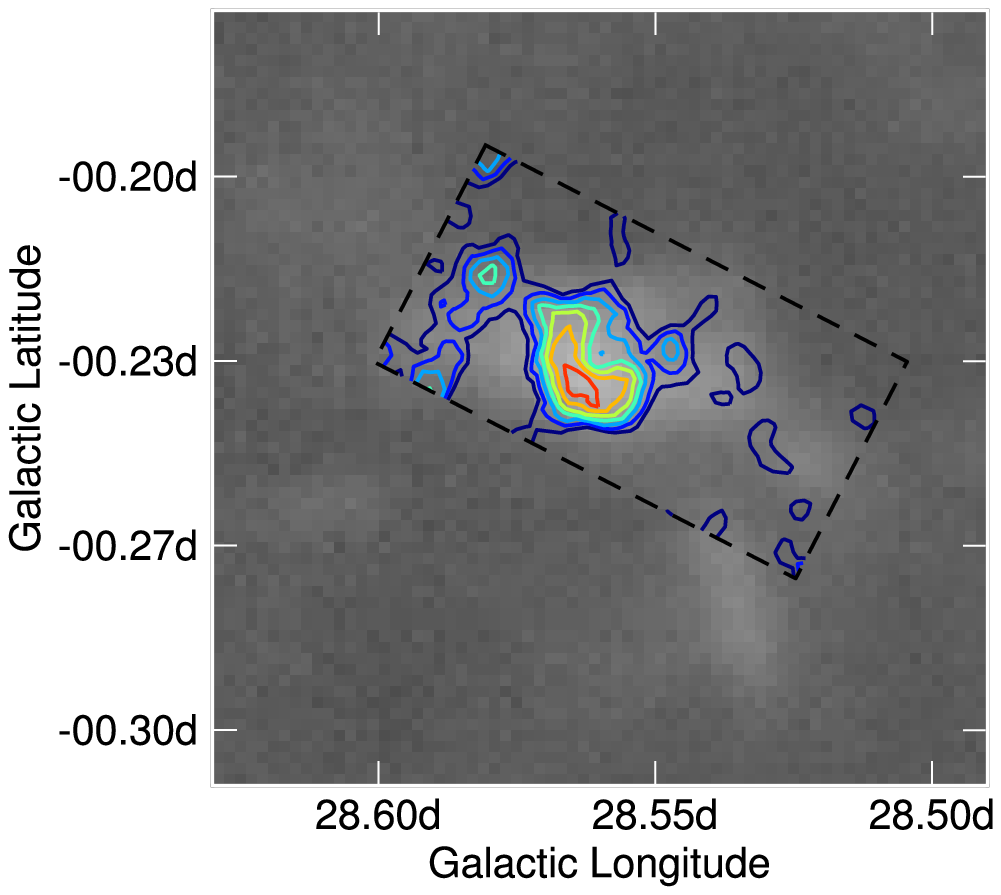}}\\
  \vspace{-.1in}
  \subfigure{
    \label{fig:irdc30_apertures}
    \includegraphics[width=.46\textwidth]{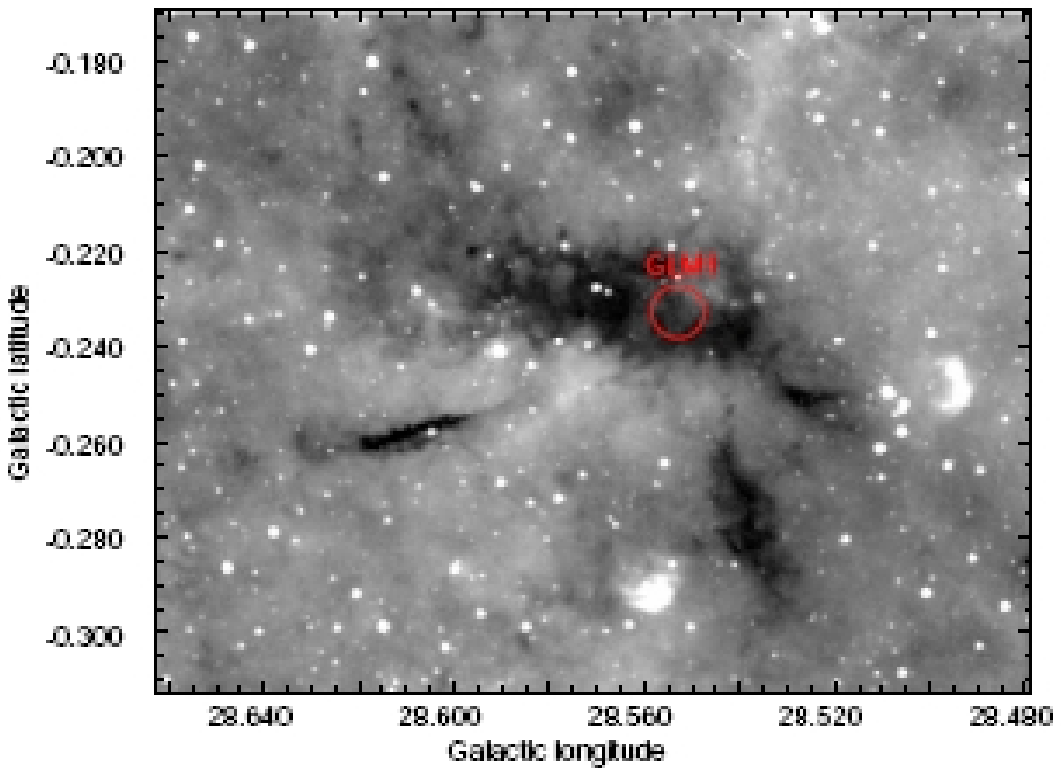}}
  \subfigure{
    \label{fig:irdc30_allspectra}
    \includegraphics[width=.46\textwidth]{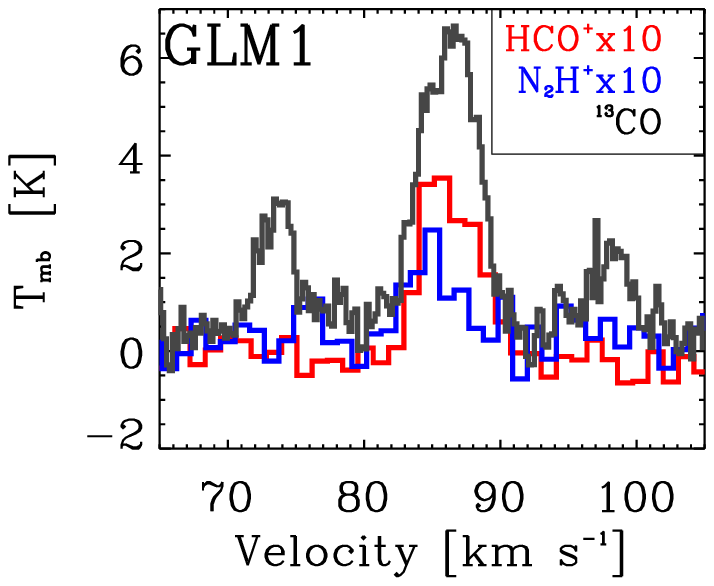}}
  \caption{G028.53-00.25: 
    Top left: BGPS 1.1 mm continuum dust emission overlaid with linear \hcop~
    contours from 0.43 to 7 K \kms.  Top right: BGPS 1.1 mm continuum
    dust emission overlaid with linear \n2hp~contours from 0.9 to 
    3.6 K \kms
    Bottom left: GLIMPSE 8 $\mu$m 
    overplotted with red BGPS beam-sized apertures that were
    used to determine clump masses.  Bottom right: \hcop, \n2hp~and \13CO
    spectra in clump GLM1 (Stage 2).  The \hcop~and \n2hp~spectra have been
    multiplied by a factor of ten, in order to see them more clearly
    relative to the \13CO.}
  \label{fig:irdc30_hcop_n2hp}
\end{figure*}

%\clearpage
\begin{figure*} 
  \centering
  \subfigure{
    \label{fig:irdc43_glm3_iram}
    \includegraphics[width=0.46\textwidth]{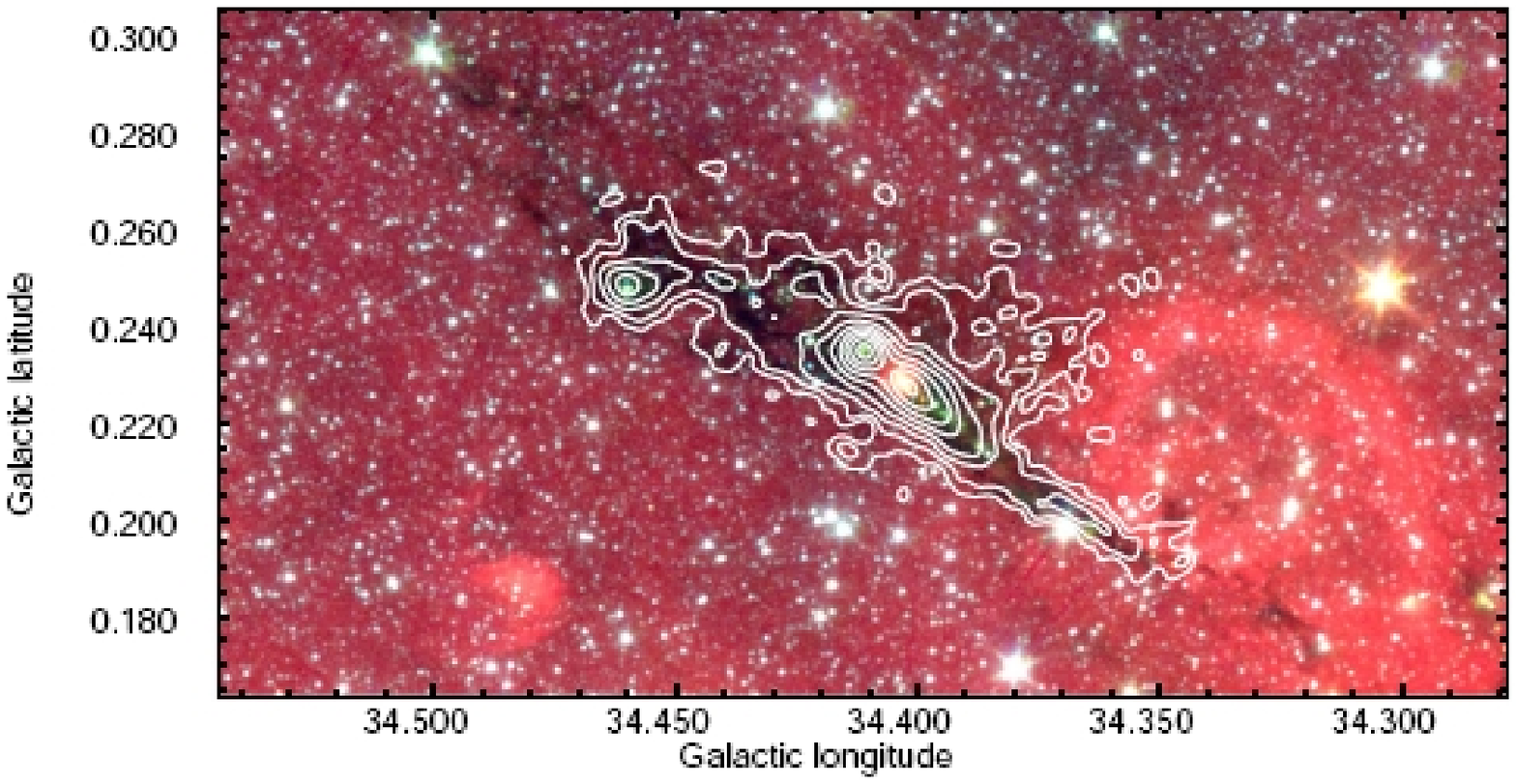}}
  \subfigure{
    \label{fig:irdc43_mips_bgps}
    \includegraphics[width=0.46\textwidth]{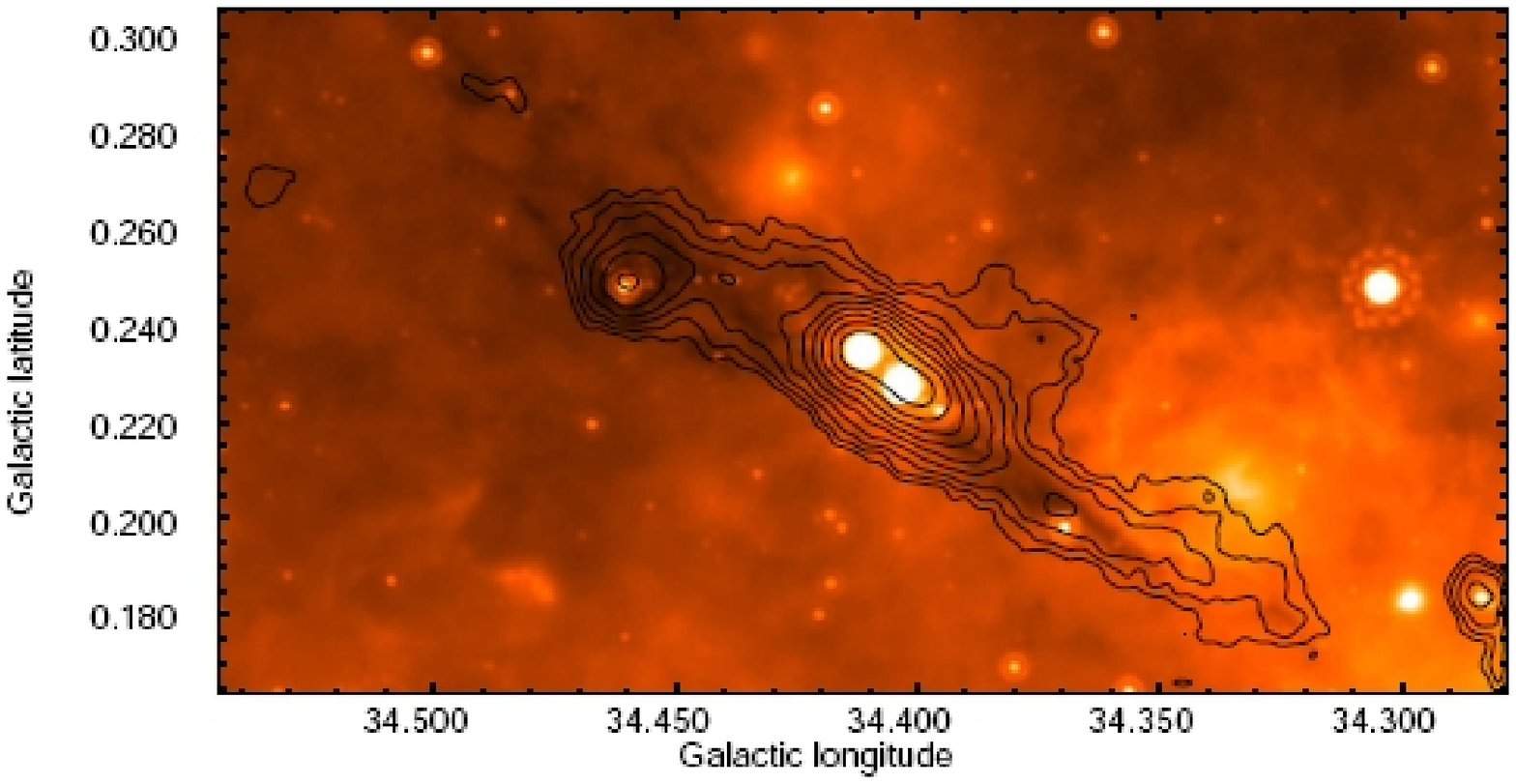}}\\
  \caption{G034.43+00.24: Left: GLIMPSE three-color image, red is 8~\micron, green is
    4.5 \micron~and blue is 3.6 \micron~with MAMBO 1.2 mm contours 
    overlaid.  The contours are on a log scale from 30 to 2400 mJy beam$^{-1}$.
    Right: MIPSGAL 24 \micron~image with BGPS 1.1 mm contours.  The contours
    are on a log scale from 0.12 to 4.3 Jy beam$^{-1}$. }
  \label{fig:irdc43_glm_mips}
\end{figure*}
%\newpage
\begin{figure*}
  \centering
  \subfigure{
    \label{fig:irdc43_bgps_hcop}
    \includegraphics[width=.46\textwidth]{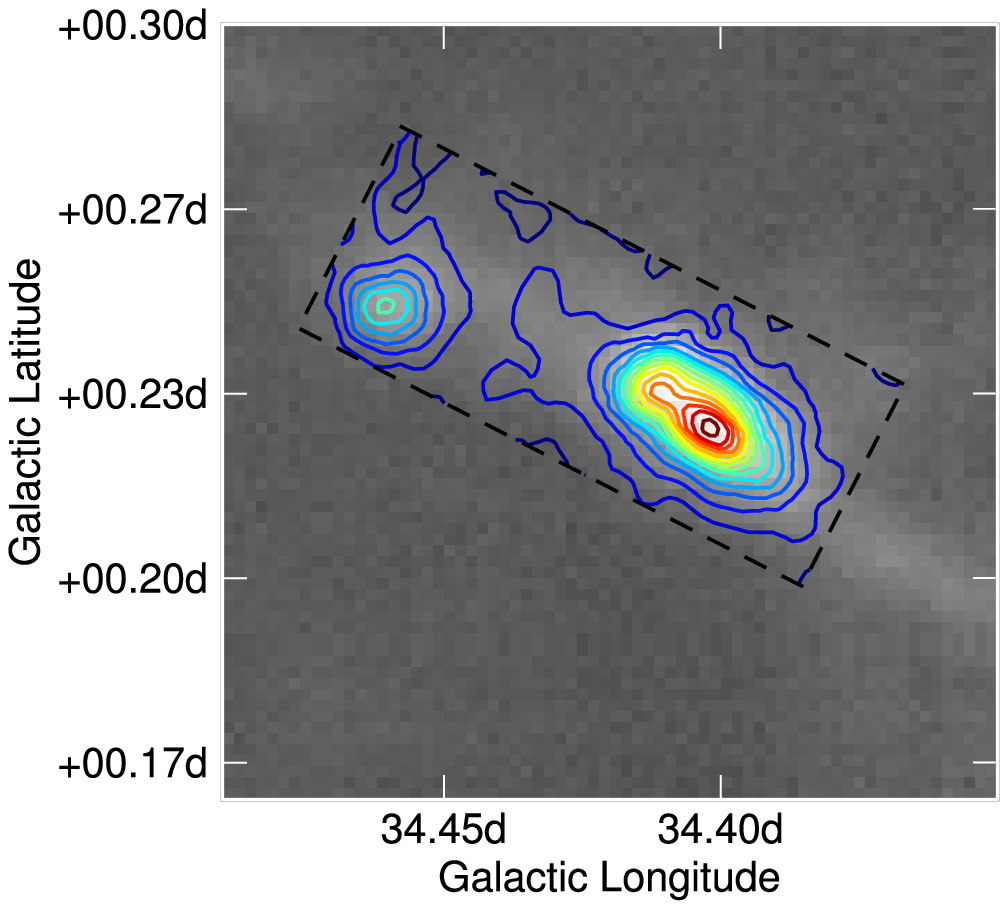}}
   \subfigure{
    \label{fig:irdc43_bgps_n2hp}
    \includegraphics[width=.46\textwidth]{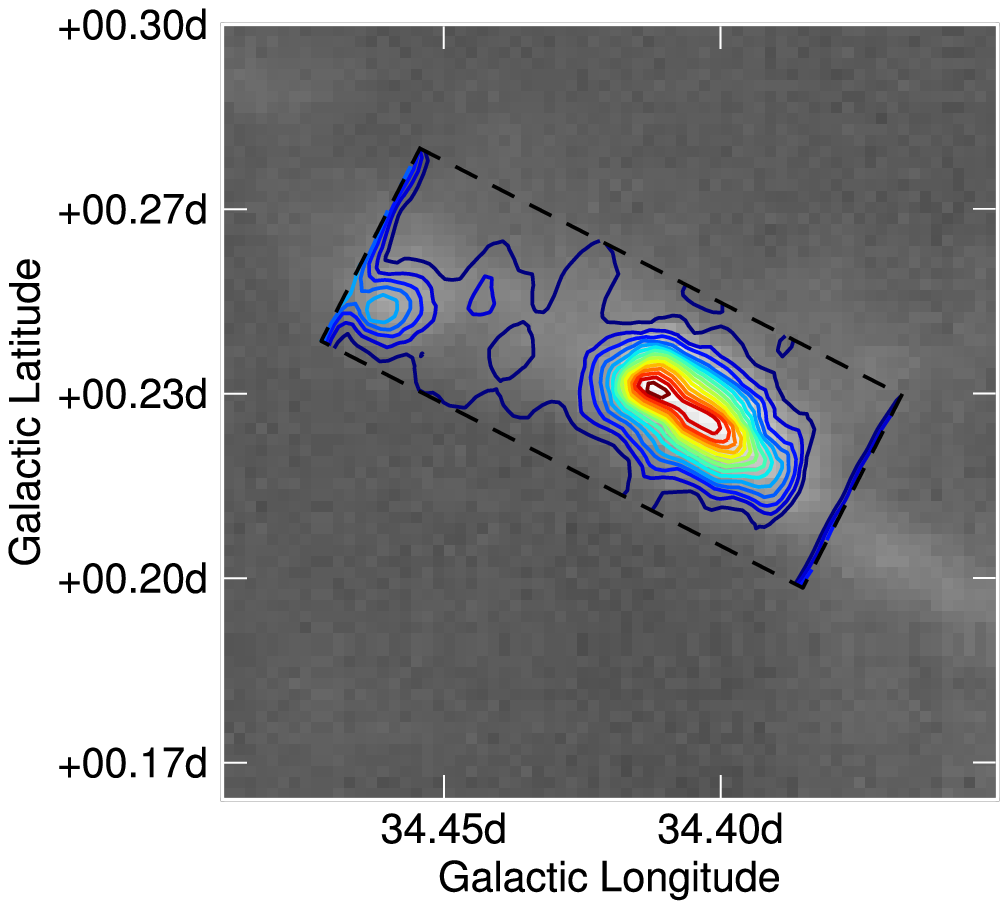}}\\
  \subfigure{
    \label{fig:irdc43_apertures}
    \includegraphics[width=1\textwidth]{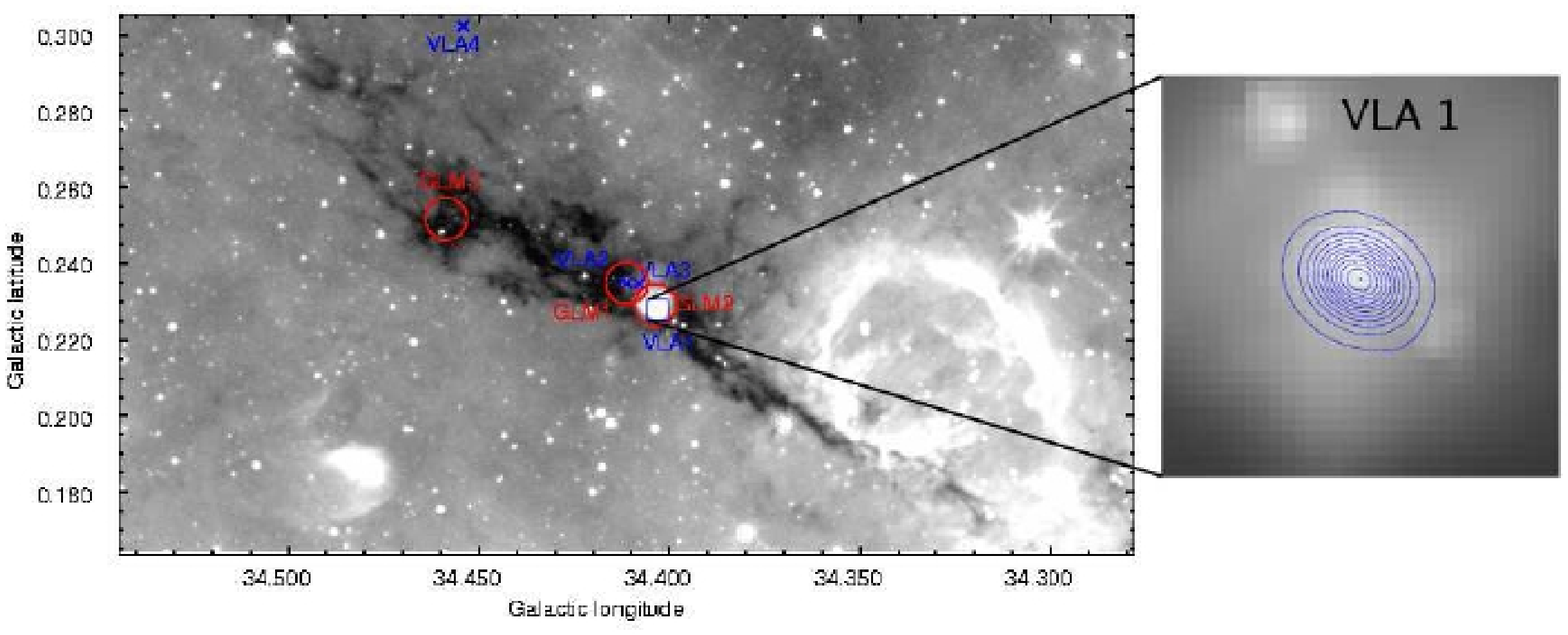}}
  \vspace{-.1in}
  \subfigure{
    \label{fig:irdc43_allspectra}
    \includegraphics[scale=0.8]{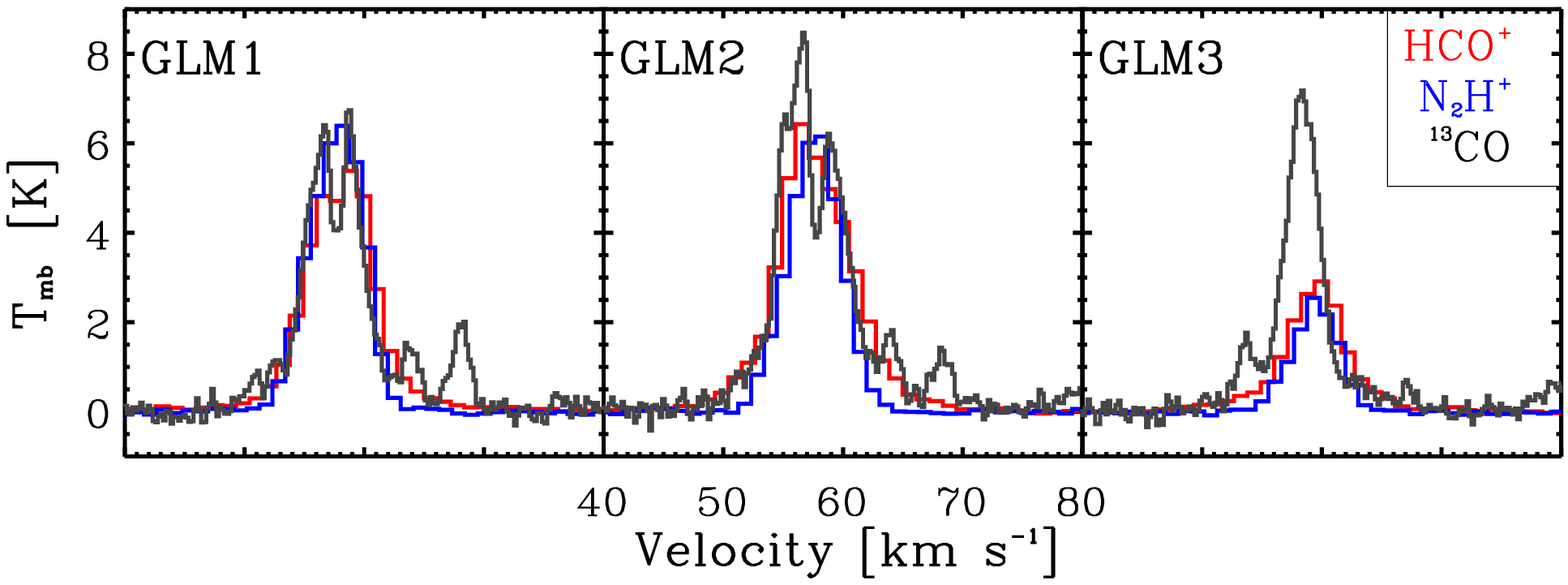}}
  \caption{G034.43+00.24: 
    Top left: BGPS 1.1 mm continuum dust emission overlaid with linear 
    \hcop~contours from 0.55 to 38 K \kms.  Top right: BGPS 1.1 mm continuum
    dust emission overlaid with linear \n2hp~contours from 0.48 to 
    27 K \kms
    Middle: GLIMPSE 8 $\mu$m 
    overplotted with the BGPS beam-sized apertures that were
    used to determine clump masses.  The blue X's are VLA 3.6 cm point
    sources, and extended 3.6 cm sources are depicted as blue contours in
    the adjacent box.  Right:  \hcop, \n2hp~and \13CO
    spectra in clumps GLM1 (Stage 3), GLM2 (Stage 4), and GLM3 (Stage 2).}
  \label{fig:irdc43_hcop_n2hp}
\end{figure*}

%\clearpage
\subsection{G028.53-00.25}
G028.53-00.25 (see Figures \ref{fig:irdc30_glm_mips}~and
\ref{fig:irdc30_hcop_n2hp}) is a curious example of a source with a very
small extinction column as compared with the BGPS 1.1 mm column.  This
cloud either has an exceptionally low background or high foreground.  The
region surrounding G028.53-00.25 is mostly devoid of extended H~II regions, PDRs,
and other strong emitters of diffuse 8~\micron~emission.  Due to the
paucity of nearby diffuse 8~\micron~emission, the background estimate is
decreased, and the extinction column estimate too low.  

G028.53-00.25 is a fairly quiescent cloud with few indicators of star formation.
The clump, GLM1, is identified as intermediate due to the presence of a
``green fuzzy" (though a fairly compact, faint one) and 24 \micron~point
source.  GLM1 shows very weak \hcop~and \n2hp~emission (the OTF-maps suffer
from low signal to noise).  G028.53-00.25:~GLM1 was observed 
on the Submillimeter Array by \citet{rat08} who found that GLM1 contains
at least three compact (less than $\sim$ 0.06 pc) cores.  
This quiescent cloud could be the host of low
mass star formation, or it could be a transient density enhancement or a 
quiescent proto-cluster. 

\subsection{G034.43+00.24}
G034.43+00.24 (see Figures \ref{fig:irdc43_glm_mips} and
\ref{fig:irdc43_hcop_n2hp}) is a stunning example of an active IRDC.  This
object is just northeast of the enormously bright H~II region at G34.3+0.2,
RH85 \citep{rei85}.  This complex contains the bright H~II region, RH85,
and the shells of supernova remnants.  G034.43+00.24, along with a few other
IRDCs not discussed in the paper, are
extremely filamentary structures, streaming radially out of this 
bright H~II region complex, increasing in width with
distance from the complex.  This is an excellent source in which to 
study the triggered formation mechanism of IRDCs \citep[e.g][]{red03}.

South of G034.43+00.24:~GLM2 is long train of ``green fuzzies,'' along the IRDC.
\citet{rat08} observed G034.43+00.24:~GLM1 with the Submillimeter Array 
and found that GLM1 remains unresolved and contains a core smaller than 
the $\sim$ 0.03 pc beam.  GLM2 itself is an excellent example of a 
``diffuse red clump,'' offset
from the millimeter peak.  The
bright 8~\micron~emission nullifies the use of an extinction mass in this
clump; the extinction mass is 60 \Msun~while the BGPS 1.1 mm mass is 780
\Msun.  This clump is one of the most ``massive'' sources in our sample
according to the BGPS 1.1 mm mass, which is almost certainly biased by a higher
dust temperature than the assumed 15 K.  This spectacular clump is an example where all of
our conventional mass tracers fail.  The \hcop~and \n2hp~are extremely
bright in all the clumps, especially GLM1 and GLM2 (the brightest by far in
our sample).  There is also strong
self-absorption in \13CO toward these clumps, further indication of the
high density and temperatures of these active clumps.

\end{document}